\documentclass[aps,prd,twocolumn,nofootinbib,preprintnumbers,superscriptaddresss,showpacs,amsmath,amssymb]{revtex4-1}
\usepackage{graphicx}
\usepackage{amsmath,amssymb}
\usepackage{amsfonts}
\usepackage{blindtext}
\usepackage{xspace} 
\usepackage[usenames]{color}
\usepackage{booktabs,chemformula}
\usepackage{dcolumn}
\usepackage{bm}
\usepackage{mathrsfs}
\usepackage[colorlinks=true]{hyperref}
\usepackage[all]{hypcap} 
\usepackage[utf8]{inputenc} 
\usepackage{slashed}

\begin{document}
%
%
%
%-----------------------Title-----------------------------%

\title{Bosonic Dark Matter in Light of the NICER Precise Mass-Radius Measurements}

\author{Soroush Shakeri$^{1,2,3}$} \email{s.shakeri@iut.ac.ir}  
\author{Davood Rafiei Karkevandi$^{1,3}$} \email{d.rafiei@alumni.iut.ac.ir} 

\affiliation{$^1$ Department of Physics, Isfahan University of Technology, Isfahan 84156-83111, Iran}
\affiliation{$^2$ICRANet, Piazza della Repubblica 10, I-65122 Pescara, Italy}
\affiliation{$^3$ ICRANet-Isfahan, Isfahan University of Technology, 84156-83111, Iran}

\date{\today}
 \begin{abstract}
We explore the presence of self-interacting bosonic dark matter (DM) within neutron stars (NSs) in  light of the latest multi-messenger observations of the Neutron Star Interior Composition Explorer (NICER) and LIGO/Virgo detectors. The bosonic DM is distributed as a core inside  the NS or as a halo around it  leading to formation of a DM admixed NS. We  focus on the variation of the visible and dark radius of the mixed object due to DM model parameters and fractions. It is shown that DM core formation reduces the visible radius and the total mass pushing them below observational limits while halo formation is in favor of the latest mass-radius observations.  Moreover, we scan over  the parameter space of the bosonic DM model considering two nuclear matter equation of states by applying the radius, maximum mass and tidal deformability constraints. Our investigation allows for the exclusion of a range of DM fractions, self-coupling constant  and sub-GeV boson masses, which limits the amount of accumulated DM to  relatively low values to be consistent with astrophysical bounds. In this paper, we introduce main features of the pulse profile corresponding to the DM admixed NS as a novel observable quantity. We find that the depth of  minimum fluxes in the pulse profiles crucially depends on the amount of DM  around NS and its compactness. The current/future astrophysics missions  may test the possibility of the existence of DM within NSs and break the degeneracies between different scenarios via multiple observations.

\end{abstract}

\maketitle

\section{{\bf Introduction}} \label{sec:intro}

The  advent of multi-messenger observations of compact astrophysical objects has provided a  unique opportunity to  explore the dense matter equation of state (EOS) and to answer the question, whether neutron stars (NSs) are composed  of normal nuclear matter, or they contain more exotic type of matter. In this regard, precise measurements of  mass, radius and tidal deformability of NSs by electromagnetic (EM) and gravitational waves (GWs), have led to remarkable improvements in our understanding of their interior structures \cite{Most:2018hfd,Fattoyev:2017jql,Annala:2017llu,Rezzolla:2017aly}. Among different possibilities, dark matter (DM)  either as  a bosonic or fermionic particle could be mixed with ordinary matter leading to a new type of compact object so-called DM admixed NS \cite{Tolos:2015qra,Ellis:2018bkr,Nelson:2018xtr,PhysRevD.102.063028,Dengler:2021qcq,Karkevandi:2021ygv}. Depending on the DM model parameters, i.e. mass, self-coupling constant and its fraction within the mixed star, DM could reside as a dense core inside NS or form an extended halo around it. 
The distribution of DM in the DM admixed NS causes some noticeable impacts on the mass-radius profile and the deformation of this configuration in a binary system, which encodes in the tidal deformability parameter \cite{Karkevandi:2021ygv,Karkevandi:2021ske,Collier:2022cpr,Lourenco:2022fmf,Rutherford:2022xeb,Hippert:2022snq,Giangrandi:2022wht,Das:2021hnk,Shirke:2023ktu,Vilhena:2023fic,Thakur:2023aqm,Bramante:2023djs,Routaray:2023spb,Parmar:2023zlg,Zollner:2023myk,Husain:2021hrx,Cronin:2023xzc,Rezaei:2023iif}.  

Among different DM candidates the
scalar or pseudoscalar bosons such as axions are of great interest from various aspects in astrophysics and cosmology \cite{Marsh:2015xka,Shakeri:2022usk,Lopes:2022efy,Shahrbaf:2022upc,Shahrbaf:2023uxy}. In this regard, self-gravitating objects made of bosonic DM could be formed whether as non self-interacting scalar fields or with repulsive/attractive self-interactions \cite{Agnihotri:2008zf,Eby:2015hsq,Wystub:2021qrn,Visinelli:2021uve,Chavanis:2022fvh,Pitz:2023ejc,Angulo:2022gpj}. In the absence of interaction between bosons, the Heisenberg uncertainty principle  provides the required quantum pressure to avoid from a gravitational collapse.  The stable self-gravitating configurations of  complex scalar fields called boson stars (BSs) were first introduced by Kaup and Ruffini-Bonazzola in the 1960s \cite{Kaup:1968zz,Ruffini:1969qy}.  Afterwards, Colpi et al. considered a repulsive self-interaction among bosons which alters the mass and radius of BSs drastically and become comparable to the ones of  NSs \cite{Colpi:1986ye}. Following these efforts, the
attention has been directed to the possible role of BSs in astrophysics  whether as a DM candidate \cite{Jetzer:1991jr,PhysRevD.68.023511,AmaroSeoane:2010qx,Liebling:2012fv,Chavanis:2019bnu}  or as black hole mimickers  \cite{Khlopov:1985jw,Mielke:2000mh,Guzman:2009zz,Bustillo:2020syj,Chavanis:2019amr}.

Asymmetric Dark Matter (ADM)  which usually do not annihilate to standard model (SM) particles  can be substantially accumulated in NSs through different processes \cite{Kouvaris:2010jy,Kouvaris:2011fi,Li_2012,Nelson:2018xtr,PhysRevD.102.063028,DelPopolo:2019nng,universe6120222,Ciancarella:2020msu,Berezhiani:2020zck,Bell:2020obw, Kalashnikov:2022txs,Liang:2023nvo,Rezaei:2016zje,Alvarez:2023fjj,Nguyen:2022zwb,Deliyergiyev:2023uer,Avila:2023rzj}.  Accretion of bosonic ADM by stars has been discussed by nonlinear numerical relativity methods, where it was showed that bosonic cores may  develop inside the star and these configurations are stable throughout most of the parameter space \cite{Brito:2015yga,Brito:2015yfh}. We may  also have gravitationally stable configurations composed entirely of fermionic or bosonic ADM named ``dark stars" \cite{Narain:2006kx,Kouvaris:2015rea,Eby:2015hsq,Maselli:2017vfi,Visinelli:2017ooc,Chavanis:2017loo,Cassing:2022tnn,Croon:2022tmr}. ADM stars  could also serve as DM cores and accretion centers for baryonic matter (BM) which may lead to the formation of a DM admixed NS with high fractions of DM \cite{Goldman:2013qla,Xiang:2013xwa,Ellis:2017jgp,Ellis:2018bkr}. Furthermore, a large amount of DM may be provided in a binary system of a dark compact object and a NS, by capturing of DM particles  or the coalescence of these objects \cite{Sandin_2009,Ciarcelluti:2010ji,Giudice:2016zpa,Gresham:2018rqo,Dietrich:2018jov}. 

 On the other hand, the existence of self-annihilating DM in NS lead to some evidences such as the variation of luminosity,  effective temperature and cooling of the mixed object \cite{Kouvaris:2007ay,deLavallaz:2010wp,AngelesPerez-Garcia:2022qzs,Acevedo:2019agu,Joglekar:2019vzy,Joglekar:2020liw,Garani:2020wge,Bramante:2021dyx,Coffey:2022eav}.  These observable effects could potentially be
detected by James Webb Space Telescope (JWST) and other infrared/optical/UV telescopes specially  for old NSs which are located in solar neighborhood    \cite{McKeen:2020oyr,McKeen:2021jbh,Baryakhtar:2017dbj,Raj:2017wrv,Baryakhtar:2022hbu,Chatterjee:2022dhp}.

Astrophysical objects which are a mixture of fermions and bosons,  interacting only gravitationaly, have been discussed based on two different methods: i) fermion-boson star \cite{HENRIQUES198999,Henriques:1989ez,HENRIQUES1990511,Valdez-Alvarado:2012rct,Valdez-Alvarado:2020vqa,DiGiovanni:2020frc,Nyhan:2022pda,Diedrichs:2023trk,Jockel:2023rrm} and ii) DM admixed NS \cite{Ellis:2018bkr,Nelson:2018xtr,Karkevandi:2021ygv,Rutherford:2022xeb,Giangrandi:2022wht}. In both above cases,  the energy momentum tensor of the  Einstein equation takes  separate contributions of fermionic and bosonic components. For the former, fermionic component is described by a perfect-fluid while the scalar bosonic field is governed by Klein–Gordon equation.  In order to compute equilibrium configurations of  fermion-boson stars, one needs to solve the coupled Einstein-Klein–Gordon equations \cite{Liebling:2012fv}. In the DM admixed NSs, fermions and bosons are considered as  perfect fluids  where the two-fluid formalism of Tolman-Oppenheimer-Volkof (TOV) equations are utilized to  describe the mixed compact star. 
Recently, the equilibrium configurations of DM admixed NSs have been extensively considered in   \cite{Karkevandi:2021ygv,Karkevandi:2021ske}, where the DM  component was modeled by self-repulsive bosons. 
In  a related study for the fermion-boson stars  \cite{DiGiovanni:2021ejn}, which was done following a similar scheme  in \cite{Karkevandi:2021ygv} but with a vanishing   bosonic  self-interaction,
the dependence of DM core/halo formation on the particle mass  is in agreement with the results presented in \cite{Karkevandi:2021ygv}.

During the last decade the measurement  techniques of the mass and radius of
NSs have been much improved.  The Shapiro delay method or orbital period measurements of binary systems  determine the gravitational mass, such as  the measured mass of the heaviest NSs, PSR J0348+0432 ($2.01\pm 0.04M_{\odot}$) \cite{Antoniadis:2013pzd} and PSR J0740+6620 ($2.14^{+0.10}_{-0.09}M_{\odot}$)   \cite{NANOGrav:2019jur}. In this direction, the Keck-telescope optical spectrophotometry and imaging of the companions of PSR J1810+1744 \cite{Romani:2021xmb} and PSR J0952-0607 \cite{Romani:2022jhd} provide novel mass results  $2.13\pm 0.04M_{\odot}$ and $2.35\pm 0.17M_{\odot}$, respectively. Furthermore, the GW data of LIGO-Virgo collaboration leads to the determination of tidal deformability of $1.4M_{\odot}$ NS ($\Lambda_{1.4}$)  to be less than 580 for GW170817 event \cite{Abbott_2017,Abbott:2018exr}. This measurement enables us to constrain the corresponding radius, hereafter called $R_{1.4}$ \cite{Fattoyev:2017jql,Most:2018hfd,Annala:2017llu,Tews:2018iwm,De:2018uhw,Radice:2018ozg,Coughlin:2018fis,Capano:2019eae,Essick:2019ldf,Breschi:2021tbm,Gamba:2020wgg,Pang:2022rzc}. 
 Afterwards, the Neutron Star Interior Composition Explorer (NICER) announced the joint precise measurements of  mass and radius  for  PSR J0030+0451 and  PSR J0740+6620
\cite{Miller:2019cac,Riley:2019yda,Miller:2021qha,Riley:2021pdl}, which opens a new window towards better understanding of NSs structure. Ever since, many attempts have been made to impose limits on dense matter EoSs and NS properties specially $R_{1.4}$  via combined analyses of GW, X-ray observations and nuclear theory or experiments \cite{Raaijmakers:2019qny,Raaijmakers:2019dks,Raaijmakers:2021uju,Bogdanov:2019ixe,Dietrich:2020efo,Reed:2021nqk,Pang:2021jta,Al-Mamun:2020vzu}. Considering the aforementioned results from multi-messenger observation of NSs, gives the value of $R_{1.4}$ to be $11.75^{+0.86}_{-0.81}\, \text{km} $ at $90\%$ confidence level at $1\sigma$ uncertainty \cite{Dietrich:2020efo} or more recent work indicating that $R_{1.4}=12.01^{+0.78}_{-0.77}\, \text{km} $ at $95\%$ confidence level \cite{Huth:2021bsp}.  Moreover, there is an estimation for the radius of a typical $1.4M_{\odot}$ NS to be $11.94^{+0.76}_{-0.87}$ at 90\% confidence \cite{Pang:2021jta} and in \cite{Breschi:2021tbm}, an average over all current estimations  has been done, giving $R_{1.4}=12.0^{+1.2}_{-1.2}$  with the 90\% credible region.

In this paper, we mainly focus on the  precise radius measurements in addition to the maximum mass and tidal deformability parameters to probe  DM admixed NSs. We take the  maximum mass constraint $M_{max}\gtrsim 2M_{\odot}$,  
$R_{1.4}\gtrsim11 $ km and   $\Lambda_{1.4}\lesssim 580$  as  reliable limits for NSs composed of a fraction of  bosonic self-interacting  DM (SIDM). In the following, we investigate the effect of DM core/halo formation on the  visible and dark radius of the mixed object in the light of the latest multi-messenger constraints.  We examine different DM model parameters such as the particle mass, self-coupling constant and also the amount of DM within the mixed NS in order to check their consistency with the allowed parameter space given by NICER and GW detectors. We  show that the existence of bosonic DM as a core of a compact object leads to a reduction in the  radius and mass of the mixed star which is disfavored by recent measurements. However, the DM halo formation increases the tidal deformability  significantly, which is not consistent with the latest constraint. Considering the aforementioned triple bounds, the allowed DM model parameter space  will be obtained through precise scans over coupling constant, DM fraction and boson mass.  This  also impose upper limits on DM fraction in NSs for sub-GeV bosonic SIDM  applying two different stiff and soft BM EoSs.

In fact, the precise mass-radius measurements by X-ray telescopes (e.g.  NICER/XMM-Newton) rely on the surface emission 
of the star. Depending on the compactness of a NS,  the path of the photons is distorted by the gravitational warping of spacetime in the vicinity of the object, leading to some modifications in the visibility of hot spots  and the corresponding pulse profiles \citep{Beloborodov:2002mr,Turolla:2013tba,Ozel:2015ykl,Miao:2022rqj}. As a new observable, we  extensively consider the variation of the EM pulse profile in a NS with a DM halo around it for different DM model parameters. We present a detailed analysis of Pulse Profile Modeling (PPM)  taking into account different amount of DM, boson masses and coupling constants. Moreover, the compactness of DM admixed NSs, which is a crucial parameter for PPM, has been investigated for a wide range of boson masses considering various DM fractions and self-coupling constants. Our results demonstrate the necessity of inclusion of the DM admixed NS among other possibilities  in order to obtain posterior distributions of samples during the analysis of NICER/XMM-Newton.

This paper is organized as follows. In Sec.~\ref{sec2} we present a framework to  model the DM admixed NS based on two-fluid TOV formalism which includes  two EoSs describing bosonic DM and BM components.  The equilibrium configurations of the mixed compact object through the mass-radius profiles are compared
 with the latest measurements of NICER ( Sec.~\ref{SEC3}). We consider the variation of the visible and dark radius of DM admixed NSs under the influence of the boson mass and self-interaction
strength as a function of DM fractions in Sec.~\ref{sec4}. The pulse profile as a new observable will be introduced in Sec.~\ref{sec5} where the effect of DM halo will be taken into account in light bending  in order to track the surface emission of a star with a baryonic core surrounded
by a DM halo. In Sec.~\ref{sec6}  a scan over  bosonic DM mass, DM fraction and coupling constant is presented  regarding three key observable features $M_{max}\gtrsim 2M_{\odot}$, $R_{1.4}\gtrsim11$ km and  $\Lambda_{1.4}\lesssim 580$.  Finally, in the last
section, we summarize our results.  In this paper we use units in which $\hbar=c=G=1$.

\section{{\bf Modeling of the DM admixed NS}} \label{sec2}
In this work to model DM, we use bosonic SIDM which  is described through the following  Lagrangian 
\begin{eqnarray}
\mathcal{L}=\frac{1}{2}\partial_\mu\phi^*\partial^\mu\phi
-\frac{m_\chi^2}{2}\phi^*\phi-\frac{\lambda}{4}(\phi^*\phi)^2,
\end{eqnarray}
we assume mean-field approximation and  expand the interaction term $\lambda(\phi^*\phi)^2/4$   in terms of $\phi^*\phi-\langle\phi^*\phi\rangle$ which leads to a linearized mean-field Lagrangian as
\begin{eqnarray}\label{MF}
\mathcal{L}_{MF}
=\frac{1}{2}\partial_\mu\phi^*\partial^\mu\phi-\frac{m_\chi^{*2}}{2}\phi^*\phi+\frac{\lambda}{4}\langle\phi^*\phi\rangle^2, 
\end{eqnarray}
as a result the effective mass $m_\chi^{*}$ depends on the expectation value of $\phi^*\phi$ as  $m_\chi^{*2}=m_\chi^2+\lambda\langle\phi^*\phi\rangle$. 

The  pressure arise from the DM perfect fluid described by this Lagrangian is

\begin{eqnarray}
P=\zeta^2(\mu_\chi^2-m_\chi^{*2})+\frac{\lambda}{4}\langle\phi^*\phi\rangle^2 \ ,
\end{eqnarray} 
where  $\zeta$ represents the amplitude of zero mode and $\mu_\chi$ is bosonic chemical potential. After some straightforward computations presented previously in Appendix of \cite{Karkevandi:2021ygv}, it turns out that the total pressure of SIDM particles is given by

\begin{equation}\label{e1}
P=\frac{m_{\chi}^{4}}{9\lambda}\left( \sqrt{1+\frac{3\lambda}{m_{\chi}^{4}}\rho}-1\right)^{2}.
\end{equation}
 This EoS has been originally obtained in \cite{Colpi:1986ye}  for BSs in an alternative way by solving Klein-Gordon-Einstein equations for a spherically symmetric configuration of the scalar fields. 
In the strong-coupling limit  $\lambda \gg 4\pi (m_{\chi}/M_{Pl})^{2}$  applied in \cite{Colpi:1986ye}, the spatial derivatives  of the scalar fields have been neglected leading to  a local solution of  field equations. In this regime, the system can be approximated as a perfect fluid and the anisotropy of pressure is ignored. The maximum mass of a BS  with the quartic self-interaction was found  to be
\begin{eqnarray}\label{emax}
M_{\text{max}}^{\text{BS}} \approx 10\, M_{\odot} \lambda^{1/2}  \left( \frac{100\,\text{MeV}}{m_{\chi}} \right)^{2} ,
\end{eqnarray}
it was shown \cite{AmaroSeoane:2010qx,Chavanis:2011cz,Chavanis:2021tva,Karkevandi:2021ske} that the  maximum compactness $\mathcal{C}_{max}=(M/R)_{max}$  grows with the coupling constant $\lambda$  and reaches to a saturated value 0.16 at strong coupling limit. Working in this regime,  we can find the minimum radius $R_{\text{min}}$ corresponding to  $M_{\text{max}}$  as 

\begin{eqnarray}\label{emax}
R_{min}^{\text{BS}} \approx 92.3 \, \text{km} \, \lambda^{1/2}  \left( \frac{100\,\text{MeV}}{m_{\chi}} \right)^{2} ,
\end{eqnarray}

To model the BM component, we utilize two reliable hadronic EoSs. One of them includes induced surface tension known as  IST  where both the short-range repulsion and long-range attraction interactions between baryons have been considered \cite{Sagun:2018cpi,Sagun11:2018sps}. This EoS is in agreement with the latest NS observations providing $M_{\text{max}}=2.08 M_{\odot}$ and the radius $11.37$ km, and tidal deformability $285$  for a  $1.4 M_{\odot}$  NS \cite{Sagun:2020qvc}. The corresponding crust part  of this EoS  is described via the polytropic EoS  with adiabatic index $\gamma=4/3$ \cite{PhysRevD.102.063028}. Moreover, we use  another nuclear EoS  obtain from the relativistic mean-field  model with density-dependent couplings called DD2 \cite{Typel:1999yq,Typel:2005ba}. The density dependence of the couplings captures the essential properties of atomic nuclei and nuclear matter
around saturation density  \cite{Klahn:2006ir}. The interactions are described as baryon-meson couplings taking into account  $\sigma$, $\omega$, and $\rho$ mesons  as degrees of freedom. In order to model the outer crust  BPS EoS \cite{Baym:1971pw} is applied, and the inner crust is gained by  Thomas-Fermi  approach \cite{Grill:2014aea,Fortin:2016hny}. A NS with  $M_{max}=2.4M_\odot$, $R_{1.4}=13.15$ km and $\Lambda_{1.4}=681$ is reproduced by this stiff EoS which  fulfill the lower bounds for the maximum mass and radius of NSs even when hyperons are included \cite{Shahrbaf:2022upc,Shahrbaf:2023uxy}.
Regarding the uncertainties governing nuclear EoSs, we select soft IST and stiff DD2 EoSs which represent different boundary values of maximum mass, radius and tidal deformability.

In order to model DM admixed NSs,  we assume an energy-momentum tensor composed of BM and DM fluids $T^{\mu \nu}=T^{\mu \nu}_{DM}+T^{\mu \nu}_{BM}$ with only gravitational interaction between them, then the Einstein equation turns out to be

\begin{eqnarray}\label{emax}
G_{\mu \nu} = R_{\mu\nu}-\frac{1}{2}R g_{\mu\nu} = 8 \pi ( T^{\mu \nu}_{DM}+T^{\mu \nu}_{BM} ) \ .
\end{eqnarray}

Each energy-momentum tensor is 
conserved separately giving rise to a set of equations of
motion called two-fluid formalism of Tolman-Oppenheimer-Volkof (TOV) \cite{Tolman:1939jz,Oppenheimer:1939ne,Sandin_2009,Ciarcelluti:2010ji}. 

\begin{eqnarray}\label{e6}
\frac{dp_{\text{B}}}{dr}&=& -\left( p_{\text{B}} +\epsilon_{\text{B}} \right) \frac{M+4\pi r^{3}(p_{B}(r)+p_{D}(r))}{r(r-2M)}\,,\\ 
\frac{dp_{\text{D}}}{dr}&=& -\left( p_{\text{D}} +\epsilon_{\text{D}} \right) \frac{M+4\pi r^{3}(p_{B}(r)+p_{D}(r))}{r(r-2M)}\,,\\  \label{e6d}
\frac{dM}{dr}&=& 4\pi r^{2}(\epsilon_{B}+\epsilon_{D})\, ,
\end{eqnarray}
where $M=M_{D}(r)+M_{B}(r)$ and B and D indices stand for BM and DM components. The solution of the above equations describe  the hydrostatic equilibrium configuration of DM admixed NSs which express the DM distribution within NSs leading to three possibilities (i) DM resides as a core inside NS, (ii) DM forms a halo around NS; (iii) DM dispersed through the whole NS. The gravitational radius of the DM admixed NS is defined by  the radius of BM fluid ($R_{B}$) for the DM core formation, and is determined based on the DM halo radius ($R_{D}$) for the DM halo formation, while for the third case $R_{B}\approx R_{D}$. The total mass of the object in all the cases is defined by
\begin{eqnarray}\label{e11}
M_{T}=M_{B}(R_{B})+M_{D}(R_{D})\, ,
\end{eqnarray}
where $M_{B}(R_{B})$ and $M_{D}(R_{D})$ are the enclosed masses of BM and DM, respectively. The fraction of DM inside the mixed object is defined as  $F_{\chi}=\frac{M_{D}{(R_{D})}}{M_{T}}$ which is realized as a model parameter in our computation. In principle, there are various accumulation scenarios for DM in NSs  introducing different values of $F_{\chi}$ (see Sec. VII of \cite{Karkevandi:2021ygv}). Furthermore, in order to obtain the tidal deformability
of DM admixed NSs, $\Lambda=\frac23 k_2 \left(\frac{R}{M}\right)^5$, M and R are considered as the total mass and outermost radius of the object, respectively. Moreover, $k_{2}$ is the tidal love number which should be calculated based on the two-fluid TOV equations and influenced by both of the applied DM and BM EoSs \cite{Karkevandi:2021ygv}.

It is noteworthy that since the radius of BM component $R_{B}$ is detectable through surface EM radiations, in this work the total mass of the DM admixed NS is shown as a function of the visible radius $R_{B}$. 
This approach has been utilized extensively in previous studies \cite{Tolos:2015qra,Ellis:2018bkr,Deliyergiyev_2019,PhysRevD.102.063028,DiGiovanni:2021ejn,Dengler:2021qcq,Lee:2021yyn}. However, in a recent study presented in \cite{Karkevandi:2021ygv}, the mass-radius (M-R) relation were plotted in terms of the outermost radius of DM admixed NSs instead of visible radius $R_{B}$. Knowing that the outermost radius can be whether $R_B$ or $R_D$ along with stable M-R sequence depending on DM model parameters and the amount of DM in the star. Both  the above mentioned approaches are identical in the case of DM core formation where both the outermost radius of the object and the radius of BM component are the same.

\section{{\bf the Mass-Radius profile of the DM admixed NS and NICER constraints}}\label{SEC3}

\begin{figure}[h]
    \centering
    \includegraphics[width=3.4in]{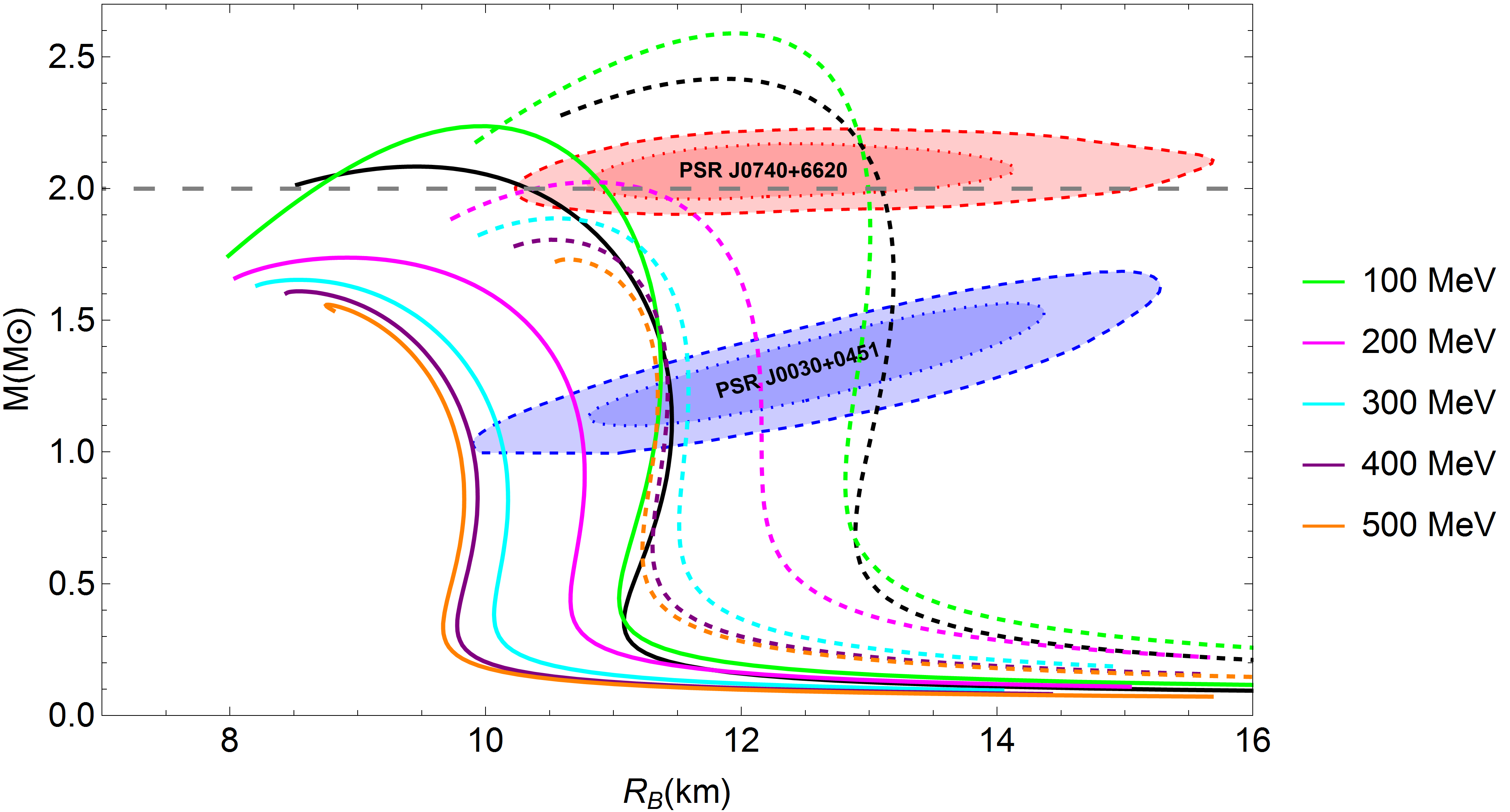}
    \caption{M-R profile of DM admixed NSs for various boson masses at $\lambda=\pi$ and $F_{\chi}=20\%$, by considering IST (solid lines) and DD2 (dashed lines) EoSs as the BM fluid. Red and blue regions correspond to the latest NICER results and the gray dashed line shows the maximum mass limit for NSs.}
    \label{fig1}
\end{figure}

\begin{figure}[h]
    \centering
    \includegraphics[width=3.4in]{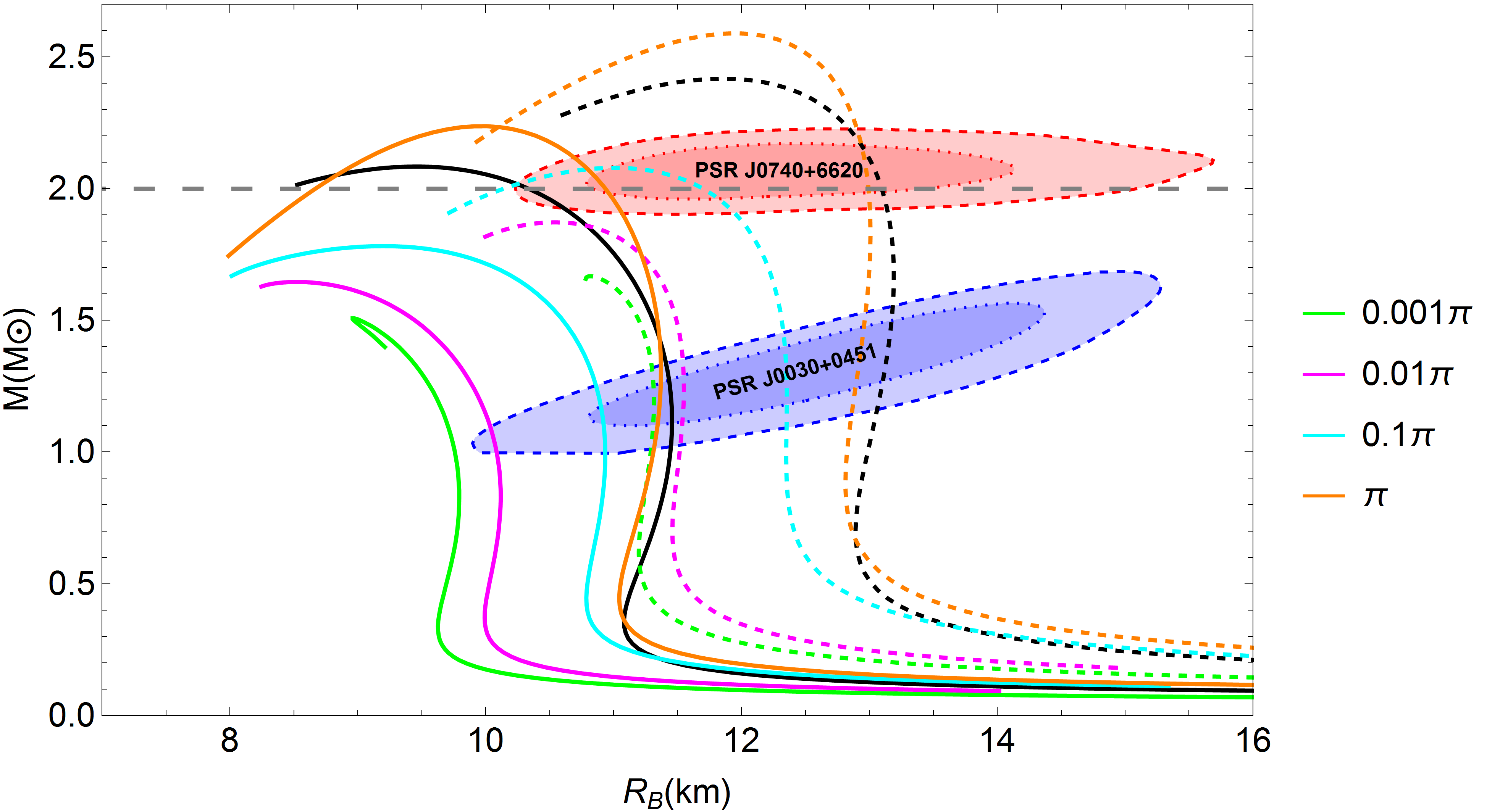}
    \caption{M-R profile of DM admixed NSs for $m_{\chi}=100$ MeV and $F_{\chi}=20\%$ considering various coupling constants as labeled for IST (solid lines) and DD2 (dashed lines) BM EoSs.}
    \label{fig2}
\end{figure}

\begin{figure}[h]
    \centering
    \includegraphics[width=3.4in]{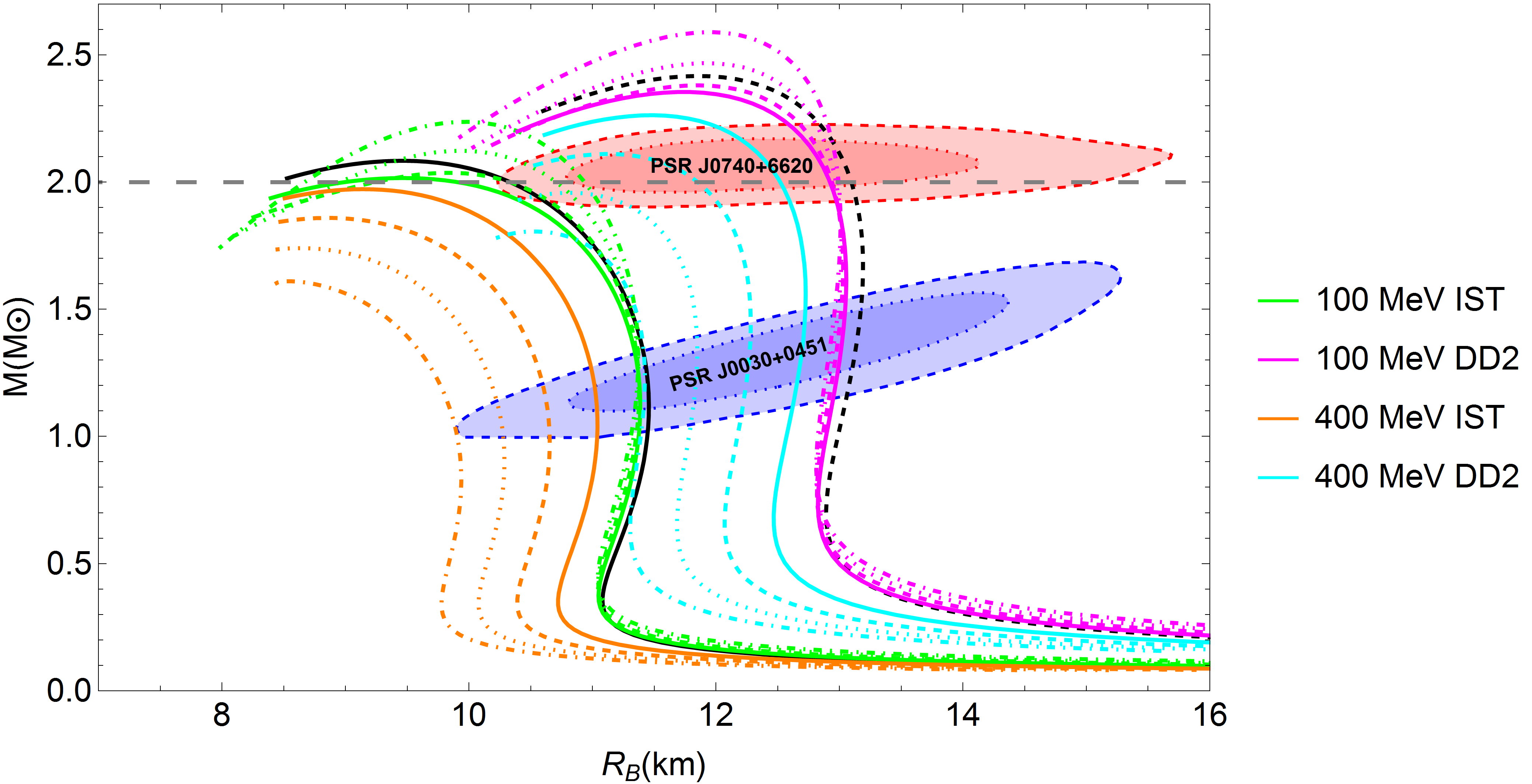}
    \caption{M-R profile of DM admixed NSs for various DM fractions $F_{\chi}=5\%$ (solid lines), $F_{\chi}=10\%$ (dashed lines), $F_{\chi}=15\%$ (dotted lines) and $F_{\chi}=20\%$ (dashed-dotted lines). Here $m_{\chi}=100$ MeV and $m_{\chi}=400$ MeV are considered as DM halo and DM core configurations, respectively,  at $\lambda=\pi$ for IST and DD2 EoSs.}
    \label{M-R3}
\end{figure}

In this section, we consider the bosonic DM model parameters, i.e. mass of bosons ($m_{\chi}$), self-coupling constant ($\lambda$) and the DM fraction  ($F_{\chi}$) via the M-R profile of DM admixed NSs taking into account two BM EoSs, IST and DD2. In the following, we compare our results with the latest mass-radius measurements of NICER  \cite{Miller:2019cac,Riley:2019yda,Miller:2021qha,Riley:2021pdl}. In figures \ref{fig1} - \ref{M-R3}, M-R credible regions ( dark: $68\%$ ($1\sigma$), light: $95\%$ ($2\sigma$) ) for PSR J0030+0451 and PSR J0740+6620 measured by NICER team are colored in blue and red, respectively. As it is shown, according to the NICER results  larger radii (i.e. $R\gtrsim 11 \text{km}$) are more favorable. The gray dashed lines illustrate the maximum mass limit for NSs, $M=2M_\odot$ and the black solid (dashed) lines show the M-R relation for pure NS described by IST (DD2)  BM EoS.

In Fig. \ref{fig1}, the coupling constant and  DM fraction are fixed at $\lambda=\pi$ and $F_{\chi}=20\%$, respectively,  for a range of boson masses 100 - 500 MeV. It is seen that by increasing $m_{\chi}$ which corresponds to DM core formation, the DM admixed NS with IST EoS disfavors NICER results, more specifically, for $m_{\chi}=400,500$ MeV, the M-R profiles are completely outside the red and blue regions. Note that in this case light bosons, for which a DM halo is formed, e.g. $m_{\chi}=100$ MeV, are in agreement with NICER data and even cause the M-R stable sequence to be more compatible with the red part in comparison with IST. Notice that for DM halo formation  specially for large $R_{D}$, gravitational effects such as the tidal deformability should be taken into account which will be discussed in Sec. \ref{sec6}. In the case of DD2 BM EoS, we see that  those M-R profiles with  $m_{\chi}\geq300$ MeV are not compatible with the red region defined by PSR J0740+6620, while mixed compact objects composed by light bosons are consistent with NICER results.

In Fig.~\ref{fig2}, we examine the impact of various self-coupling constants from  $0.001\pi$ to $\pi$ for fixed $m_{\chi}=100$ MeV and $F_{\chi}=20\%$ in DM admixed NSs. It can be seen that by decreasing the coupling constant, in all cases M-R curves move towards left and gradually go out of the NICER permitted parameter spaces. In fact for low self-coupling constant, even light bosons such as $m_{\chi}=100$ MeV would reside as a DM core inside NSs which decreases the mass and radius of the object in comparison with pure BM NS. 

The influence of DM fraction on  M-R sequence of mixed objects  with DM core/halo configurations for $\lambda=\pi$ is shown in  Fig. \ref{M-R3}. The boson mass $m_{\chi}=100$ MeV corresponds to DM halo and $m_{\chi}=400$ MeV relates to DM core  as labeled by different colors for both IST and DD2. Here different types of lines stand for various fractions i.e. $F_{\chi}=5\%$ solid lines, $F_{\chi}=10\%$ dashed lines, $F_{\chi}=15\%$ dotted lines and $F_{\chi}=20\%$ dashed-dotted lines. We can see that by increasing $F_{\chi}$ from $5\%$ to $20\%$ for $m_{\chi}=400$ MeV,  the mass and radius of the mixed object are decreased (due to the DM core formation) and consequently  M-R curves are shifted away from the allowed parameter spaces. It is found that for $m_{\chi}=100$ MeV and  higher $F_{\chi}$,  maximum masses of  M-R diagrams  are  increased while $R_{B}$ remains approximately unchanged.

In summary, we show that sub-GeV DM bosons with smaller masses and higher self-coupling constants cause the DM admixed NS to be more compatible with the latest NICER mass-radius data. In addition, if a DM core is formed inside NS,  low DM fractions are in more agreement with PSRJ0030+0451 and PSR J0740+662 regions, while for DM halo configuration those regions are respected in the whole considered range of fractions. Thus, regarding the above-mentioned conditions, the NS could contain bosonic DM as a core or halo and be consistent with NICER results. Furthermore, we see that DM admixed NSs comprised of self-interacting bosonic DM and IST or DD2 as BM EoS could potentially interpret exotic compact objects such as the most massive one with $2.6 M_{\odot}$ in  GW190814 event \cite{LIGOScientific:2020zkf} and the lightest NS with mass around $0.77 M_{\odot}$ known as HESS J1731-347 \cite{2022NatAs...6.1444D}.

It is instructive to compare our results  with those obtained in a recent study  \cite{Rutherford:2022xeb} where the authors explored a different  DM model with vector bosons as mediators between DM particles \cite{Agnihotri:2008zf,Nelson:2018xtr} to describe the DM admixed NS. The role of DM particles only as a core component was considered in \cite{Rutherford:2022xeb} while in our study we investigate both halo and core distributions of bosonic SIDM. They showed that  lighter bosons with high self-interaction strength and low fractions are favored by the latest NICER  mass-radius measurements which is compatible with our results.

\section{{\bf Dark and Visible Radius of DM admixed NS}}\label{sec4}

Concerning the distribution of DM in NSs as a core or halo, in this section we extensively examine the behaviour of the visible and dark radius ($R_{B}$ and $R_{D}$) for various DM model parameters (i.e. $m_{\chi}$ and  $\lambda$) along with a range of DM fraction. Note that the visible radius of the mixed compact object is $R_{B}$, and  the corresponding radius of a DM admixed NS with $M_{T}=1.4M_\odot$ is compared with the lower observational limit of radius  $\simeq11$  km \cite{Dietrich:2020efo,Huth:2021bsp,Pang:2021jta,Breschi:2021tbm}. The analyses presented in this section  will be used in Sec. \ref{sec6} in order to produce a scan plot over DM parameter space for the radius constraint $R_{1.4}\gtrsim11$.

\begin{figure}[h]
    \centering
    \includegraphics[width=3.3in]{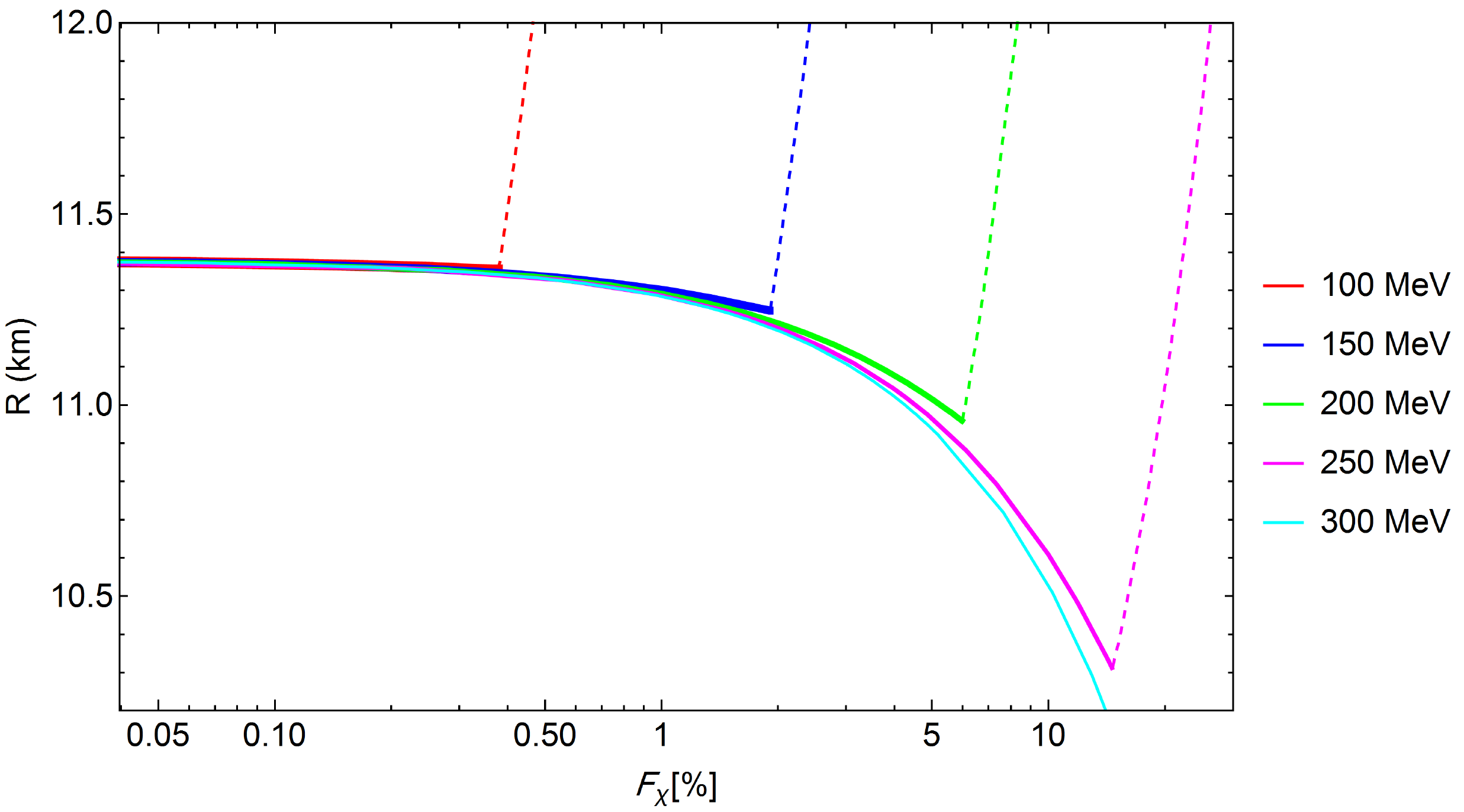}
    \includegraphics[width=3.3in]{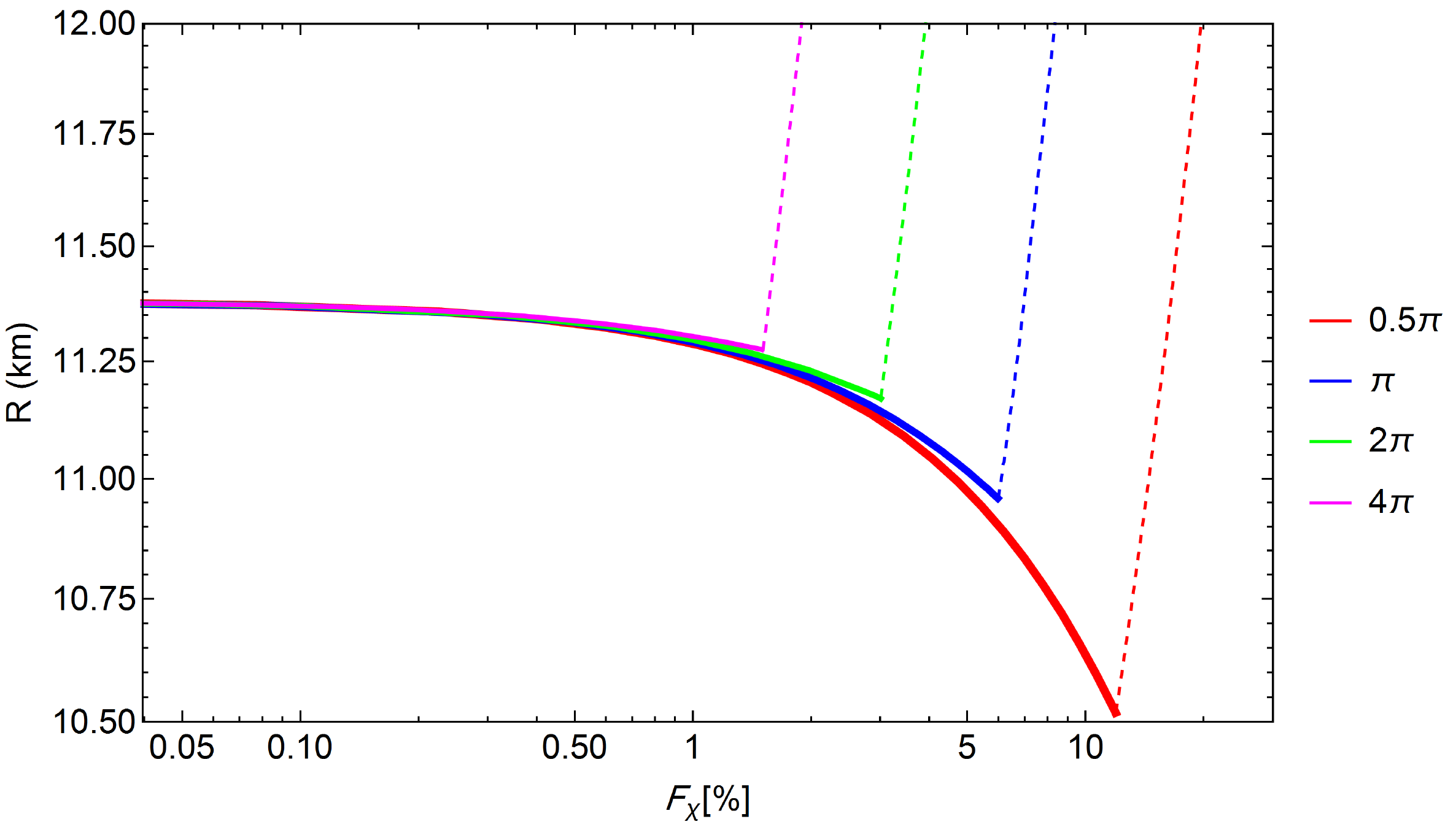}
    \caption{Variation of the outermost radius of  DM admixed NSs with total mass  $1.4M_\odot$ as a function of DM fractions for IST EoS. Solid curves show the variation of $R_B$ (visible radius) for DM core formation. Dashed lines indicate the change of $R_D$  where a DM halo is formed. Different boson masses are considered as labeled for $\lambda=\pi$ in the upper panel while the lower panel is related to $m_{\chi}=200$ MeV and various coupling constants.}
    \label{R-fx}
\end{figure}

\begin{figure}[h]
    \centering
    \includegraphics[width=3.3in]{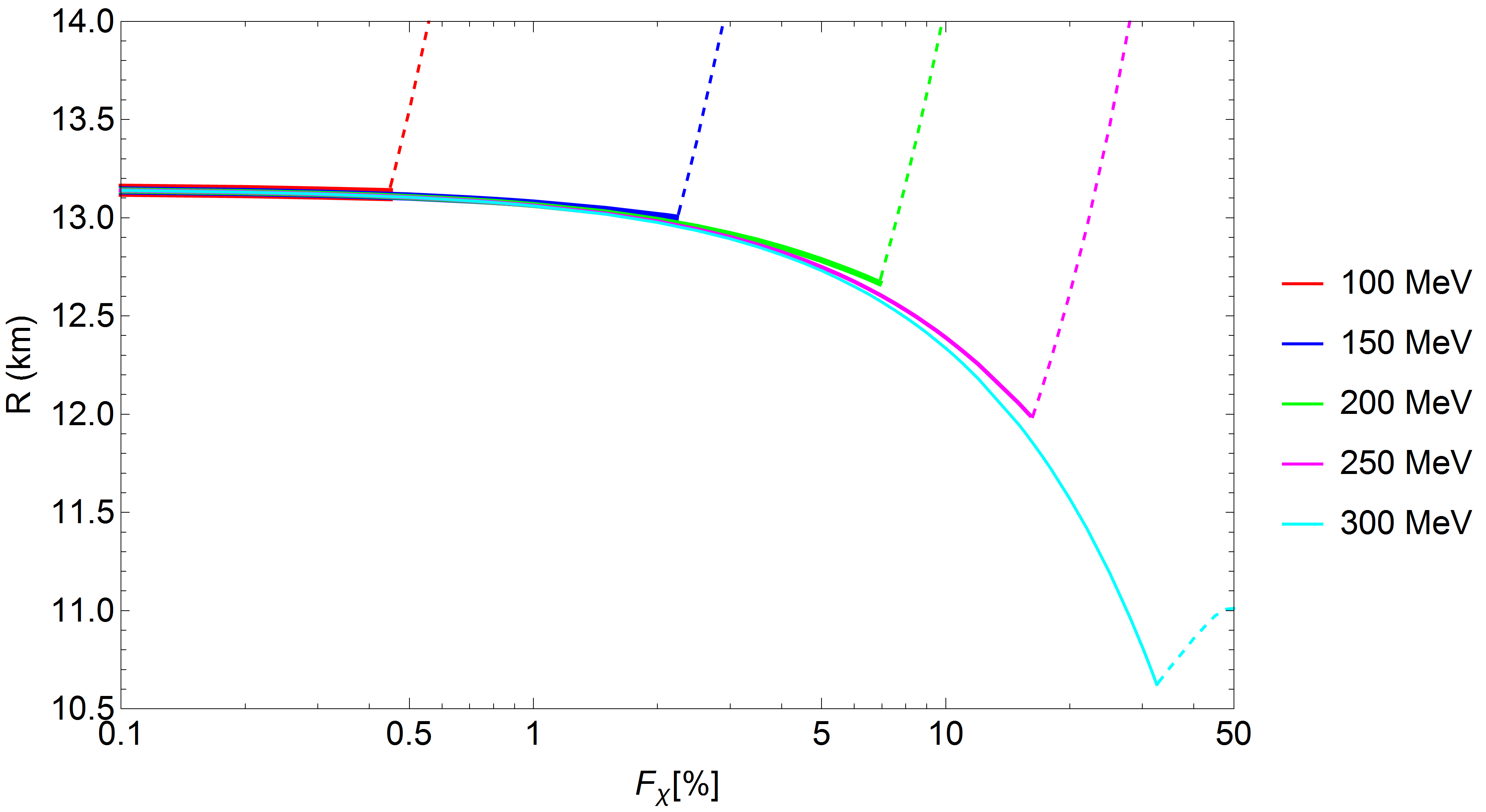}
    \includegraphics[width=3.3in]{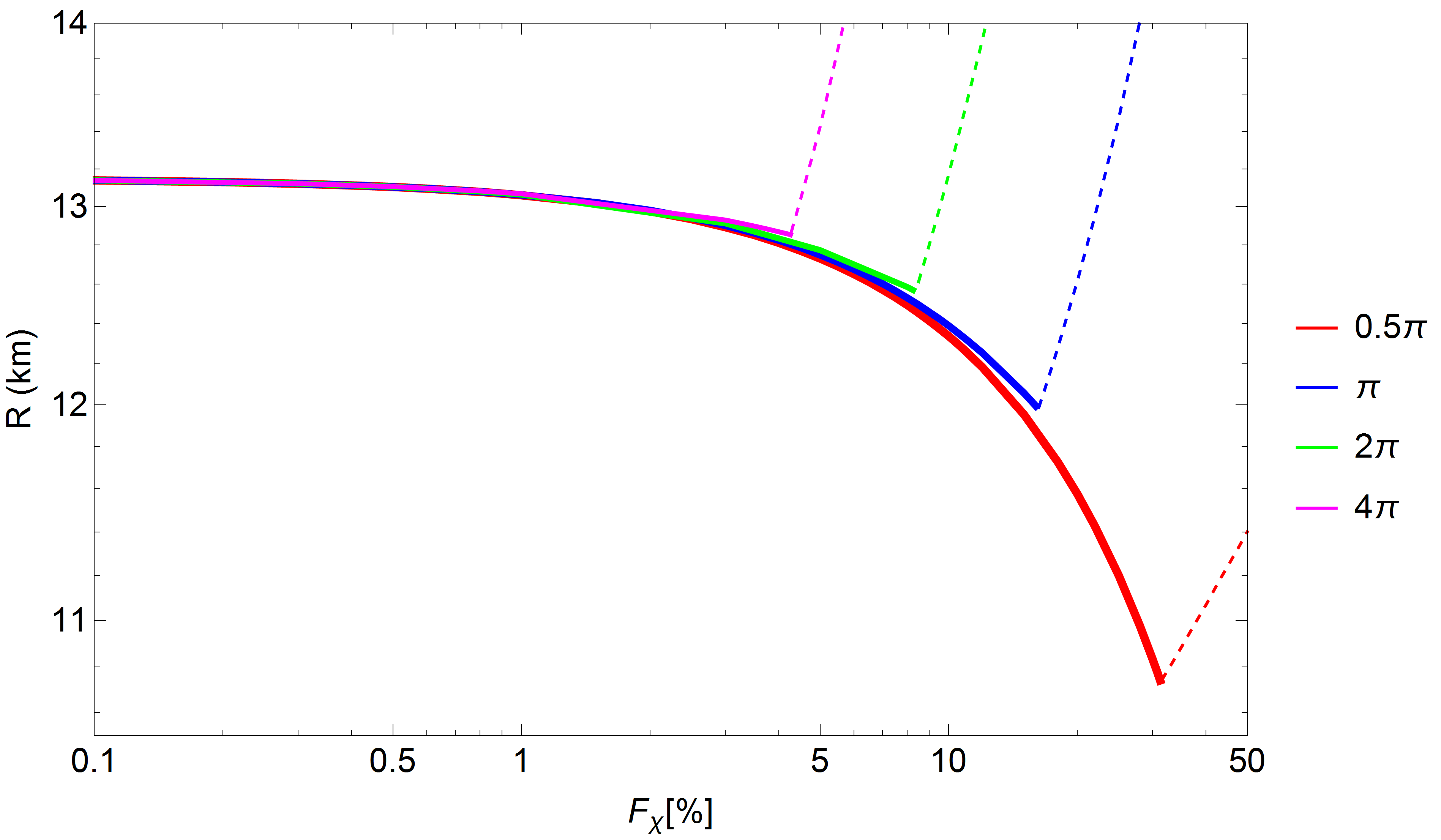}
    \caption{Similar to Fig. \ref{R-fx}, but for DD2 EoS and in the lower panel for $m_{\chi}=250$ MeV. }
    \label{R-fx2}
\end{figure}

In Figs. \ref{R-fx} and \ref{R-fx2}, we show the variation of the outermost radius of DM admixed NSs with $M_{T}=1.4M_\odot$ as a function of $F_{\chi}$ for different boson masses and self-coupling constants applying IST and DD2 BM EoSs. It is seen that by increasing the DM fraction, the outermost radius decreases up to a turning point and then it goes sharply upwards. In these plots, the radius  of the mixed object will change from the visible radius $R_{B}$ (solid lines) to the dark radius $R_{D}$ (dashed lines) where a DM core to DM halo transition occurs. In fact, we see that depending on  $m_{\chi}$ and $\lambda$, a DM core will be formed generally in low $F_{\chi}$ while by increasing the amount of DM, the dark radius $R_{D}$ will grow and a DM halo is formed around NS.
The transition takes place at the turning point of the  diagrams where  $R_{B}\approx R_{D}$ that is equivalent to the formation of a DM admixed NS in which DM is dispersed throughout the whole object.  In the upper panels of Figs. \ref{R-fx} and \ref{R-fx2}, $m_{\chi}$ is varied from 100 MeV to 300 MeV for fixed $\lambda=\pi$, we see that for lighter bosons, the DM halo is forming in very low fractions, however, one can show that for the massive bosons, the DM halo will not form even up to relatively high fractions. In the lower panels, the impact of the coupling constant on the outermost radius is considered where for higher $\lambda$ the transition point and the DM halo is appeared at smaller values of $F_{\chi}$.

In the following Figs. \ref{RBD-Fx} and \ref{RBD-Fxdd2}, the variation of $R_{B}$ and $R_{D}$ for  the DM admixed NS with total gravitational mass $1.4M_\odot$ are shown separately as a function of $F_{\chi}$ for both DM core (solid lines) and DM halo (dashed lines) formations.  In these plots, black dashed horizontal line indicates  $R_{1.4}=11$ km as a lower bound for the visible radius of the mixed object.

In Figs. \ref{RBD-Fx} and \ref{RBD-Fxdd2}, the impact of different boson masses $m_{\chi}\in [100,300]$ \text{MeV}   are examined at a fixed coupling constant $\lambda=\pi$ for a given range of DM fractions. In the upper panels, we see that for both DM core and DM halo formations, $R_{B}$ is a decreasing function of $F_{\chi}$ where for higher $m_{\chi}$ the more reduction rate is observed. The variation rate for light bosons   becomes slow down reaching a constant value for larger $F_{\chi}$. It is shown that while for massive bosons one can define an upper limit for the DM fraction to satisfy a given radial limit (e.g. 11 km), for light enough  bosons $R_{B}$  lies above the radius limit for the whole considered range of DM fractions.  It turns out  that  by increasing $F_{\chi}$ for  massive bosons, $R_{B}$ becomes less than 11 km which is not in favour of the lower observational limit of $R_{1.4}$. As it is depicted, the DM core inside NSs (solid line) is formed up to relatively high fractions for heavy bosons while at low fractions we may have core formations even for the light bosons $m_{\chi}\lesssim150$ MeV. 
By doing precise calculations, we find  $m_{\chi}\simeq153.5$ MeV  and $m_{\chi}\simeq230$ MeV  as the heaviest bosons for IST and DD2, respectively, which  $R_{B}\gtrsim11$ km condition is met for the whole considered range of DM fraction. From the lower panels of Figs. \ref{RBD-Fx} and \ref{RBD-Fxdd2}, we see that $R_{D}$  is  an increasing function of $F_{\chi}$ for all the considered boson mass range, meanwhile, the DM radius of lighter particles increases with steeper slopes in low fractions compare to more massive ones. Regarding the mass range $m_{\chi}\in [100,300]$ \text{MeV} for DM fractions  $F_{\chi}\lesssim5\%$ (IST) and $F_{\chi}\lesssim28\%$ (DD2), visible radii of the mixed objects are higher than 11 km.

\begin{figure}[!h]
    \centering
    \includegraphics[width=3.35in]{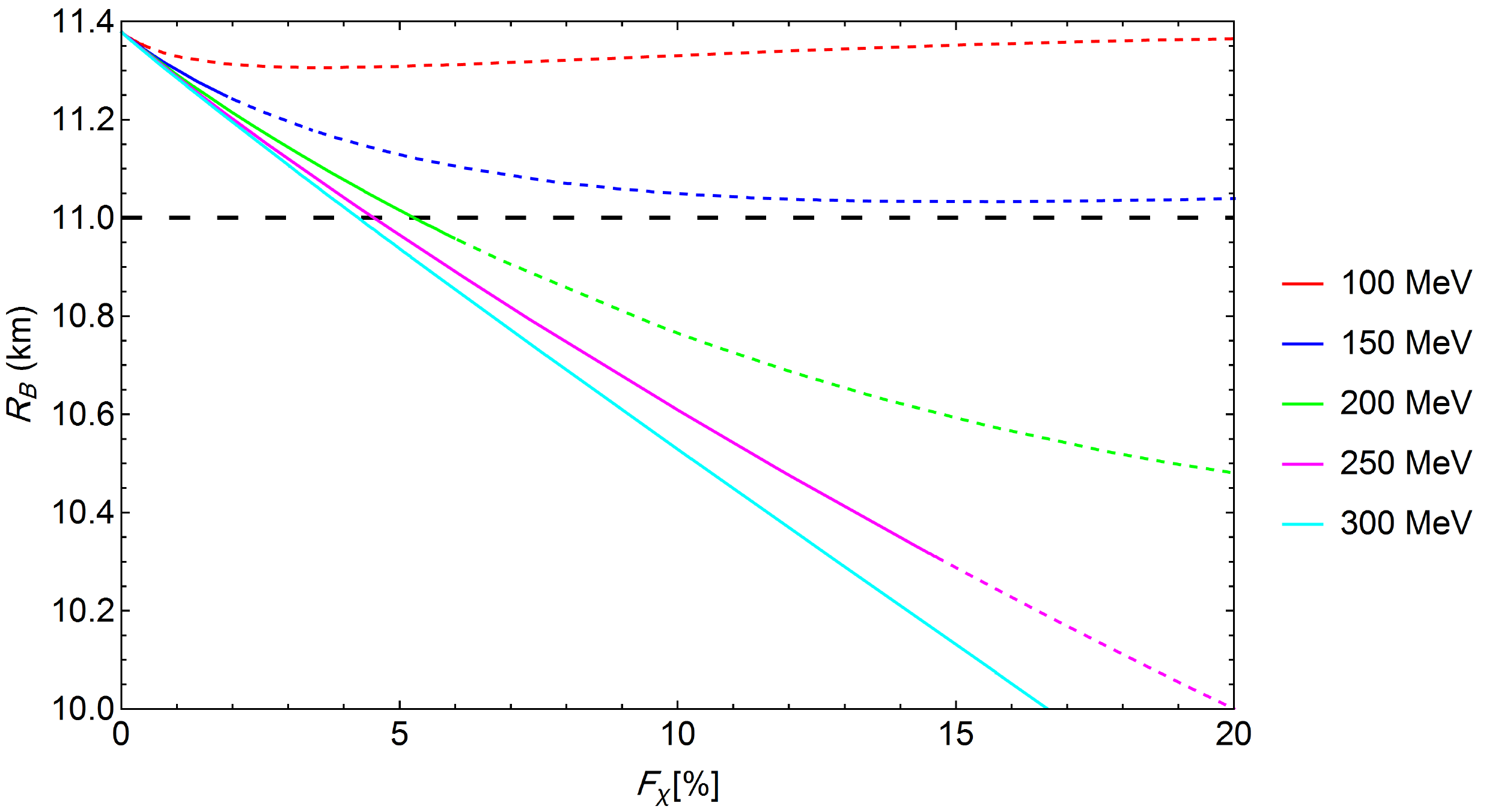}
    \includegraphics[width=3.35in]{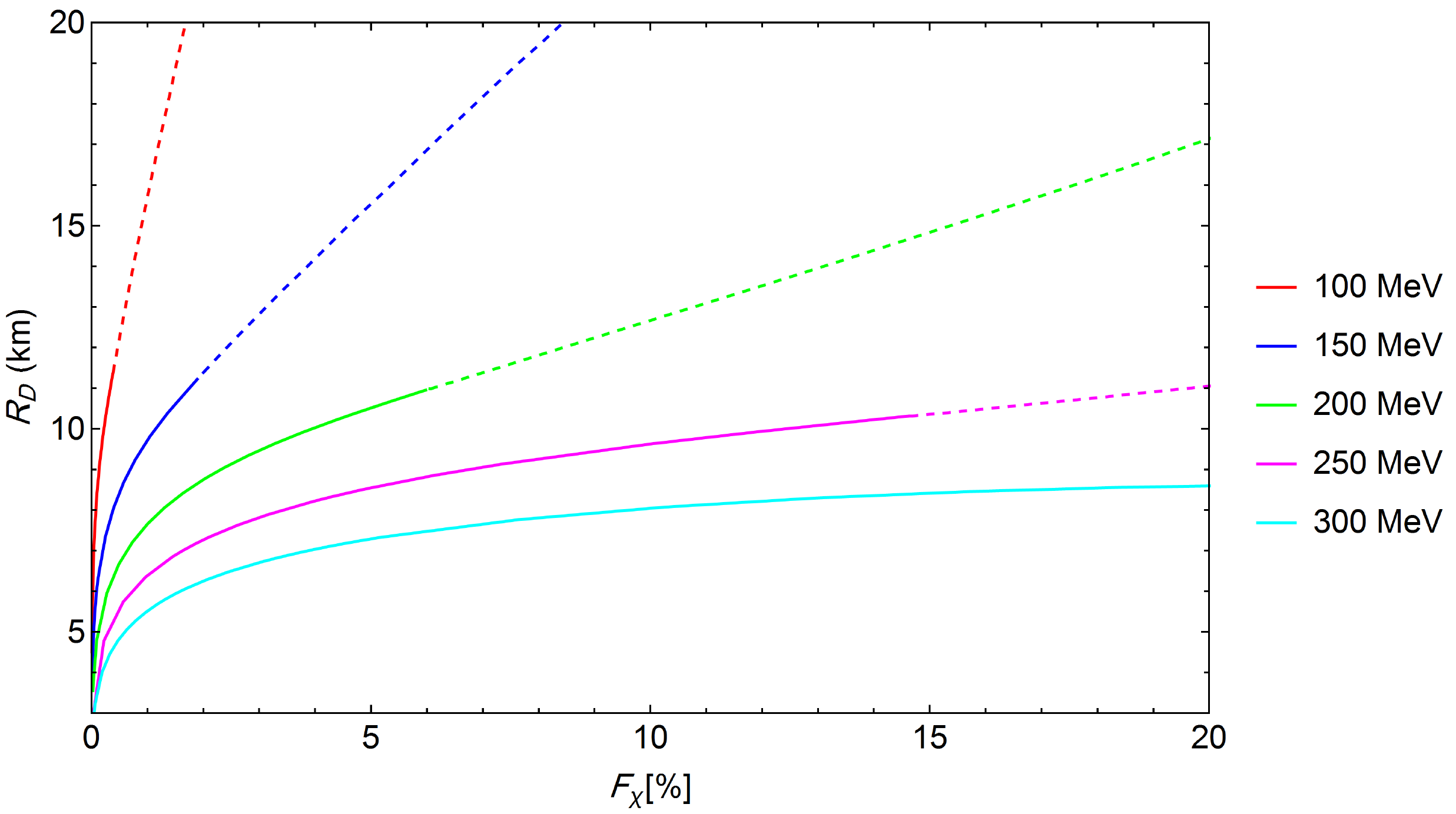}
    \caption{Variation of $R_B$ (up) and $R_D$ (down) for IST EoS are indicated separately with respect to $F_{\chi}$ for $M_{T}=1.4M_\odot$. Different boson masses are considered as labeled for $\lambda=\pi$. The solid and dashed curves depict the DM core and DM halo formations, respectively. The black dashed line in the upper panel shows the lower observational constraint for $R_{1.4}$.}
    \label{RBD-Fx}
\end{figure}

\begin{figure}[!h]
    \centering
    \includegraphics[width=3.35in]{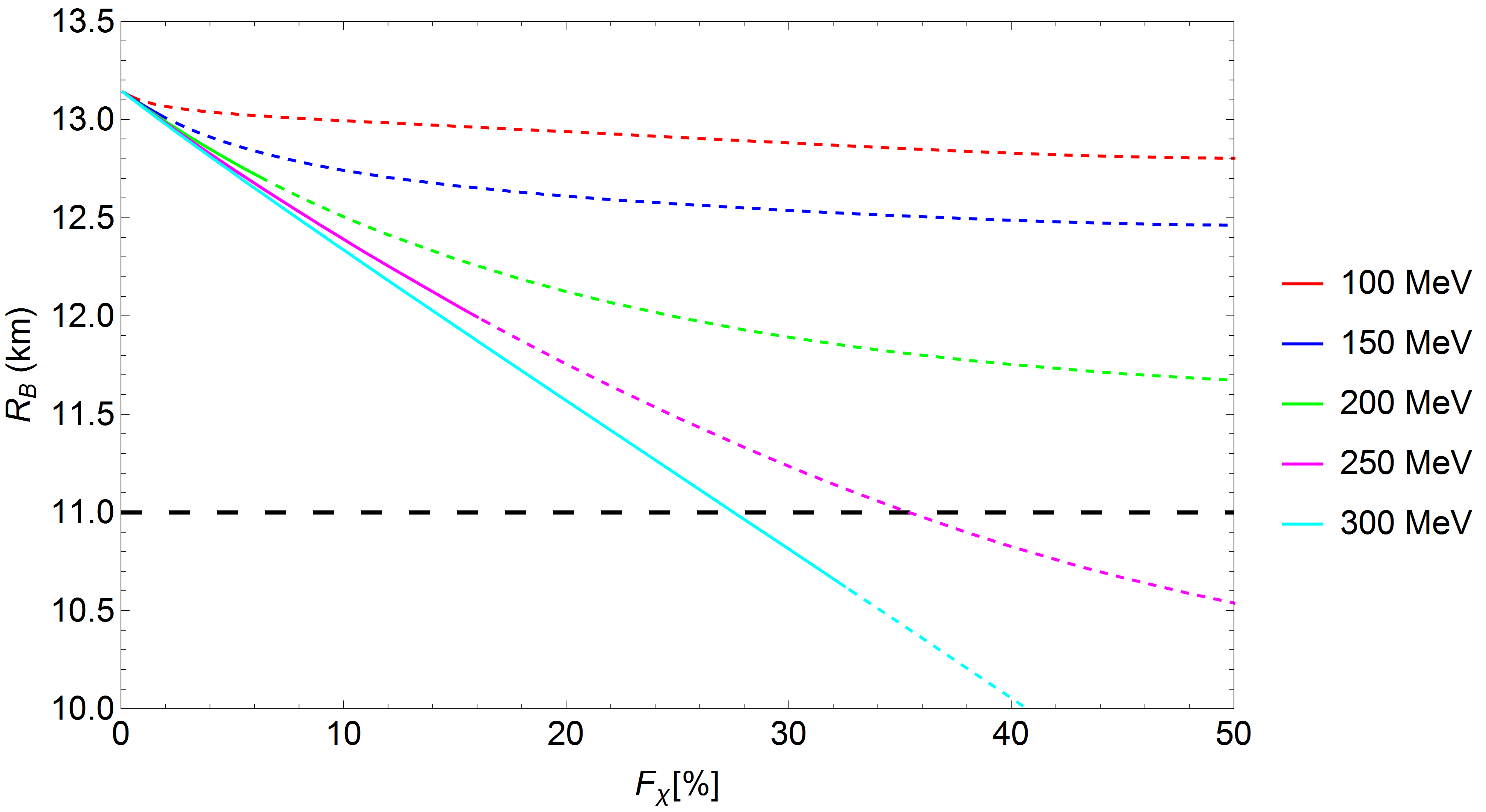}
    \includegraphics[width=3.35in]{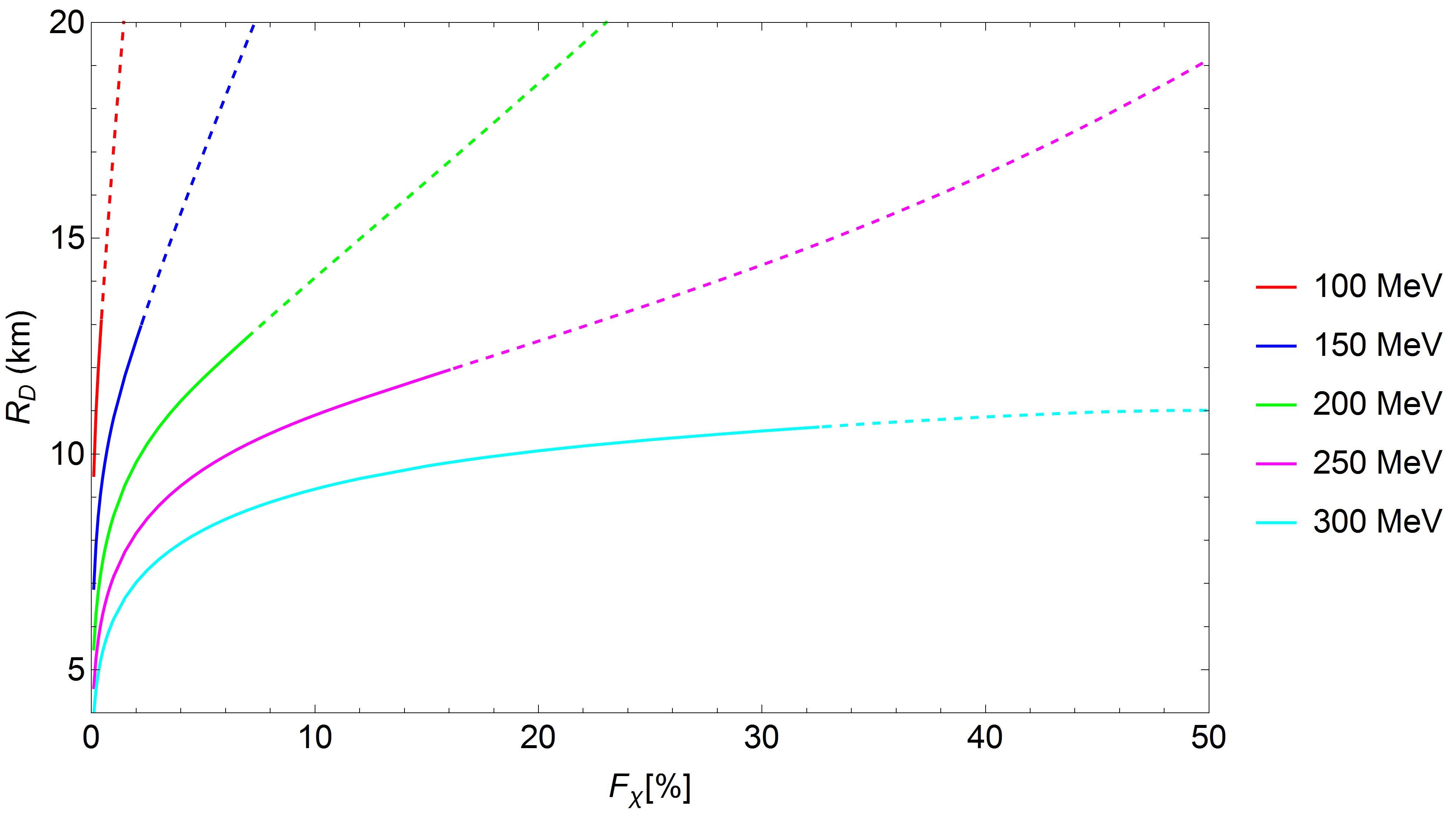}
    \caption{Similar to Fig .\ref{RBD-Fx}, but for DD2 EoS.}
    \label{RBD-Fxdd2}
\end{figure}

\begin{figure}[!h]
    \centering
    \includegraphics[width=3.35in]{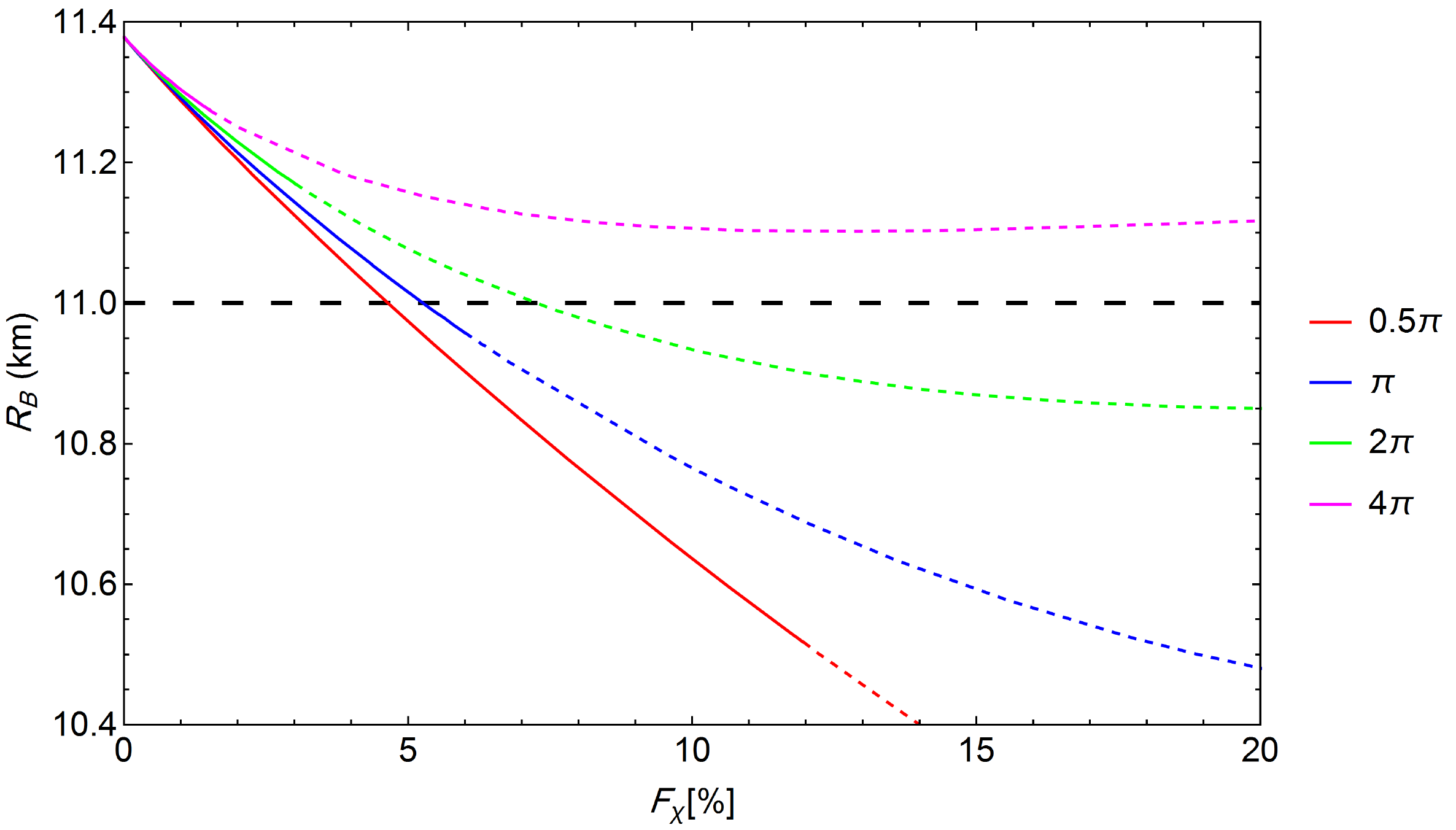}
    \includegraphics[width=3.35in]{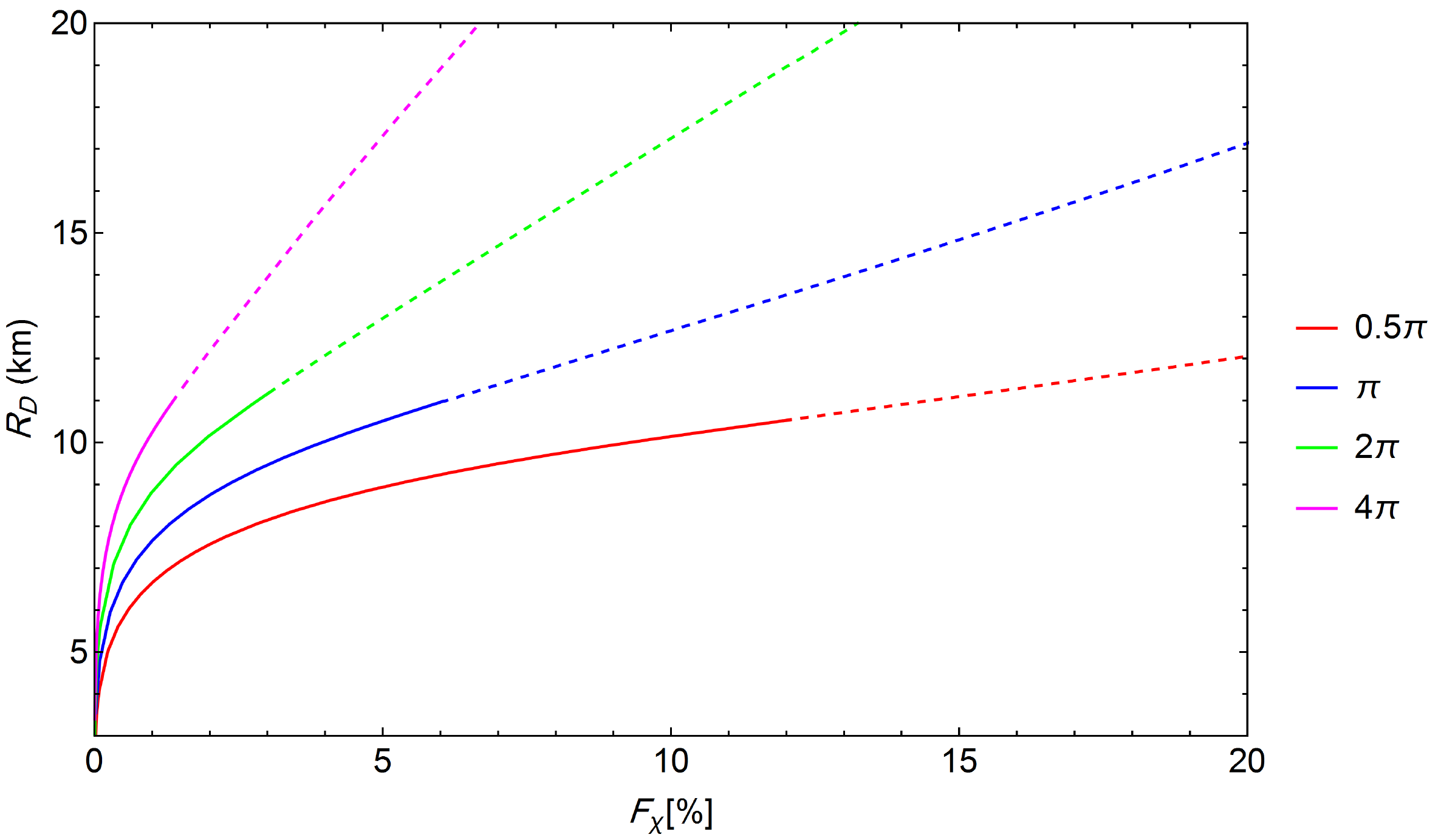}
    \caption{Variation of $R_B$ (up) and $R_D$ (down) for IST EoS are indicated separately with respect to $F_{\chi}$ for $M_{T}=1.4M_\odot$. Different coupling constants are considered as labeled for $m_{\chi}=200$ MeV. The solid and dashed curves depict the DM core and DM halo formations, respectively. The black dashed line in the upper panel shows the lower observational constraint for $R_{1.4}$.}
    \label{RBD-FxL5}
\end{figure}

\begin{figure}[!h]
    \centering
    \includegraphics[width=3.35in]{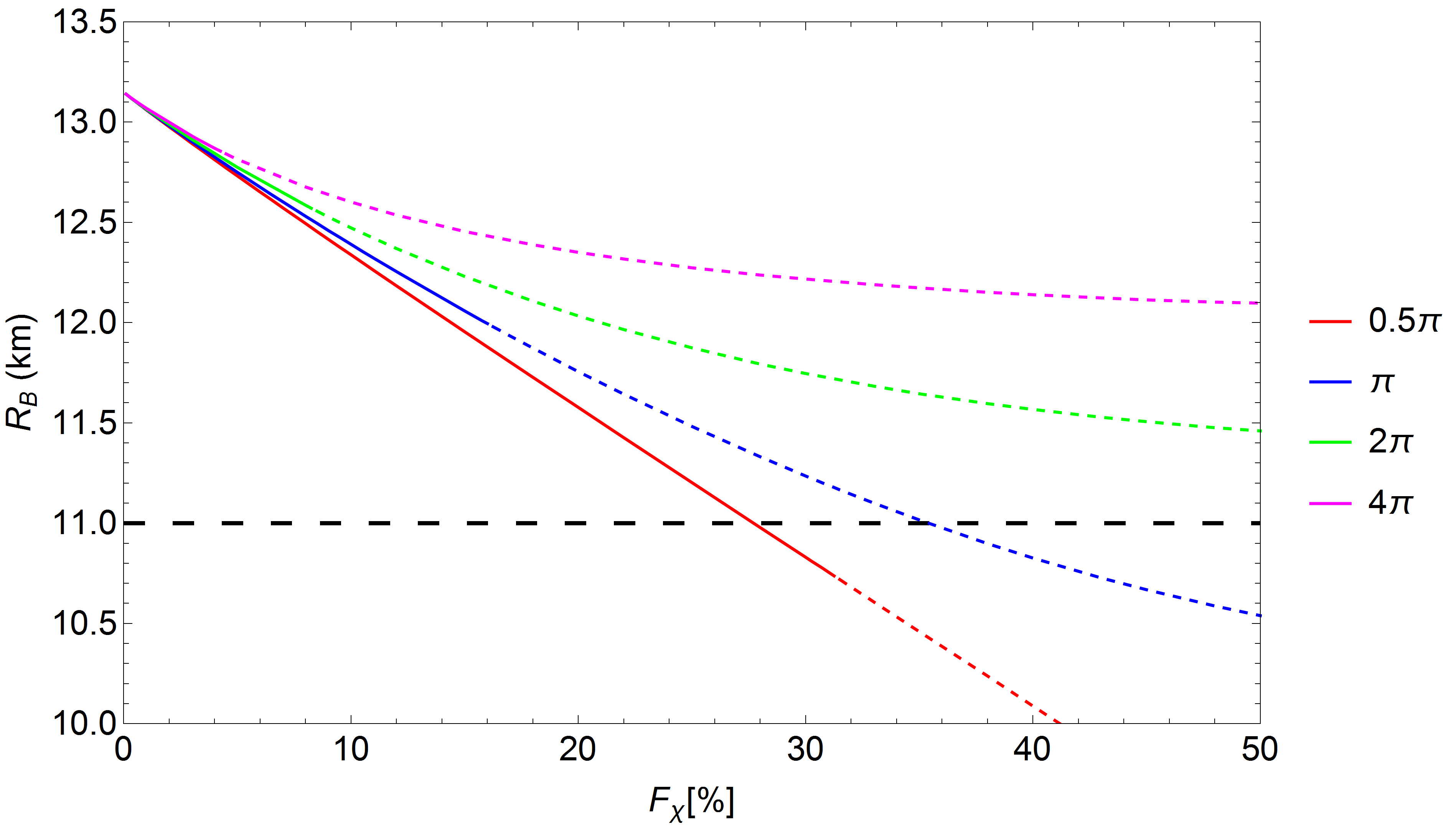}
    \includegraphics[width=3.35in]{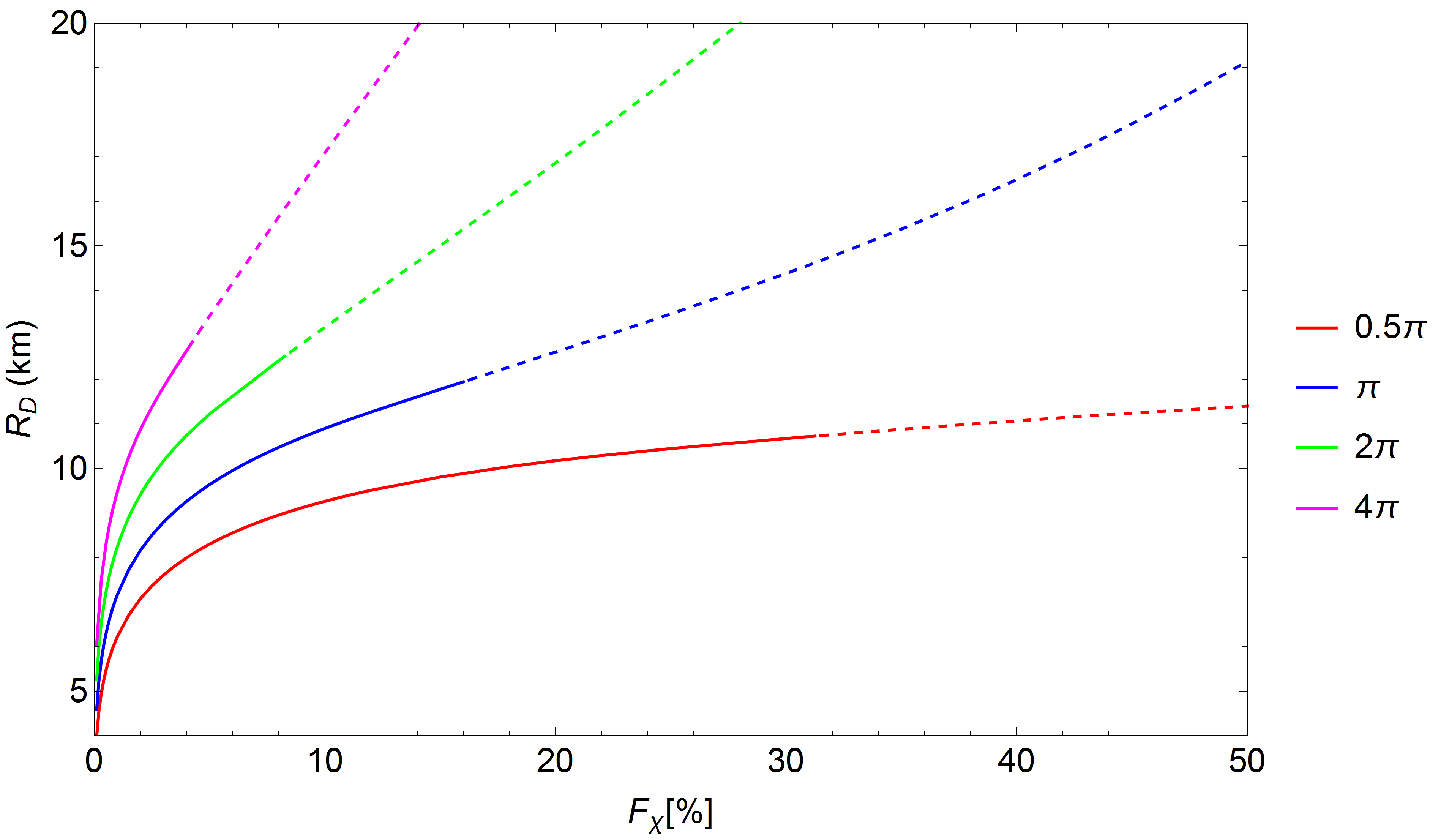}
    \caption{Similar to Fig. \ref{RBD-FxL5}, but for DD2 EoS and $m_{\chi}=250$ MeV.}
    \label{RBD-FxLDD2}
\end{figure}

We also investigate the effect of self-coupling constant on the visible and  dark radius of DM admixed NSs with different amount of DM (Figs. \ref{RBD-FxL5} and \ref{RBD-FxLDD2}) in which the mass of DM particles is fixed at $m_{\chi}=200$ MeV (IST) and $m_{\chi}=250$ MeV (DD2)  while $\lambda$ is changed between  $0.5\pi$ and $4\pi$. It is seen that for both DM core/halo formations,  $R_{B}$ decreases more towards lower $\lambda$ going faster below 11 km limit which is disfavoured by the latest observations. In Figs. \ref{RBD-FxL5} and \ref{RBD-FxLDD2} (upper panels)  one can denote an upper bound  for $F_{\chi}$,  where below this value the radius constraint is satisfied, note that for some coupling constants (e.g. $\lambda=4\pi$)  the curve is always above 11 km. The lowest values of self-coupling constants for which the condition $R_{B}\gtrsim11$ km is fulfilled for the given range of DM fractions are $\lambda\simeq2.9\pi$ (IST) and  $\lambda\simeq1.4\pi$ (DD2).
As it is illustrated in the lower  panels, the DM radius increases with  $F_{\chi}$ which for  higher values of $\lambda$ a sharper rise can be seen. From both upper and lower panels, it turns out that higher coupling constants lead to the DM halo formation at lower $F_{\chi}$, however, for smaller $\lambda$, DM resides as a core for a wider range of DM fractions.

\begin{figure*}
 \includegraphics[width=3in]{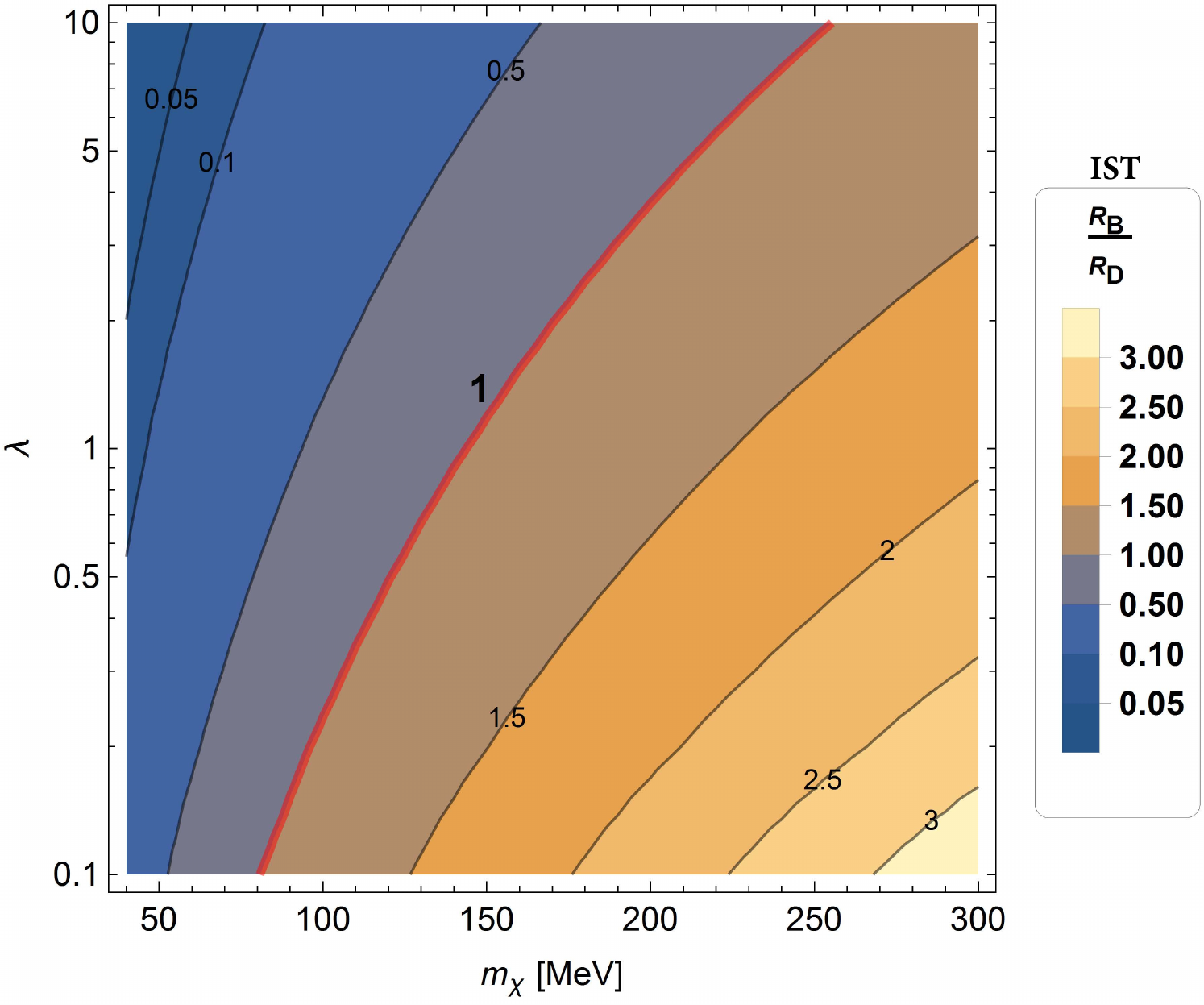}
\includegraphics[width=3in]{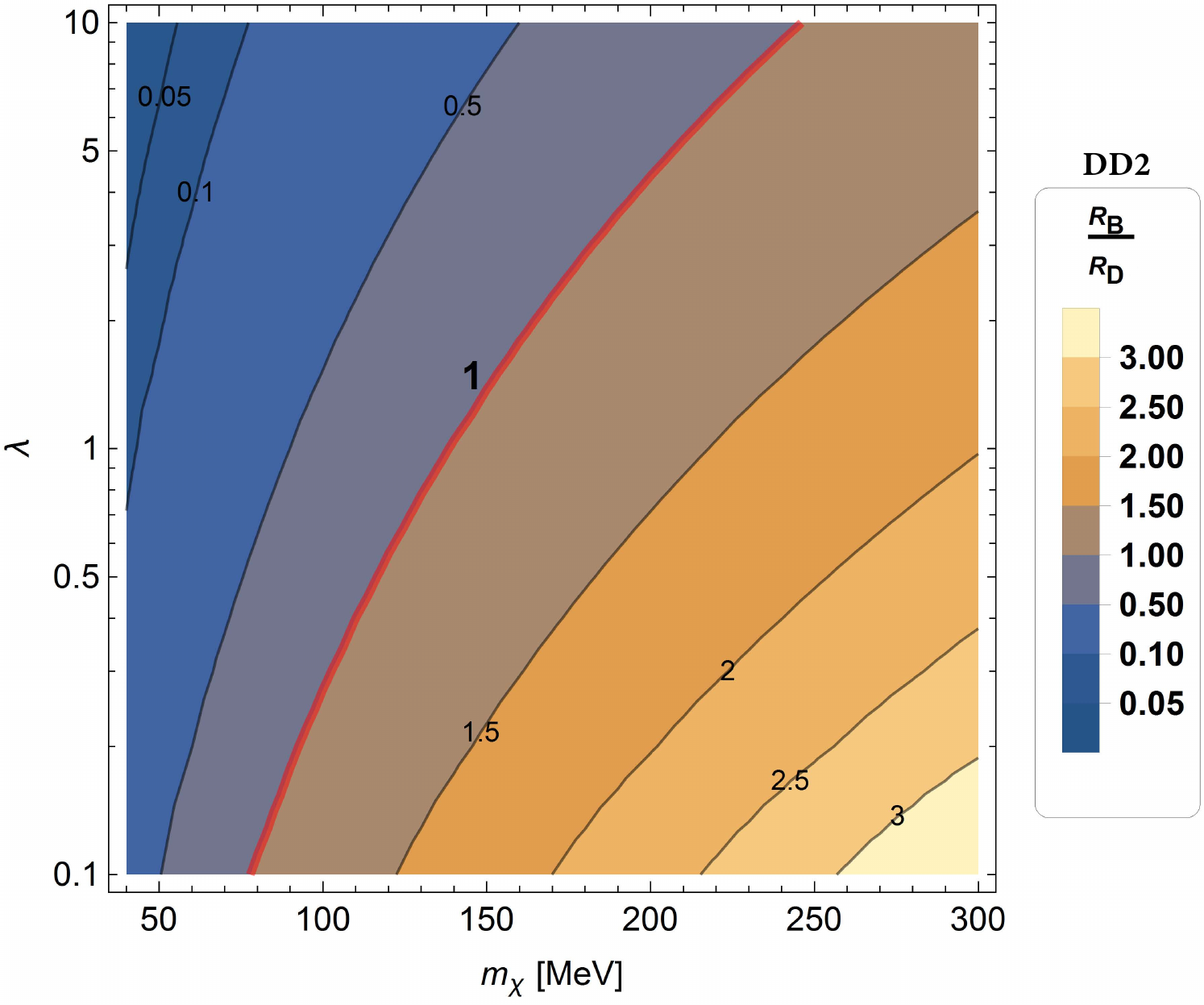}
\includegraphics[width=3in]{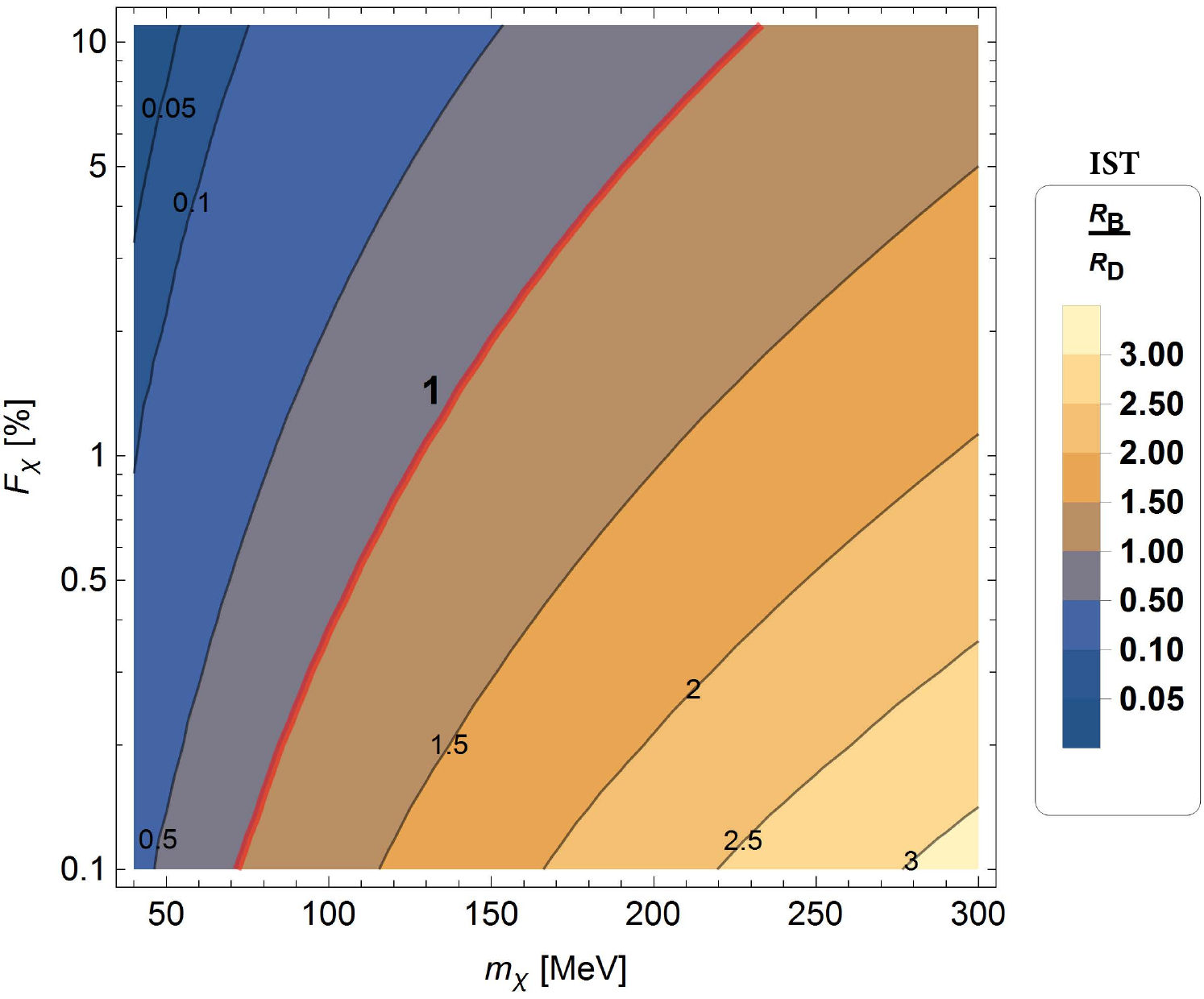}
\includegraphics[width=3in]{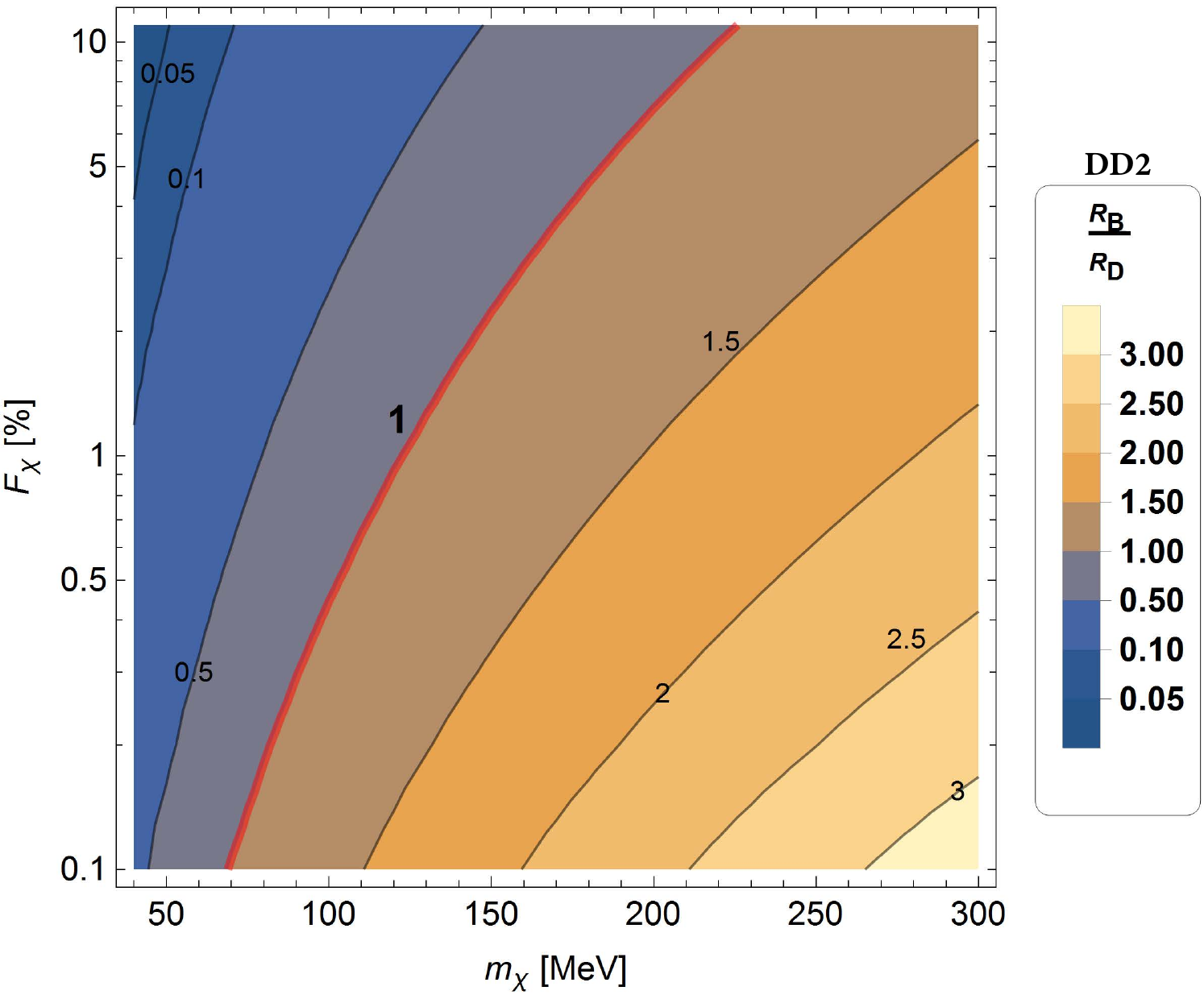}
  \caption{DM parameter space specifying different regions for which a DM core or a DM halo is  formed for a DM admixed NS with  $M_{T}=1.4M_{\odot}$. Upper panels display contour plots in terms of $\lambda-m_{\chi}$ parameter space at $F_{\chi}=5\%$ for IST (left) and  DD2 (right) EoSs describing the BM fluids. Each curves indicate the value of $R_{B}/R_{D}$ where the red line shows the special case  $R_{B}=R_{D}$ defining the boundary of DM core-halo transition. Lower panels are scans over   $F_{\chi}-m_{\chi}$ parameter space at fixed self-coupling constant $\lambda=\pi$ and different BM fluids IST and DD2 as labeled.  }\label{corehaloscan}
\end{figure*}

To clarify further, the DM model parameters  for which a DM core and a DM halo could be formed are illustrated in Fig. \ref{corehaloscan}. In the upper panels $\lambda-m_{\chi}$ parameter space for $F_{\chi}=5\%$ and in the lower panels $F_{\chi}-m_{\chi}$ parameter space for $\lambda=\pi$ are explored in order to determine DM core/halo region for a $1.4M_{\odot}$  DM admixed NS. Two different EoSs  IST (left) and  DD2 (right) are utilized for modeling BM component of the mixed object.  The density label represents a fraction of visible to dark radius $R_{B}/R_{D}$ and the red lines mark the border of the core-halo regions where $R_{B} =R_{D}$. As expected, heavy bosons tend to form a DM core at a fixed fraction or coupling constant. However,  relatively light bosons, by increasing self-coupling constant and DM fraction, tend to form a DM halo. Remarkably, the core-halo formation is independent of BM EoSs, while it crucially depends on the DM model parameters and its fraction. In the appendix, regarding some simplified assumptions, we obtain an analytic formula (See Eq. \ref{e23}) representing the boundary lines of the DM core-halo transition to demonstrate that these lines are independent of the details of  BM EoSs.
 
In  this section, we show that the outermost radius of the mixed object changes from visible radius ($R_{B}$) to dark radius ($R_{D}$)  by increasing $F_{\chi}$ for which the corresponding fraction point, where $R_{B}\approx R_{D}$,  will be increased for heavier bosons and/or lower coupling constants. Generally, it is seen that for both DM core and DM halo formations, the visible radius ($R_{B}$) of the DM admixed NS will be decreased by increasing $F_{\chi}$ while $R_{D}$ is a rising function. Therefore, the lower limit of $R_{1.4}$ (11 km), which is inferred from various observational measurements \cite{Dietrich:2020efo,Huth:2021bsp,Pang:2021jta,Breschi:2021tbm}, is disfavoured for heavier bosons and/or smaller coupling constants. In fact, we conclude that smaller $m_{\chi}$ and larger $\lambda$ are in favor of radius constraints for NSs with $M=1.4M_\odot$. It is worth saying that the reduction of the visible radius for DM particles with $m_{\chi}=300$ MeV which constitute $20\%$ of a DM admixed NS is about $14.4\%$ (IST) and $12\%$ (DD2) of the corresponding radius of a pure NS. The drop of $R_{B}$ for a mixed compact object composed of $20\%$ DM component with $m_{\chi}=200$ MeV and $\lambda=0.5\pi$  is $11.18\%$ (IST) and $10.11\%$ (DD2)  of the radius of a normal NS.

Note that the current NICER uncertainty in determination of the radius of a NS  is $\sim 10\%$ \cite{Rutherford:2022xeb}. The next generation of X-ray telescopes  such as STROBE-X are capable to reach  $5\%$  uncertainty level (and even incredible $2\%$ uncertainty for longer observations) in measurements of mass and radius of the compact objects
 \cite{STROBE-XScienceWorkingGroup:2019cyd}  and hence they will be served as  unique opportunities to probe the presence of DM within the NSs.

\section{{\bf Pulse Profile and Light Bending due to the DM halo}}\label{sec5}

X-ray telescopes measure the radius  of a NS by tracking the X-ray emission from hot spots on the surface as the star rotates. The trajectory of photons is affected by the gravitational field of compact objects leading to the gravitational light bending which is taken into account in the analysis of NICER via Pulse Profile Modeling (PPM) \citep{Beloborodov:2002mr,Poutanen:2006hw,Turolla:2013tba,Ozel:2015ykl,2012SPIE.8443E..13G}.
In this section, we consider the effect of bosonic DM halo on the self-light bending from DM admixed NSs.  
The thermal surface photons usually originate from the baryonic radius at $R_B$, while it is possible to have some non-thermal emissions produced at higher  latitude $r>R_B$ \cite{2004hpabook}. Due to  the lack of interaction between DM particles and photons, the DM halo up to radius  $R_D$ remains invisible in the light of EM interactions but it affects the light propagation  through the gravitational interaction. We  assume a spherically symmetric non-rotating space-time (Schwarzschild metric) outside $R_{B}$ as 
\begin{eqnarray}
ds^{2} = -g(r) dt^{2} + f(r) dr^{2} +r^{2} d\theta^2 +r^2 \sin^{2}\theta d\phi^{2} ,
\end{eqnarray}
where $g(r)=f(r)^{-1}=1-2M(r)/r$, the total mass $M(r)$ similar to Eq. (\ref{e11}) has two contributions coming from both BM and DM fluids at $r\ge R_{B}$. In a NS without DM halo $M(r\ge R_{B})$ is a constant value, the similar situation happens for a DM admixed NS with  DM core formation. However, the formation of the DM halo around  NS can potentially change the geometry outside the surface of  NS and also the light propagation characteristics in this region.  In this section in all considered cases we fix the total mass of DM admixed NSs at  $M_{T}(R_{D})=1.4M_{\odot}$ and we focus only on those regions of DM model parameter space in which we have the DM halo formation. 

\begin{figure}[!h]
    \centering
    \includegraphics[width=3.35in]{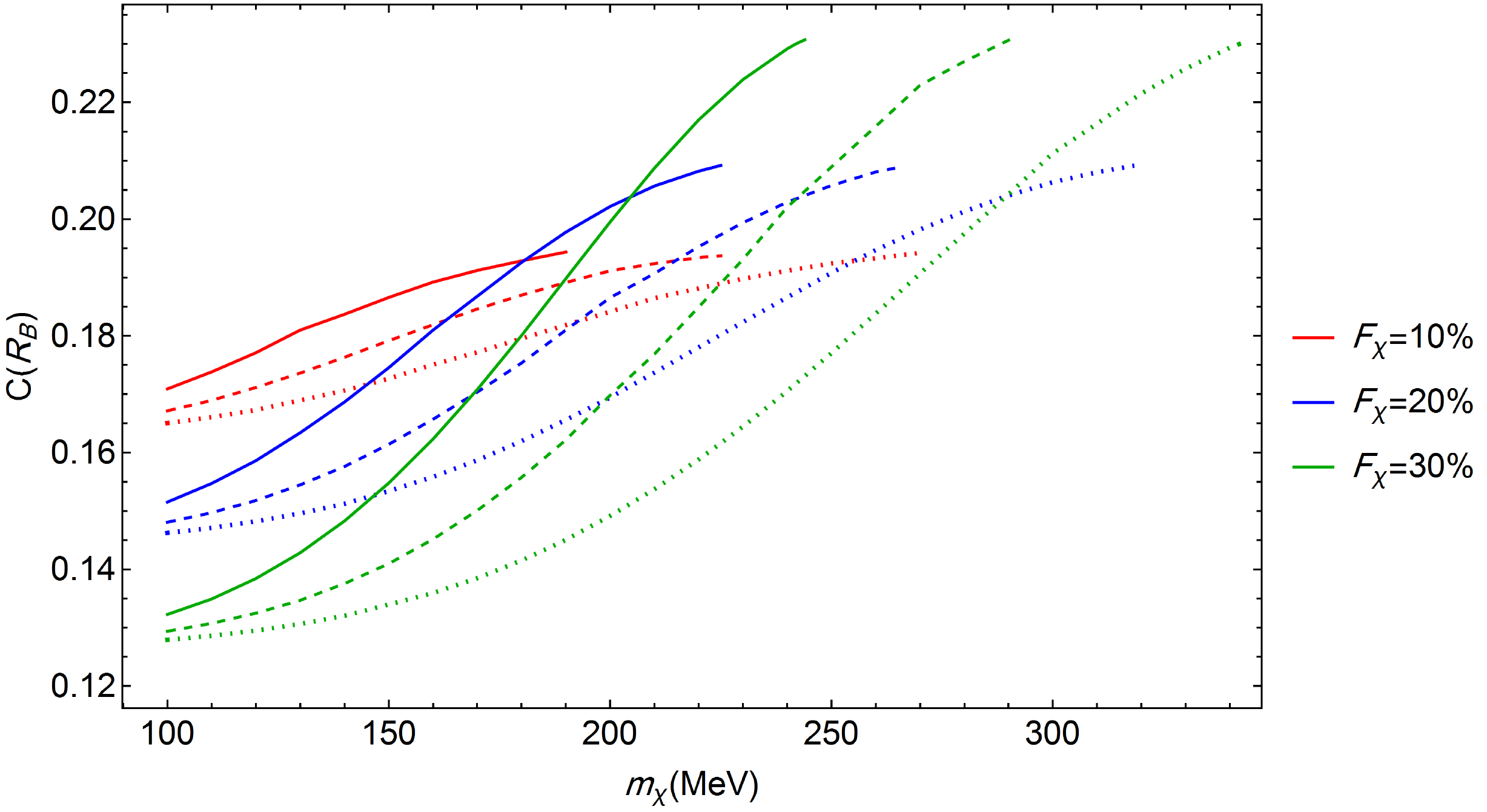}
    \includegraphics[width=3.35in]{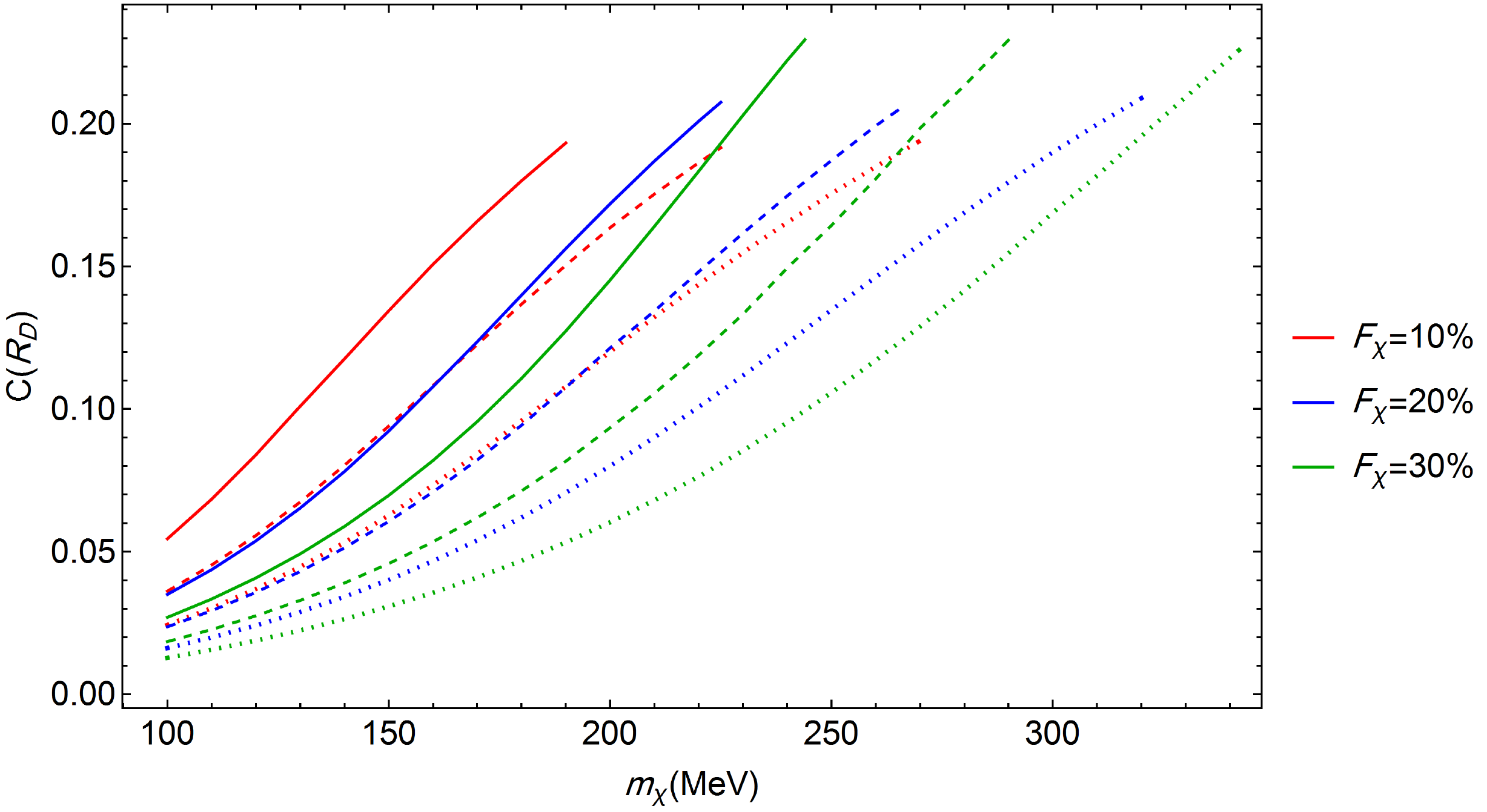}
    \caption{Compactness up to the visible radius  $C(R_{B})=M_{T}(R_{B})/R_{B}$ (top), and till the dark radius  $C(R_{D})=M_{T}(R_{D})/R_{D}$ (down), as a function of boson masses for various DM fractions and self-coupling constants. In all of the cases the total mass of the objects are $M_{T}(R_{D})=1.4M_{\odot}$ while $F_{\chi}$ is varied from 10\% to 30\% as labeled and coupling constants $\lambda=0.5\pi$, $\pi$ and $2\pi$  are shown by solid, dashed and dotted lines, respectively. All of the diagrams are plotted for IST EoS and in the presence of DM halo until the $m_{\chi}$ for which $R_{D}\approx R_{B}$ indicating a marginal radius to form DM core. 
    }
    \label{RBD6}
\end{figure}

As it was  discussed in the preceding sections, the bosonic DM is confined inside a BM shell generally for heavy DM particles  while the light DM particles usually form  halo and distributed up to large radius out of the NS. The presence of DM inside or around the NS
would change the compactness  and  
the gravitational potential of the object,  consequently, it contributes to the deflection of the light originating from the visible surface of the star. The compactness of  DM admixed NSs is displayed in Fig. \ref{RBD6} for IST EoS as a function of boson masses, self-coupling constants and for different DM fractions as labeled. Here we consider the total compactness of the mixed star up to the  visible and dark radius as $C(R_{B})=M_{T}(R_{B})/R_{B}$ and $C(R_{D})=M_{T}(R_{D})/R_{D}$, respectively. We see in Fig. \ref{RBD6}  that the total compactness at $R_{D}$ and $R_{B}$ 
increases with the boson masses, while the increasing rate of the BM core compactness is larger for higher DM fractions towards more massive bosons. Along with each curves for a constant $F_{\chi}$ towards massive bosons, the radius of DM halo and BM core decrease  and $M_{T}(R_{B})$ increases which lead to   forming a more compact object. We also examine the impact of different values of self-coupling constants (i.e. $\lambda=0.5\pi$, $\pi$ and $2\pi$)  on the compactness of the DM admixed NS in Fig. \ref{RBD6}. Assuming a constant DM fraction for a given boson mass, it is seen that for larger values of $\lambda$ the total compactness decreases, this effect is more evident for larger $m_{\chi}$. Note that the maximum compactness for various self-coupling constants is equal for each fraction which happens in $R_{B}\approx R_{D}$ at the end point of each curves where $C(R_{B})\approx C(R_{D})$. In both upper and lower panels of  Fig. \ref{RBD6}  for light bosons $m_{\chi}\lesssim200$ MeV, increasing the DM fraction for a fixed self-coupling constant lead to a decrease in the compactness. Note that DM admixed NS with DD2 EoS will follow similar behaviors in Fig. \ref{RBD6} for $C(R_{B})$ and $ C(R_{D})$.  

\begin{figure}[!h]
    \centering
    \includegraphics[width=3.35in]{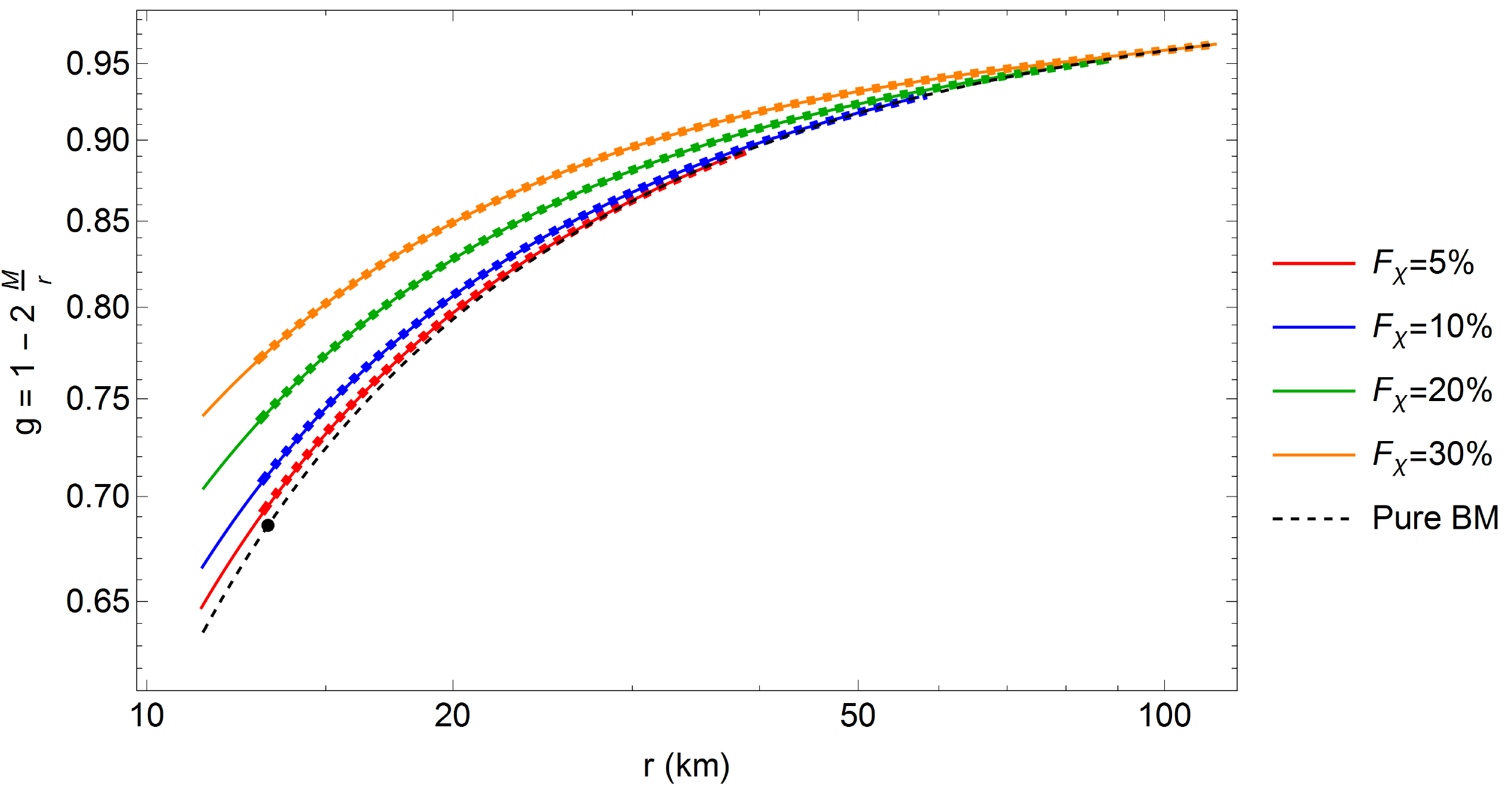}
    \includegraphics[width=3.35in]{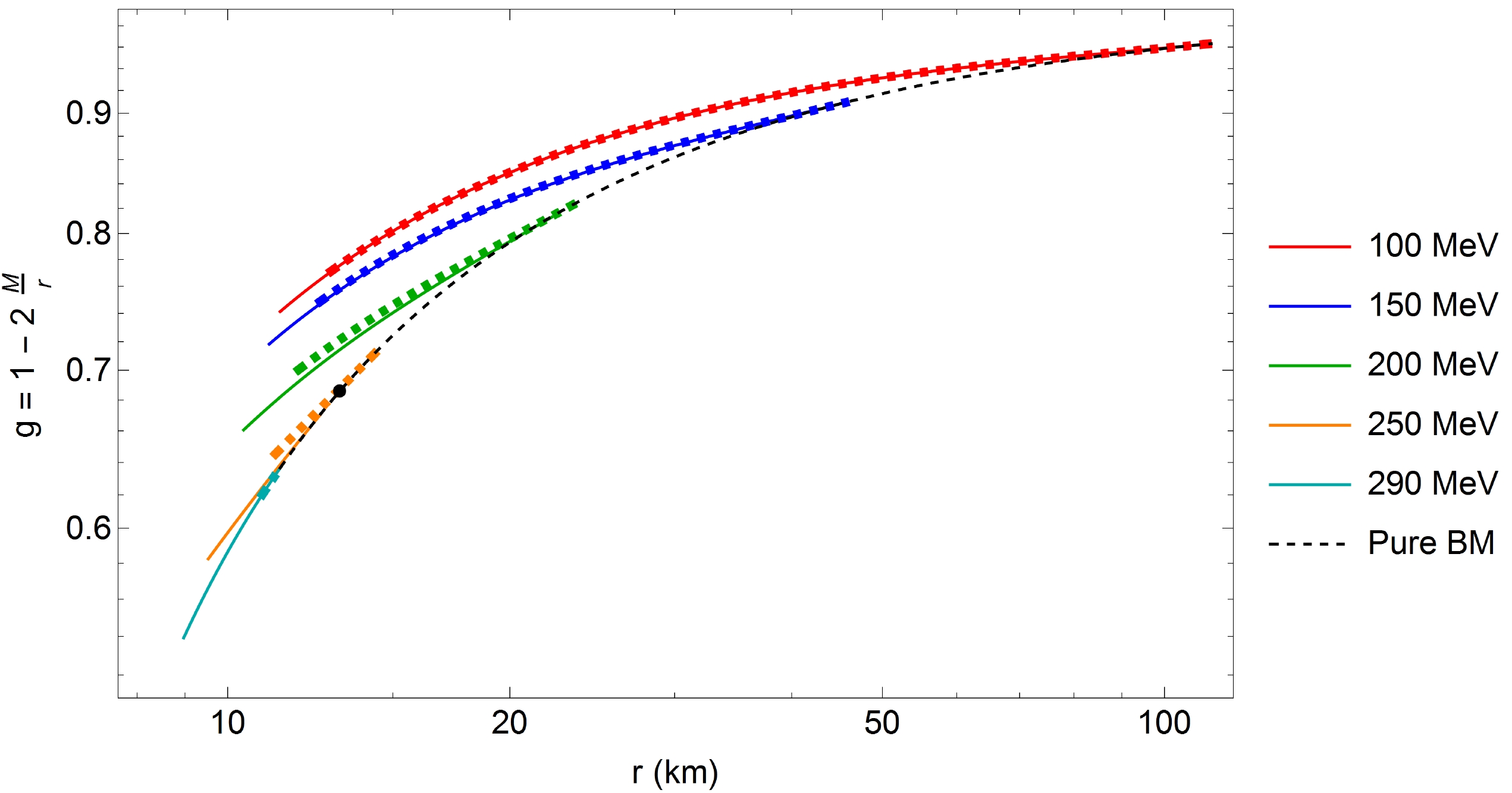}
    \includegraphics[width=3.35in]{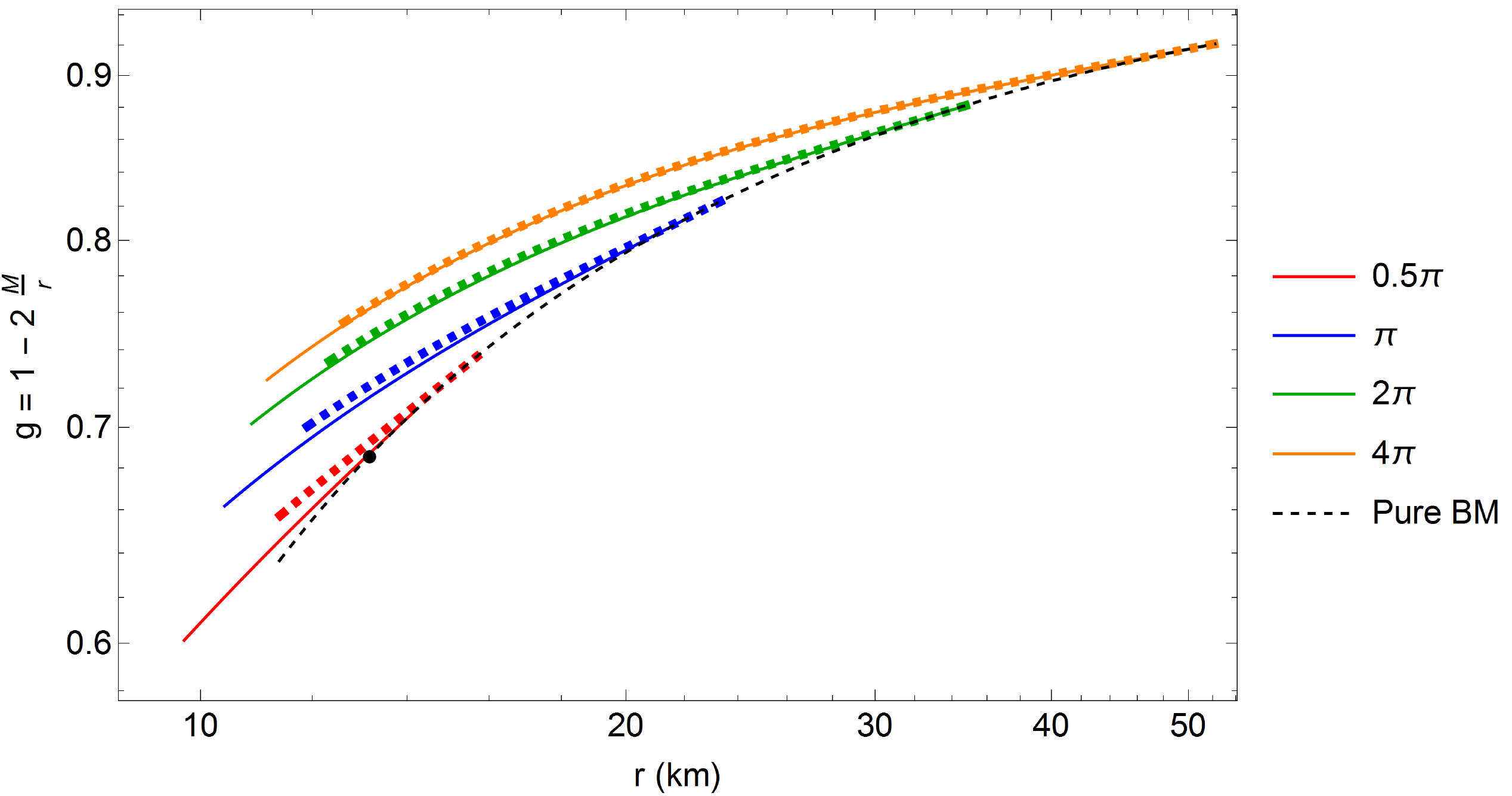}
    \caption{Variation of the metric  function $g(r)=1-2M(r)/r$ with respect to  radial distance : for various DM fractions, $m_{\chi}=100$ MeV and $\lambda=\pi$ (top); for different boson masses, $F_{\chi}=30\%$ and $\lambda=\pi$ (middle); and for $m_{\chi}=200$ MeV and $F_{\chi}=30\%$ by considering various coupling constants (down). For all the considered DM admixed NSs the variation of $g(r)$ is plotted from $R_{B}$ till $R_{D}$ where the halo is ended and $M_{T}(R_{D})=1.4M_{\odot}$. The curves related to  IST and DD2 BM EoSs are indicated by solid and dashed lines, respectively. The dashed black curves represent the pure NSs 
    where IST starts at 11.37 km and DD2 starts at 13.15 km (marked by black dots) and continues up to the largest DM halo radius in each plots.}
    \label{RBD7}
\end{figure}

In order to compute the pulse profile of a DM admixed NS, we need to determine the  metric function $g(r)$ outside the surface of NS taking into account the DM halo contribution which is shown in Fig. \ref{RBD7}. The variation of $g(r)$ in terms of different DM fractions, boson masses and self-coupling constants  are illustrated as a function of  radial distance starting from the corresponding $R_B$ for each EoSs. All curves originated from distinct points as a result of different values of $M_{T}(R_{B})/R_{B}$, and towards larger radius all the way to $R_{D}$, the lines merge to the pure BM case with $M=1.4 M_{\odot}$. Moreover, we see that the value of  $g(r)$ corresponding to IST EoS (solid lines) is started from lower values compared to DD2 EoS (dashed lines) which is due to the higher compactness of the former BM fluid. Note that for an observer outside the DM halo radius, the mixed object is similar to a star with total mass $1.4 M_{\odot}$. 

We consider the trajectory of  photons  emitting from a small spot ($dS'$) on the star surface (BM core) where the emission angle $\alpha$ is evaluated with respect to the normal vector of the emission surface. The star is spinning and the visible spot changes its position periodically and light rays would be bent by  gravity and reach an observer at a far distance $D$ appearing as observed pulse profiles.

The bolometric observed flux of a visible spot is given by \cite{Beloborodov:2002mr,Poutanen:2006hw}:
\begin{eqnarray}
\mathcal{F} = g(R_B) \delta^{5} I'(\alpha')\cos \alpha \frac{d\cos\alpha}{d\cos \Psi} \frac{dS'}{D^{2}}\,,
\end{eqnarray}
here $\delta$ is the Doppler factor and appears due to the rotation of the star, in the case of a slowly rotating star $\delta=1$.
The local intensity of radiation $I'(\alpha')$ depends on the applied emission model, in the following without loss of generality, we normalize the pulse profile to value $\mathcal{F}_{0}=I'(\alpha') dS'/D^{2}$   
. The bending angle $\Psi$ which is defined as an angle between the line of sight (LOS) and the local radial direction of a visible spot on $R_B$ is as follows
\begin{eqnarray}\label{e14}
\cos \Psi = \cos i \cos \theta + \sin i \sin \theta \cos \phi\,,
\end{eqnarray}
where $i$ measures the angle between the rotational axis of the star and LOS, $\theta$ is the colatitude of the visible spot and $\phi$ is the rotational phase of the pulsar.  The bending angle can also be extracted through general relativity computations  using the null geodesic equation taking into account modified metric function $g(r)$ in the presence of DM halo outside $R_{B}$ as \cite{Misner:1973prb}:
\begin{eqnarray}\label{psi}
 \Psi = \int_{R_{B}} ^\infty \frac{dr}{r^{2}} \Big[\frac{1}{b^2}-\frac{g(r)}{r^{2}} \Big]^{-1/2}\,,
\end{eqnarray}
where  $\it{b}$ is the impact parameter which is known as the distance between the LOS and the point that the photon hits the observer sky at infinity and is connected to  $\alpha$ as

\begin{eqnarray}\label{bimp}
b=R_{B} \sin \alpha / \sqrt{g(R_B)}\,,
\end{eqnarray}

The maximum value of $\Psi$ is obtained at $\alpha=\pi/2$ and the visibility condition of a spot is defined by $\Psi<\Psi_{max}(\alpha=\pi/2)$.  In order to compute the normalized pulse profile $\mathcal{F}/\mathcal{F}_{0}$, one  needs to determine $\Psi$ and $\alpha$ and the relation between them through Eqs. (\ref{psi}) and (\ref{bimp}) for a given $R_{B}$. Note that in Eq. (\ref{bimp}) the metric function is evaluated at the baryonic surface  where the radiation originated while in Eq. (\ref{psi}) we use $g(r)$ in which the contribution of DM halo is taken into account (see Fig. \ref{RBD7} ). In order to express the pulse profile in terms  of the pulsar phase $\phi$, we have to construct a correspondence between the pulsar phase $\phi$ and the bending angle $\Psi$ via Eq. (\ref{e14}) on one hand and a combination of Eqs. (\ref{psi}) and (\ref{bimp}) on the other hand following the approach presented in \citep{Poutanen:2006hw}.  

\begin{figure}[!h]
    \centering
    \includegraphics[width=3.5in]{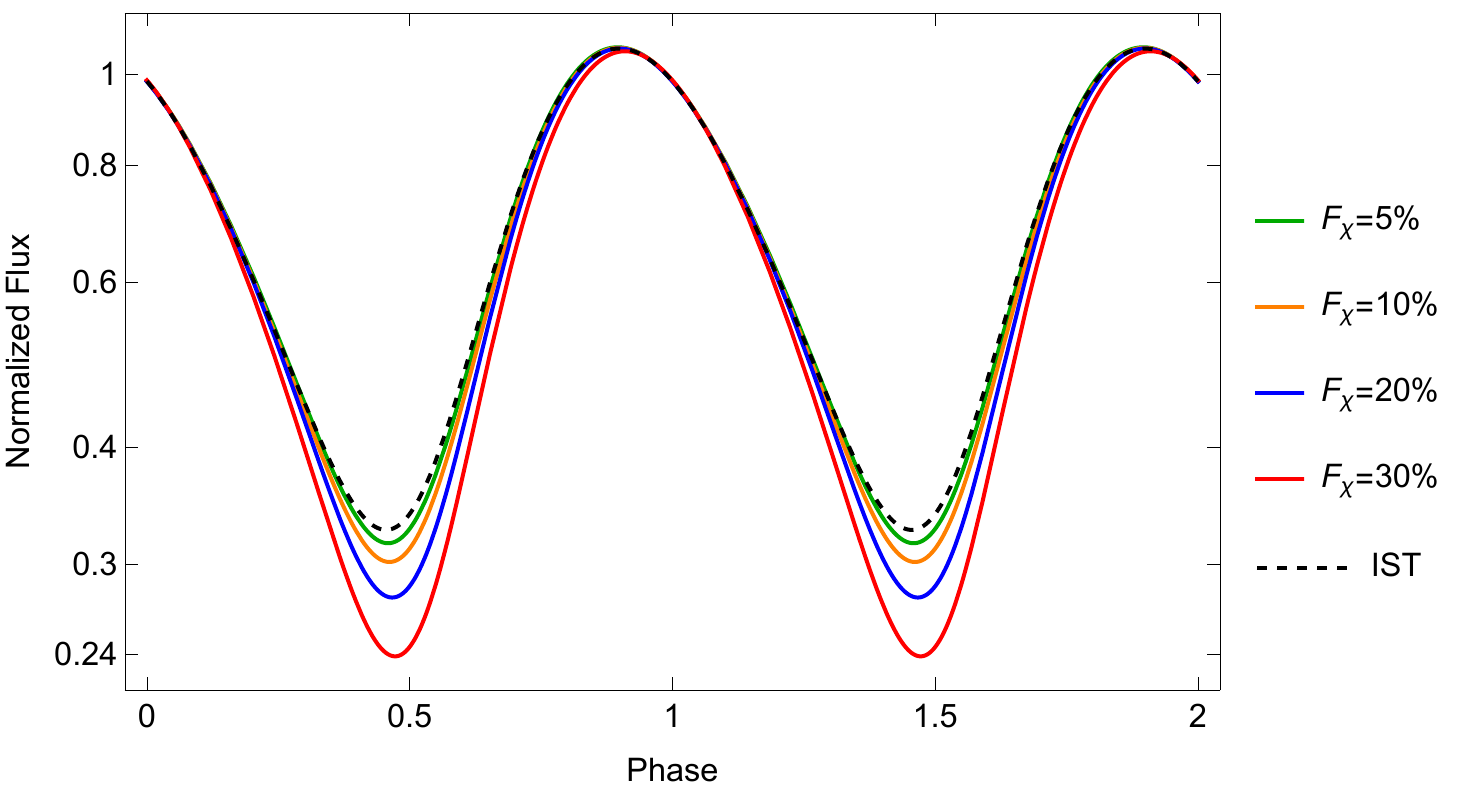}
    \includegraphics[width=3.5in]{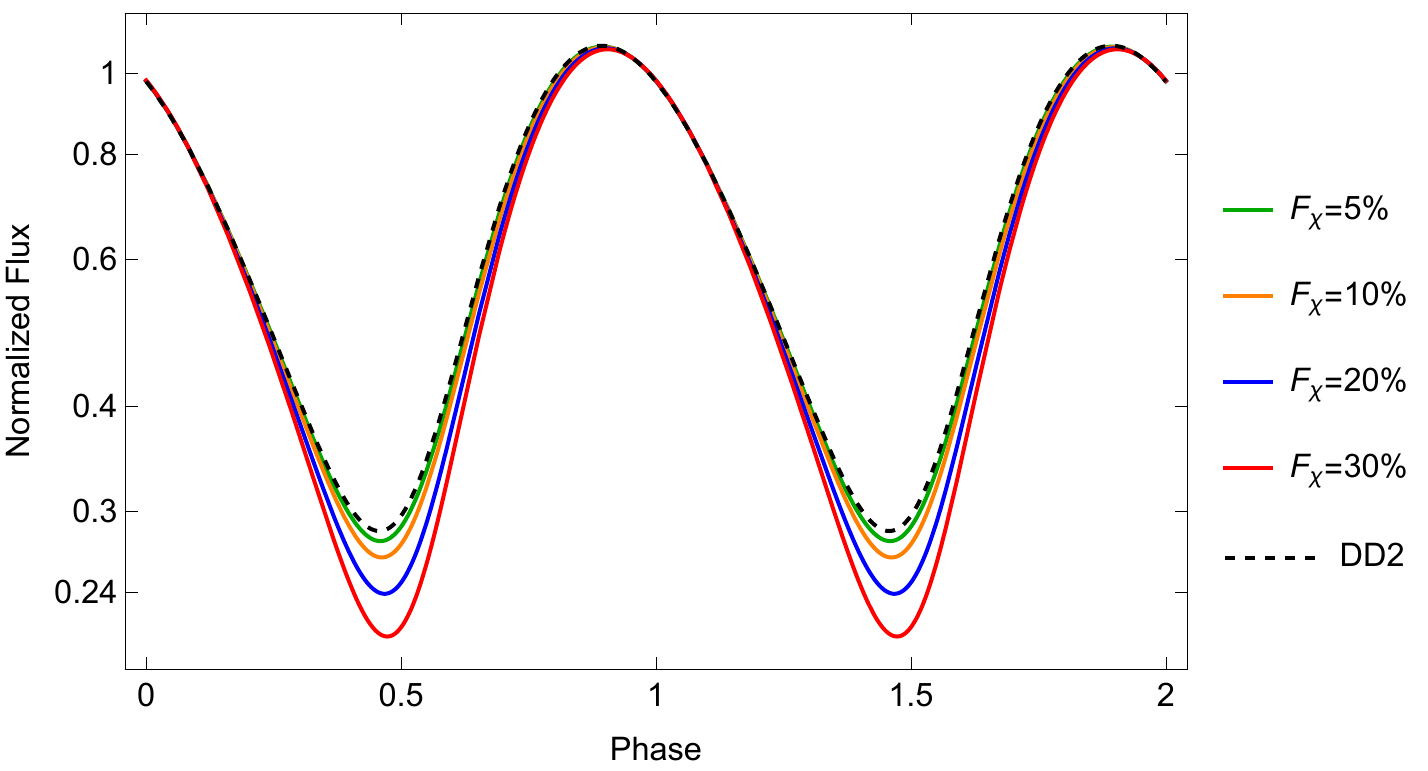}
     \caption{ The bolometric pulse profile of a visible spot on the BM  surface of a DM admixed NS as a function of observed rotational phase. The gravitational light bending has taken into account for a rotating DM admixed NS ($\nu=400 \text{Hz}$) with total mass $M_{T}(R_{D})=1.4M_\odot$.  We assume the inclination angle of the spin axis with respect to the line of sight $i=\pi/4$ and the colatitude of the visible spot $\theta=\pi/4$.  The DM fraction is changing from $5\%$ to $30\%$ for $m_{\chi} = 100 \, \text{MeV}$ and $\lambda=\pi$. The BM component is modeled with IST and DD2 EoSs in upper and lower panels, respectively. The pulse profile of the pure  NS is shown by black dashed lines.}
    \label{RBD-FxL}
\end{figure}

\begin{figure}[!h]
    \centering
    \includegraphics[width=3.45in]{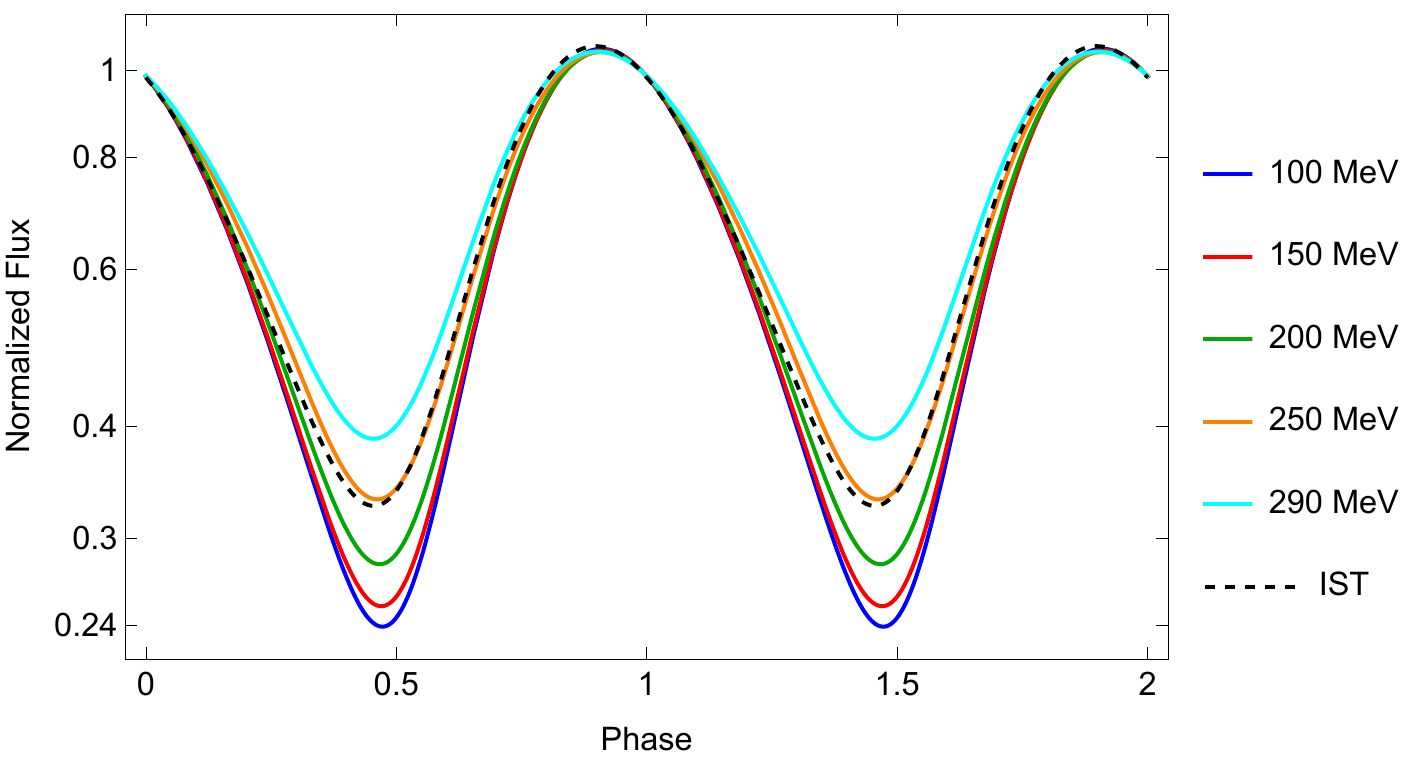}
    \includegraphics[width=3.4in]{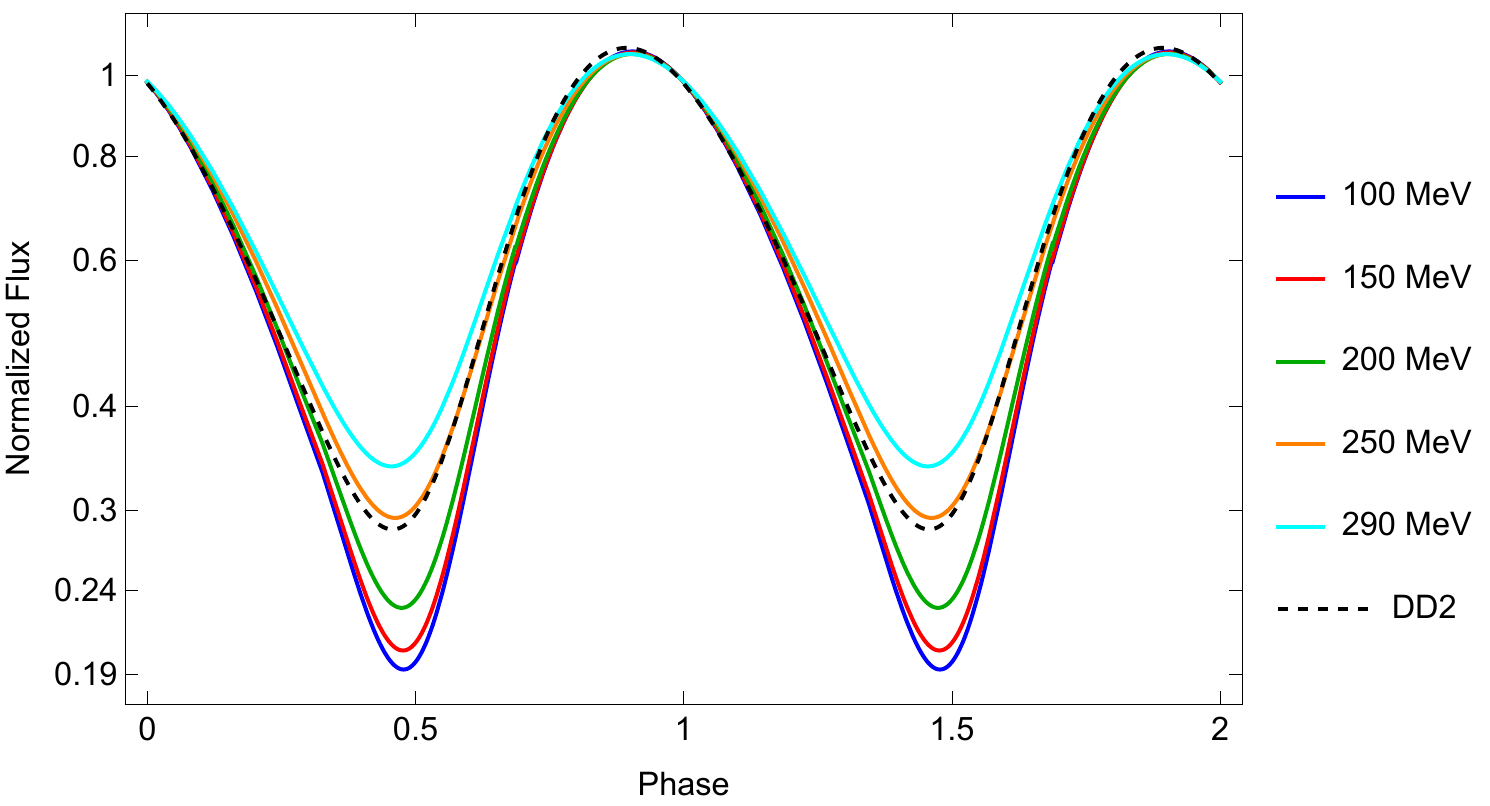}
     \caption{The pulse profiles are depicted similar to Fig. \ref{RBD-FxL},  but for various boson masses as labeled, considering $F_{\chi}=30\%$ and  $\lambda=\pi$.}
     \label{PM mx}
\end{figure}

\begin{figure}[!h]
    \centering
    \includegraphics[width=3.4in]{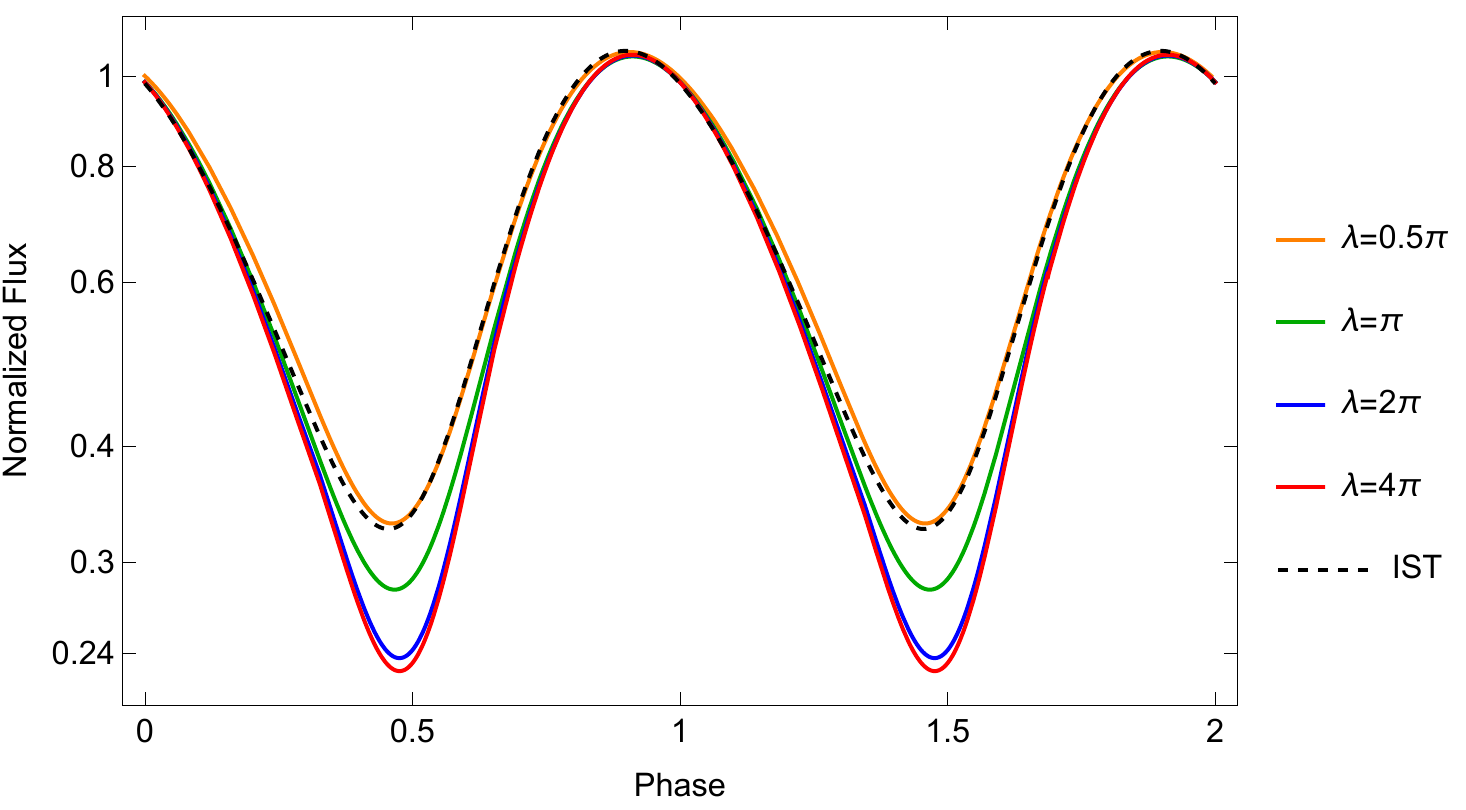}
    \includegraphics[width=3.4in]{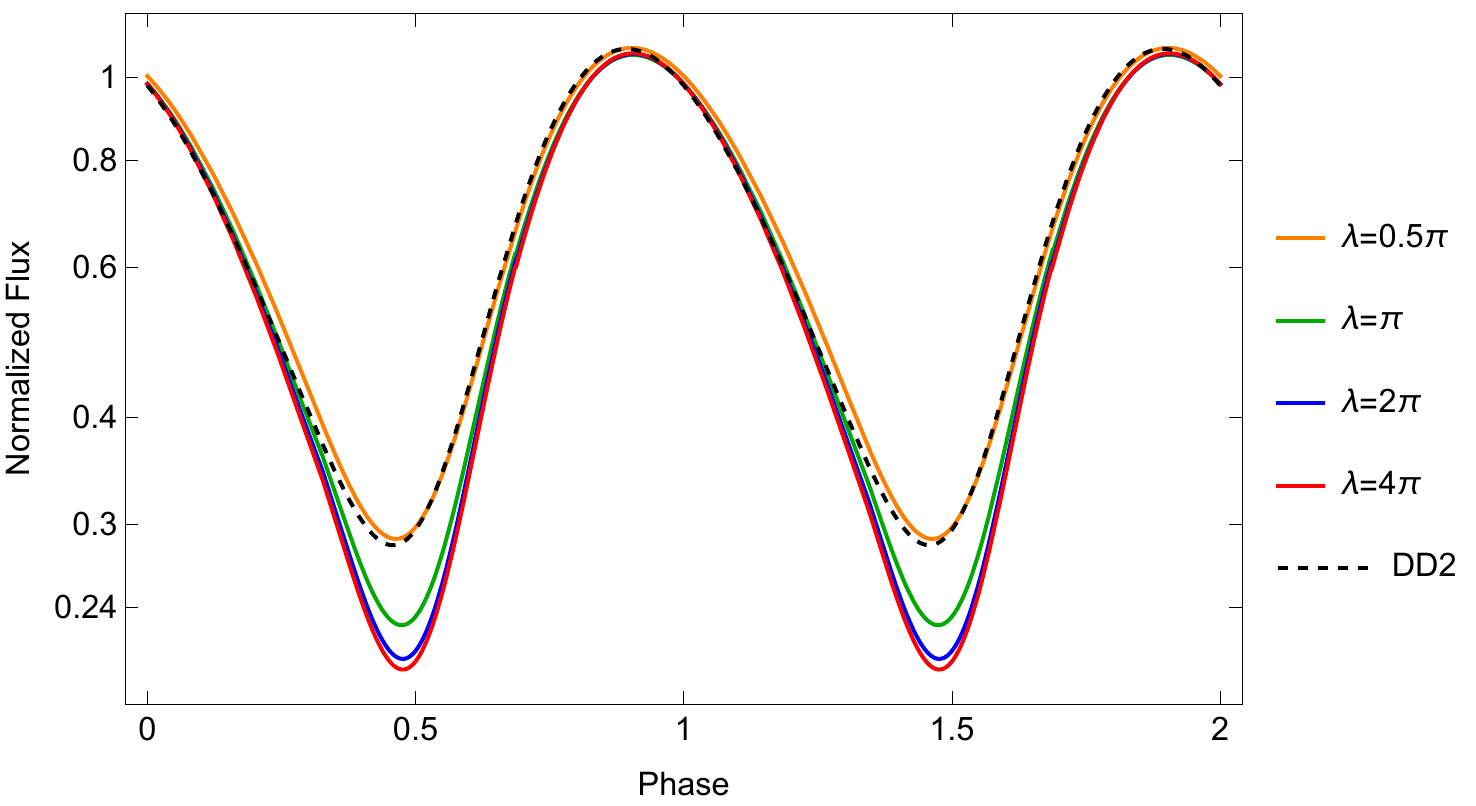}
     \caption{The pulse profiles are depicted similar to Fig. \ref{RBD-FxL},  but for various self-coupling constants as labeled, considering $F_{\chi}=30\%$ and $m_{\chi}=200$ MeV.}
    \label{RBD-FxL3}
\end{figure}

The pulse profile of DM admixed NSs for different  DM fractions, boson masses and coupling constants are displayed in Figs. \ref{RBD-FxL}- \ref{RBD-FxL3} in which upper and lower panels related to IST and DD2 EoSs, respectively. In these figures the total mass of the mixed object at $R_{D}$ is assumed to be $1.4M_\odot$, the angular position of the visible spot is defined by $i=\theta=\pi/4$ and the rotation frequency of the star is $400\text{Hz}$. The fall and growth  of the flux (even in pure BM shown by dashed line) as a function of the rotational phase is due to changing the position of the spot compared to  a distant observer, the minimum flux corresponds to the far-side position. The depth of the minimum  crucially depends on the compactness of the object which affects the  gravitational light-bending and alters the invisible surface of the star, the less compact object gives more deeper minimum. In Fig. \ref{RBD-FxL}, we consider bosonic SIDM with mass $m_{\chi}=100$ MeV and self-coupling constant $\lambda=\pi$ which distributes as a halo around the star, by increasing the amount of DM  the minimum fluxes of the profiles are decreased. As it was mentioned earlier, the compactness of the object encodes in the minimum depth, it is seen that the higher DM fraction leads to formation of less compact objects (see Fig. \ref{RBD6}). The deviation of the minimum amplitudes of the highest considered fraction $F_{\chi}=30\%$ from pure NS (i.e. $\Delta \mathcal{F}/\mathcal{F}^{\text{BM}}_{\text{min}}= (\mathcal{F}^{\text{BM}}_{\text{min}}-\mathcal{F}^{30\%}_{\text{min}})/\mathcal{F}^{\text{BM}}_{\text{min}}$) for IST (DD2) EoS is about $27\%$ (25.3\%) which is a notable effect of the presence of DM halo around NS. It is worth mentioning that the difference between the total compactness at $R_{B}$ for DM admixed NS and pure NS $(\Delta \mathcal{C}(R_{B}))$ over $\mathcal{C}(R^{\text{BM}}_{B})$  following approximately similar ratio which is about  $\Delta \mathcal{C}(R_{B})/\mathcal{C}(R^{\text{BM}}_{B})\approx28.7\%$ (27.3\%) for  DM fraction $30\%$  (see Tables \ref{table:2} and \ref{table2}).  These values for IST (DD2) EoS, $m_{\chi}=100$ \text{MeV} and $F_{\chi}=20\%$, 10\% and $5\%$ are $\Delta \mathcal{F}/\mathcal{F}^{\text{BM}}_{\text{min}},\Delta \mathcal{C}(R_{B})/\mathcal{C}(R^{\text{BM}}_{B})\approx 15.3\%,18.2\%$ $(15.8\%,17.2\%)$, $7.6\%,7.7\%$ $(7.4\%,7.6\%)$ and $3\%,2.8\%$  $(2.8\%,2.5\%)$, respectively. It is remarkable to see that there is an approximate linear correlation between the deviation in the minimum fluxes and the compactness at $R_{B}$ for different $F_{\chi}$. Such a relation may not be seen for more massive bosons due to relatively small $R_{B}$ as a result of experiencing a significant  DM core formation along with higher DM fractions before forming the halo.

The dependency of the pulse profile to boson masses is examined in Fig. \ref{PM mx} for IST (upper panel) and DD2 (lower panel) EoSs, where the DM fraction and the  self-coupling constant  are fixed at $F_{\chi}=30\%$ and  $\lambda=\pi$, respectively. We expect to have more compact objects at baryonic radius for larger $m_{\chi}$ which makes the spot more visible during star's rotation and rises the minimum flux for massive bosons. Since the compactness $\mathcal{C}(R_{B})$ associated with boson masses $m_{\chi}=250$ and $\ 290 $ MeV are 
higher than the compactness of a pure NS,  the  minimum luminosities  are observed  at upper values  compared to pure NS described by IST (DD2) EoS, $\Delta \mathcal{F}/\mathcal{F}^{\text{BM}}_{\text{min}}=2.1\%$ (2.8\%) and $19.6\%$ (19.3\%), respectively. On the other hand, the BM core of mixed objects  composed of bosonic SIDM particles with masses 100, 150 and 200  MeV have lower compactness and hence lower minimum fluxes of the pulse profiles are seen.  It is worth mentioning that for $m_{\chi}=250$ MeV with $30\%$ fraction of DM, we have approximately  the same pulse profile compared to pure NS, owing to the fact that DM is distributed in a radius relatively close to the normal NS. 

Furthermore, for the sake of completeness, the impact of the coupling constant has been investigated in Fig. \ref{RBD-FxL3} for $m_{\chi}=200$ MeV and $F_{\chi}=30\%$. It is seen that for both IST (upper panel) and DD2 (lower panel) EoSs, higher values of self-coupling constants lead to less compact DM admixed NSs which reduce the minimums of the corresponding pulse profiles. We see that for $\lambda=0.5\pi$ the pulse profiles are similar to the pure NS case because of the close values of $\mathcal{C}(R_{B})$.

It is worth noting that the general behaviour of the pulse profiles observed in Figs. \ref{RBD-FxL} - \ref{RBD-FxL3} can be 
interpreted by different compactness of the BM cores (see Fig. \ref{RBD6}, Tables \ref{table:2} and \ref{table2}). It is found that the compactness of IST EoS is higher than DD2 EoS, thus the minimum values of the pulse profiles are lower in all the corresponding DM admixed NSs with DD2 BM component.

\begin{table*}[ht]
  \centering
 \begin{tabular}{|| c | c | c | c | c | c | c | c ||} 
 \hline
 $m_{\chi}$ (MeV) & $F_{\chi}$ & $\lambda$ & $M_{T}(R_{B}) (M_{\odot}$) & $R_{B}$ (km) & $R_{D}$ (km) & $\mathcal{C}(R_{B})$ & $\mathcal{C}(R_{D})$  \\ [0.5ex] 
 \hline\hline
 100 & 5\% & $\pi$ & 1.352
 & 11.308
 & 37.378 & 0.176 & 0.055  \\
 \hline
 100 & 10\% & $\pi$ & 1.282
 & 11.331
 & 57.201 & 0.167 & 0.036  \\
 \hline
 100 & 20\% & $\pi$ & 1.139 & 11.365
 & 87.515
 & 0.148 & 0.023  \\
 \hline
 100 & 30\% & $\pi$ & 0.995 & 11.363
 & 112.133
 & 0.129 & 0.018 \\  
 \hline
 150 & 30\% & $\pi$ & 1.056 & 11.061
 & 45.121
 & 0.140 & 0.045 \\  
 \hline
 200 & 30\% & $0.5\pi$ & 1.314 & 9.729
 & 14.239
 & 0.199 & 0.145 \\ 
 \hline
 200 & 30\% & $\pi$ & 1.19 & 10.392
 & 22.222
 & 0.169 & 0.093 \\ 
 \hline
 200 & 30\% & $2\pi$ & 1.097 & 10.86
 & 34.31
 & 0.149 & 0.06 \\ 
 \hline
 200 & 30\% & $4\pi$ & 1.04 & 11.137
 & 51.747
 & 0.137 & 0.039 \\ 
 \hline
 250 & 30\% & $\pi$ & 1.347
 & 9.522
 & 12.582
 & 0.208
 & 0.164 \\
 \hline
 290 & 30\% & $\pi$ & 1.399 & 8.962
 & 9.012
 & 0.23 & 0.229 \\ [0.5ex]
\hline
\end{tabular}
\caption{Properties of DM admixed NSs with $M_{T}(R_{D})=1.4M_{\odot}$ for different boson masses, coupling constants and DM fractions as listed.   Here the BM component is described by IST EoS for which the corresponding radius of a $1.4 M_{\odot}$ pure NS is $11.37$ \text{km} and the compactness is
$\mathcal{C}=M/R=0.181$.}\label{table:2}
\end{table*}

\begin{table*}[ht]
  \centering
 \begin{tabular}{|| c | c | c | c | c | c | c | c ||} 
 \hline
 $m_{\chi}$ (MeV) & $F_{\chi}$ & $\lambda$ & $M_{T}(R_{B}) (M_{\odot}$) & $R_{B}$ (km) & $R_{D}$ (km) & $\mathcal{C}(R_{B})$ & $\mathcal{C}(R_{D})$  \\ [0.5ex] 
 \hline\hline
 100 & 5\% & $\pi$ & 1.353
 & 13.031
 & 38.561 & 0.153 & 0.053  \\
 \hline
 100 & 10\% & $\pi$ & 1.285
 & 12.997
 & 57.847 & 0.145 & 0.035  \\
 \hline
 100 & 20\% & $\pi$ & 1.142 & 12.941
 & 87.621
 & 0.13 & 0.023  \\
 \hline
 100 & 30\% & $\pi$ & 0.998 & 12.884
 & 112.064
 & 0.114 & 0.018 \\  
 \hline
 150 & 30\% & $\pi$ & 1.067 & 12.541
 & 45.601
 & 0.125 & 0.045 \\  
 \hline
 200 & 30\% & $0.5\pi$ & 1.321 & 11.38
 & 15.87
 & 0.171 & 0.13 \\ 
 \hline
 200 & 30\% & $\pi$ & 1.206 & 11.895
 & 23.315
 & 0.149 & 0.088 \\ 
 \hline
 200 & 30\% & $2\pi$ & 1.111 & 12.332
 & 34.972
 & 0.133 & 0.059 \\ 
 \hline
 200 & 30\% & $4\pi$ & 1.049 & 12.624
 & 52.194
 & 0.122 & 0.039 \\ 
 \hline
 250 & 30\% & $\pi$ & 1.35
 & 11.237
 & 14.357
 & 0.177
 & 0.143 \\
 \hline
 290 & 30\% & $\pi$ & 1.399 & 10.879
 & 11.069
 & 0.189 & 0.186 \\ [0.5ex]
\hline
\end{tabular}
\caption{Similar to Table. \ref{table:2}, but for the DD2 EoS as the BM fluid for which the radius of a pure $1.4 M_{\odot}$ NS is $13.15$ \text{km} and the compactness equals to
$\mathcal{C}=M/R=0.157$.}\label{table2}
\end{table*}

\begin{figure}[!h]
    \centering
    \includegraphics[width=3.5in]{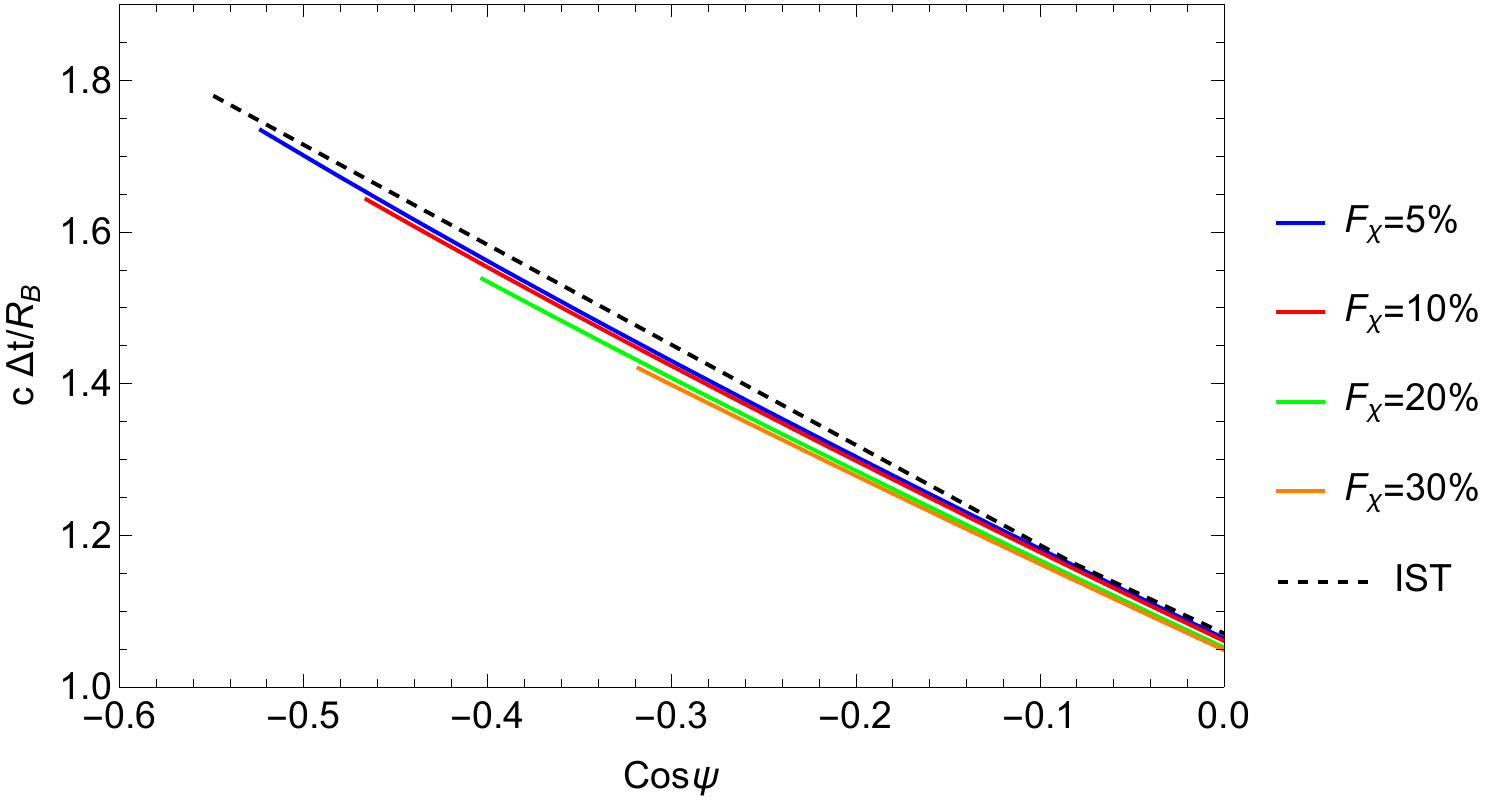}
    \includegraphics[width=3.5in]{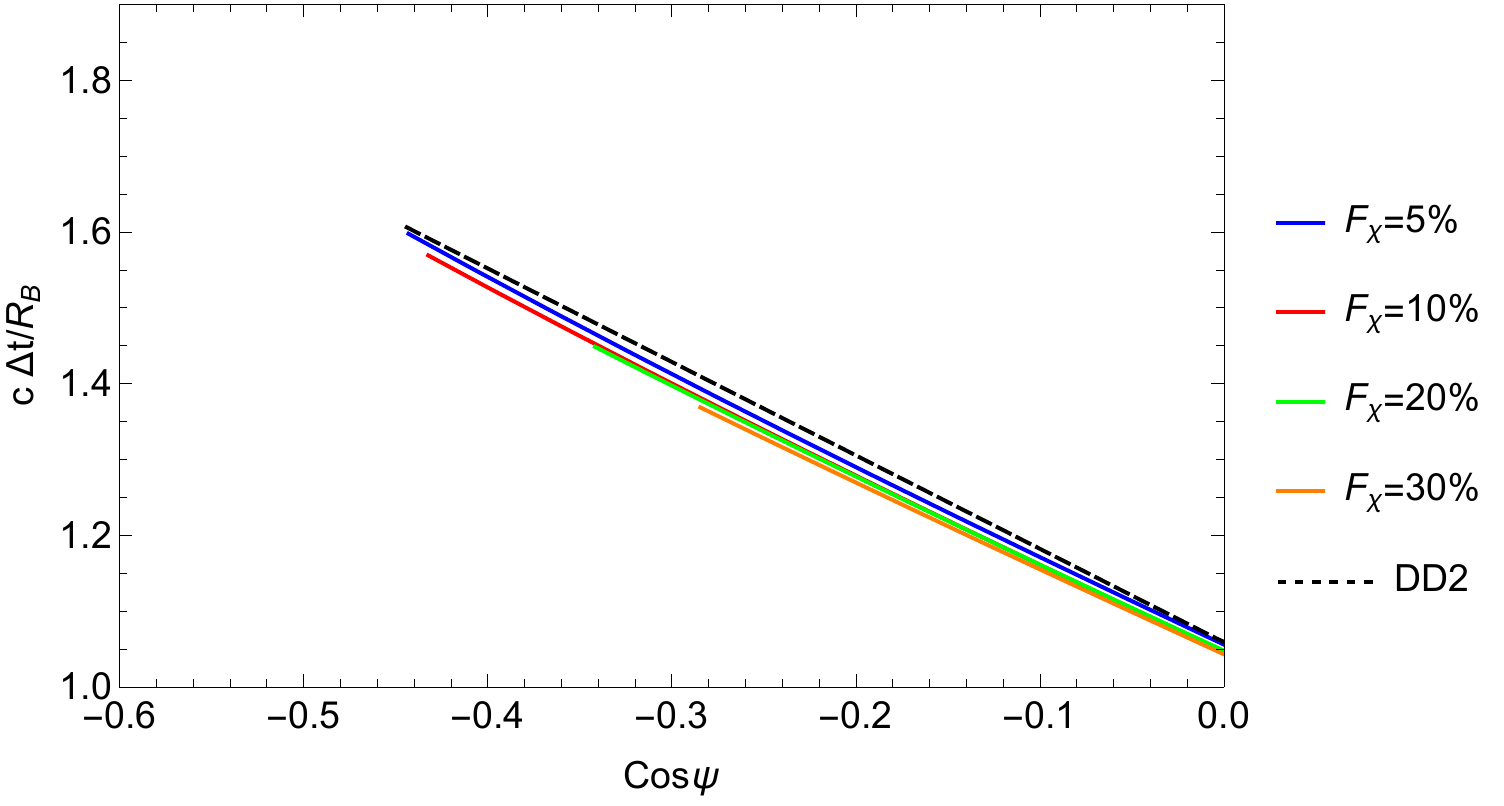}
    \caption{ Time delay of surface photons as a function of bending angle for different fractions of DM, considering IST (up) and DD2 (down) EoSs as BM fluids where $m_{\chi}=100$ MeV and $\lambda=\pi$. Time delay is scaled by $R_{B}/c$  which is the time needed for photons to propagate radially from the center of the star to the baryonic surface. The corresponding time delay for the pure NS is shown by black dashed lines.}
    \label{timedelay1}
\end{figure}

Emitted photons depending on their initial positions on NS surface will pass trajectories with different lengths. One can define the difference of photon arrival times to an observer sky at infinity with respect to a specific reference point, which is selected to be aligned with LOS \citep{Poutanen:2006hw}. In fact, the time delay of photons is scaled with the travel time of photons emitted from the closest region   to the observer  
as $\Delta t = t(b) - t(0)$, since the integral of photon geodesic equation is diverged for vanishing impact parameter. The scaled time delay is given by

\begin{eqnarray}\label{delay}
\Delta t (b)= c^{-1}\int_{R_{B}} ^\infty \frac{dr}{g(r)}
\Big[ \Big(1-\frac{b^{2}g(r)}{r^{2}} \Big)^{-1/2}-1  \Big]\,,
\end{eqnarray}
the above time delay should also be taken into account in the photon arrival phase to the observer $\phi_{obs}=\phi+2\pi\nu\Delta t [b(\phi)]$ \citep{Poutanen:2006hw}, note that for a slowly rotating star $\phi_{obs}\approx\phi$ and the phase shift is negligible.  The time delay  (\ref{delay}) scaled with $R_{B}/c$  is illustrates in Fig. \ref{timedelay1} for $m_{\chi}=100$ MeV, $\lambda=\pi$ and different $F_{\chi}$ as labeled. Due to the increasing of the compactness $\mathcal{C}(R_{B})$ for lower fractions, the higher gravitational light bending enhances the time delay. In fact the baryonic core with larger compactness produces higher values of gravitational potential around the visible surface and causes more deflection  of light  and therefore the time dilation  will be increased. Notice that for DM admixed NSs described by IST EoS (upper panel),  the time dilation is larger than the ones with DD2 EoS (lower panel), due to the fact that IST EoS provides  more compact objects. As it is shown  the time delay in our case has almost negligible effect which is appeared  as a very small phase shift in the corresponding pulse profile of the DM admixed NS compared to pure NS.

In this section, we showed that the presence of DM halo will change the geometry around  NS and therefore it modifies the light trajectory and light bending effect in the vicinity of  DM admixed NSs.  Our results show that the mixed objects should be considered as a possibility  in the numerical simulation codes based on ray tracing methods in order to interpret observations of X-ray telescopes  from compact objects. 
Therefore, by taking into account DM halo around compact objects, the observational constraints may change and this will also affect the interpretation of those compact objects and the determination of their mass and radius. It is seen that the compactness of the mixed object, which depends on DM model parameters and fractions, plays an important role to impact the observed pulse profile and the minimum flux during star rotation.
Recently, the effect of DM halo on the pulse profile of emitted photons  from DM admixed NSs has been investigated  in  \cite{Miao:2022rqj} where the total mass  up to the baryonic radius $R_{B}$ is assumed to be $M_{T}(R_{B})=1.4M_{\odot}$ and obviously the total  gravitational mass of the object (i.e. up to $R_{D}$) is more than $1.4M_{\odot}$.  In that study, the halo mass $(M_{halo})$ was  defined as the mass which occupies in range $R_{B} \leq r\leq R_{D}$  and deviation of  the peak flux compared to pure NS depends on $M_{halo}/R_{D}$ with an approximately linear relation. In \cite{Miao:2022rqj} in addition to applying different BM/DM model, contrary to our study, $\mathcal{C}(R_{B})$ is the same for all considered cases and instead the compactness of DM halo is changed. In our work we use a different definition for DM halo mass including all DM content and starting from $r=0$ all the way to $R_{D}$, besides both $\mathcal{C}(R_{B})$  and $\mathcal{C}(R_{D})$ change with $m_{\chi}$, $\lambda$ and $F_{\chi}$ while the total gravitational mass of the DM admixed NS ($M_{T}(R_{D})$) is fixed at $1.4M_{\odot}$ for all cases which is more reasonable assumption for comparison. In fact in order to consider the validity of our scenario, there could be other observational methods to measure the total gravitational mass of the object rather than the surface emission measured by X-ray telescopes. 
The effect of DM halo in  \cite{Miao:2022rqj} is more significant at the peak of the pulse profile, while in our study  the modification in the observed luminosity  is more evident at  minimums (rather than maximums) where the spot is at the far-side position. Moreover, their modeling of the pulse profile from a two point-like spot is ambiguous since they assumed $i=\theta=\pi/4$ for which according to \citep{Beloborodov:2002mr}  the primary spot is visible all the time and the antipodal one is never seen, therefore one spot and two-spot models should give the same results for the aforementioned assumption.

\section{{\bf Joint constraints from EM and GW  Observations}}\label{sec6}

In this section, we attempt to investigate the parameter space of the bosonic DM model i.e. $F_{\chi}$, $m_{\chi}$ and $\lambda$, regarding  three multimessenger observables including the visible (BM) radius, total maximum mass of object and the tidal deformability. We perform  a precise scan over the model parameters in given ranges  $m_{\chi}\in (0,1000]$ $\text{MeV}$, $\lambda\in [0.1,10]$ and $F_{\chi}\in (0,10]\%$, taking into account astrophysical constraints of NSs, $M_{T_{max}}\geq2M_{\odot}$, $R_{1.4}\geq11$ km and $\Lambda_{1.4}\leq 580$. Here, we utilize two EoSs i.e. IST and DD2 to model BM component of DM admixed NS. In fact the distribution of DM in or around a NS would affect its astrophysical properties, the DM core formation can significantly decrease the  mass, visible radius and tidal deformability which has the potential to violate the obtained constraints for NSs. However, formation of a DM halo around the NS,  slightly decline the visible radius and will increase the total gravitational mass, tidal deformability, this  may disfavour the tidal limit. 
 
 \begin{figure}[!h]
    \centering
 \includegraphics[width=3.5in]{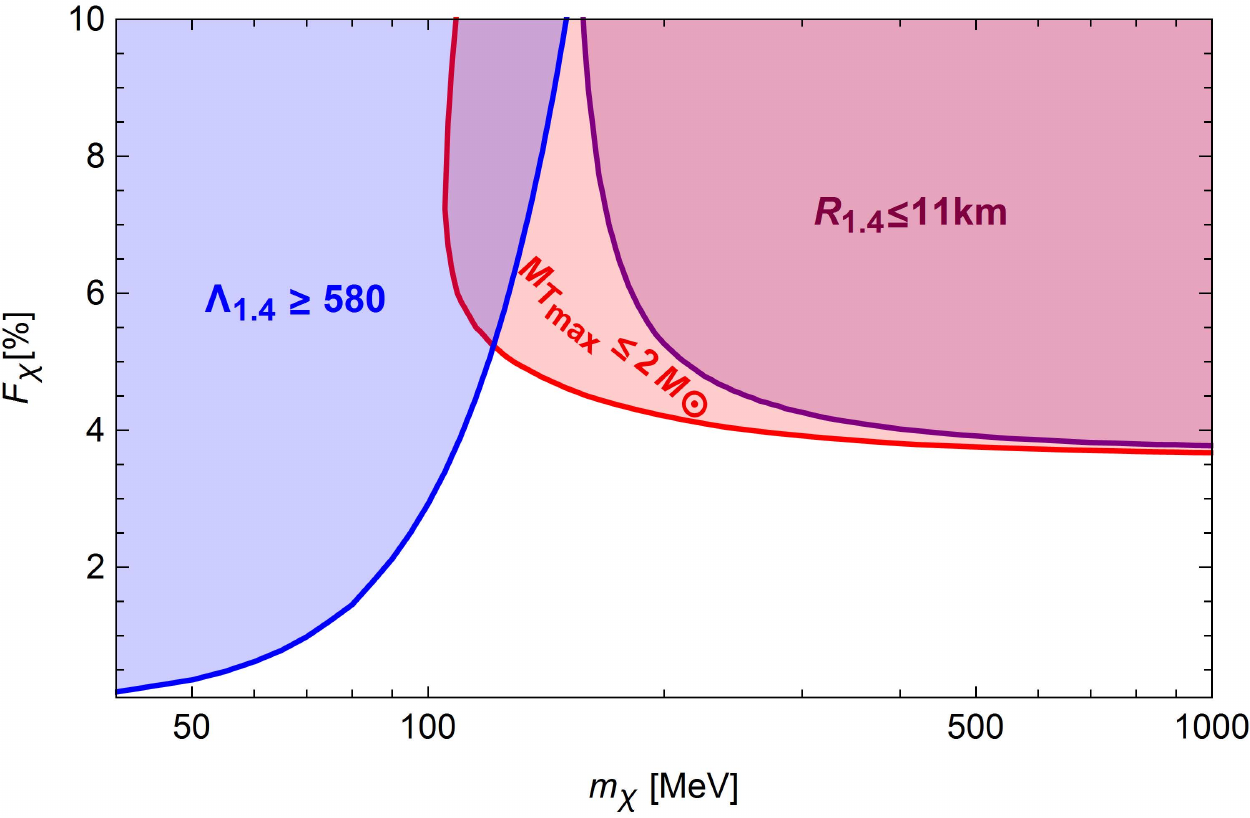}
    \caption{ The exclusion regions of $F_{\chi}-m_{\chi}$ parameter space are shown by different colors 
for $\lambda=\pi$ and IST EoS. The red area represents the maximum
total gravitational mass to be less than or equal to 2$M_\odot$. The blue part indicates $\Lambda_{1.4}\geq580$ for tidal deformability of DM admixed NSs with $M_{T}=1.4M_\odot$. The purple region shows the DM admixed NS  with  visible radius    $R_{1.4}\leq11$ km.}
    \label{fig17}
\end{figure}

  \begin{figure}[!h]
    \centering
    \includegraphics[width=3.5in]{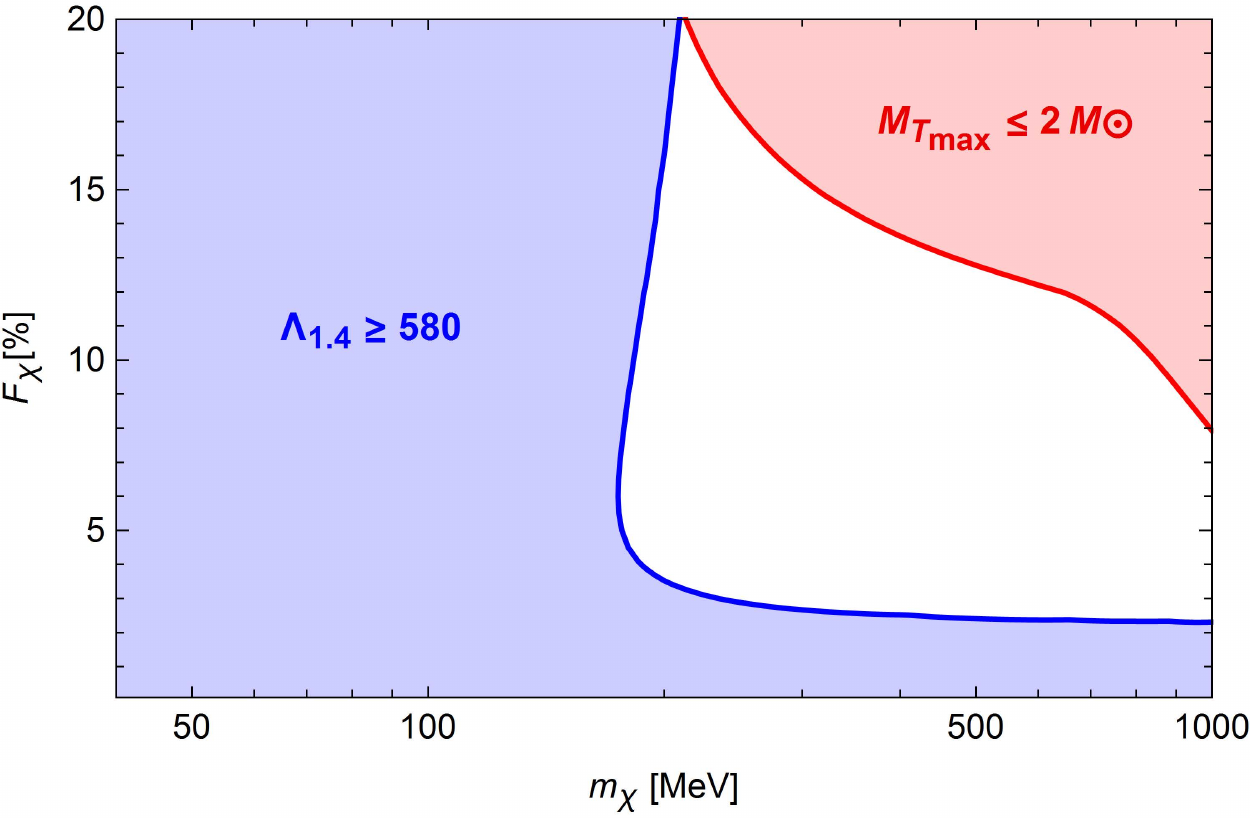}
    \caption{ The $F_{\chi}-m_{\chi}$ parameter space of bosonic DM model for DD2 EoS and $\lambda=\pi$ by considering $\Lambda_{1.4}\leq580$ and $M_{T_{max}}\geq2M_\odot$ astrophysical constraints. The blue and red regions indicate the excluded parts regarding the observational bounds. Note that  the radius constraint  is well satisfied for the whole parameter space.}
    \label{fig18}
\end{figure}

 \begin{figure}[!h]
    \centering
    \includegraphics[width=3.5in]{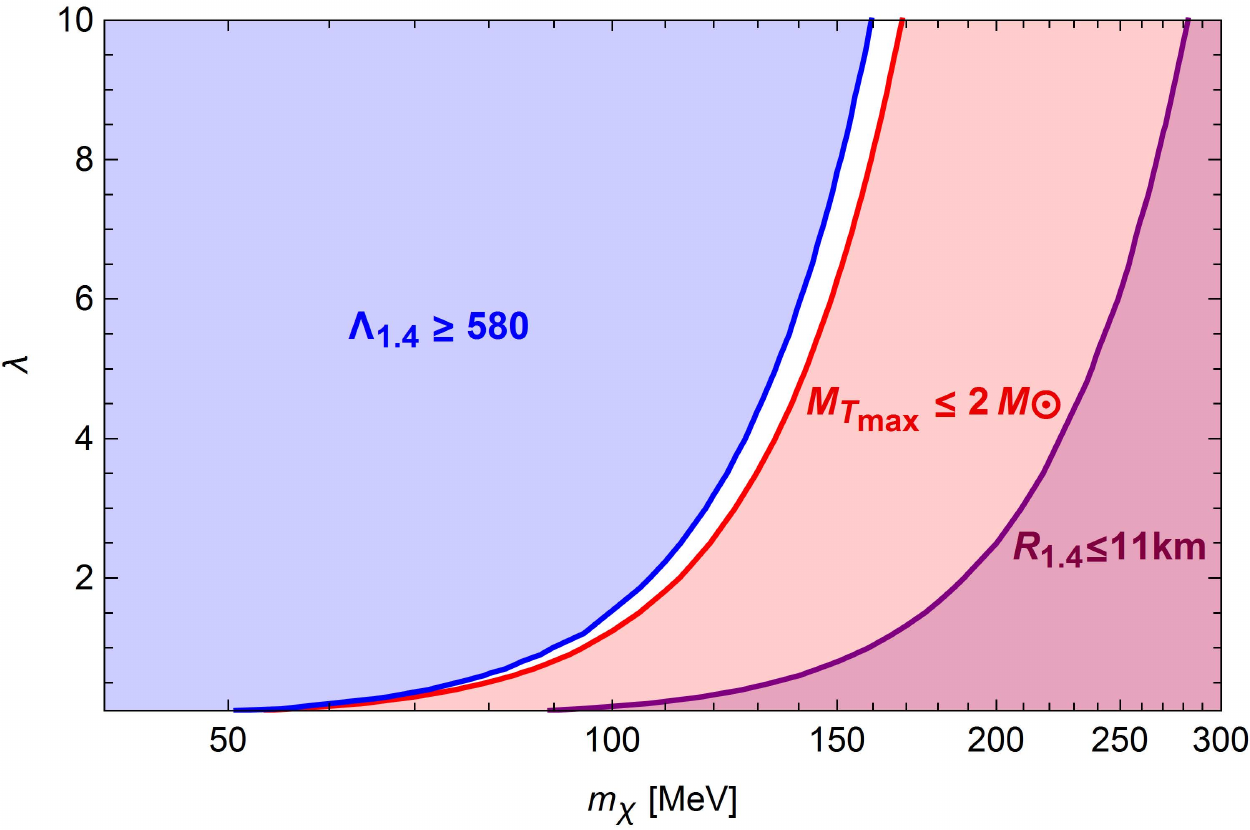}
    \includegraphics[width=3.5in]{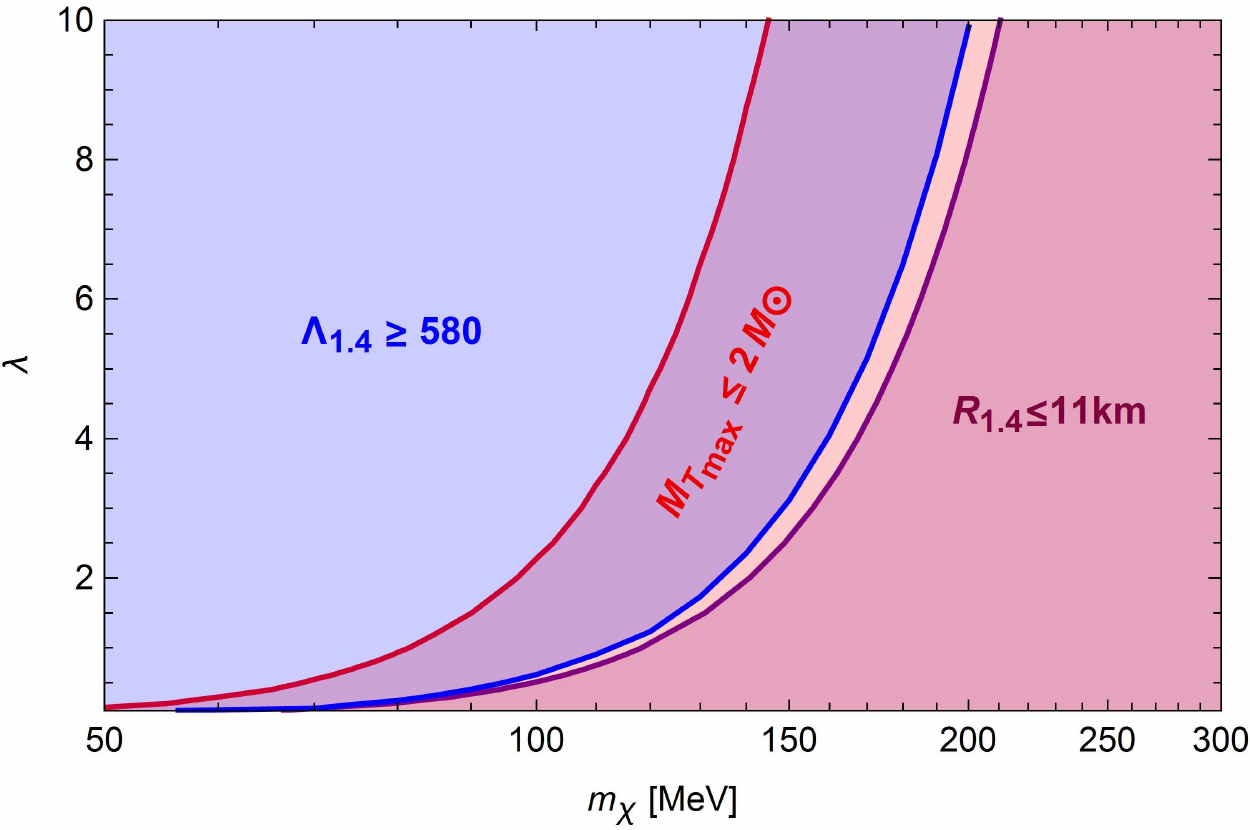}
    \caption{The $\lambda-m_{\chi}$ parameter space of bosonic DM model for $F_{\chi}=5\%$ (up) and $F_{\chi}=10\%$ (down) applying IST EoS.  The colored regions illustrate the parameter spaces which do not respect the $\Lambda_{1.4}$, $M_{T_{max}}$ and $R_{1.4}$ constraints. It is seen that the allowed region is significantly limited for the upper panel and is completely excluded in the lower one according to the astrophysical bounds.}
    \label{fig19}
\end{figure}

 \begin{figure}[!h]
    \centering
    \includegraphics[width=3.5in]{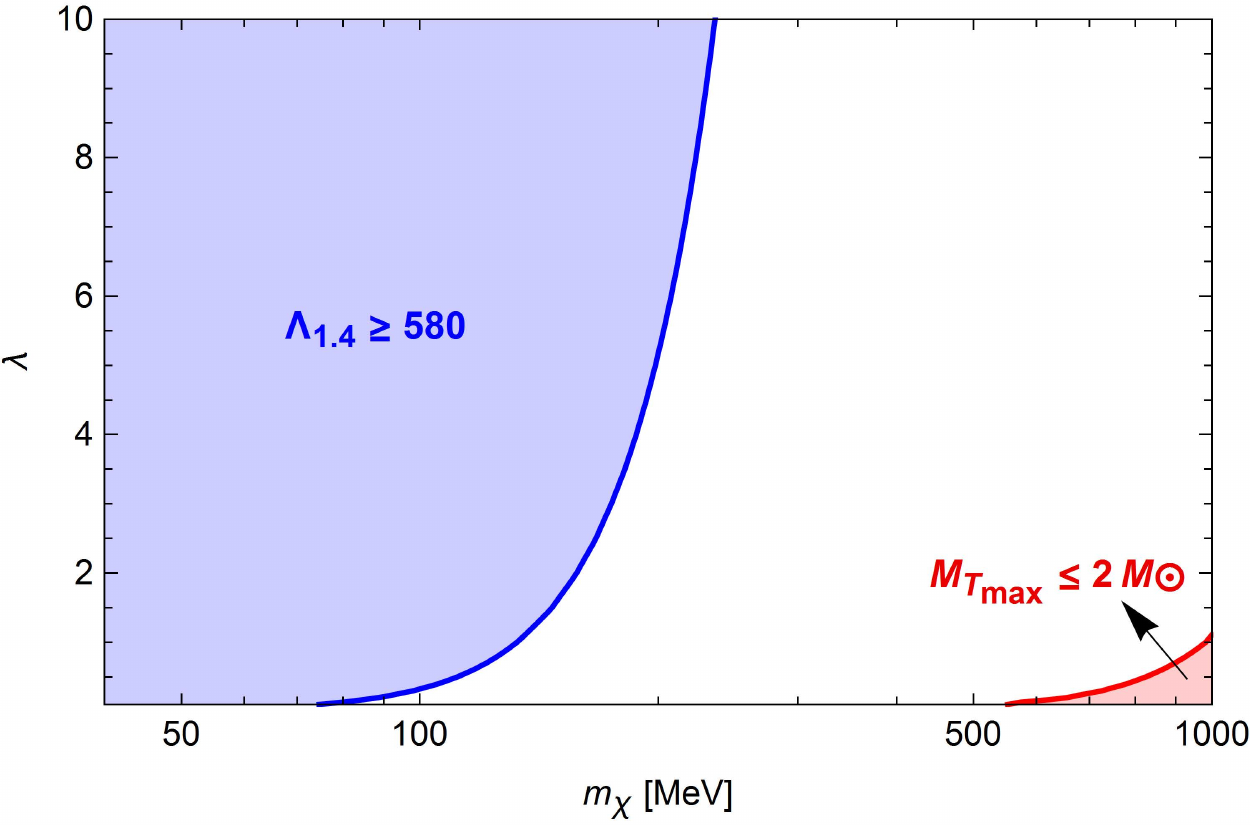}
    \includegraphics[width=3.5in]{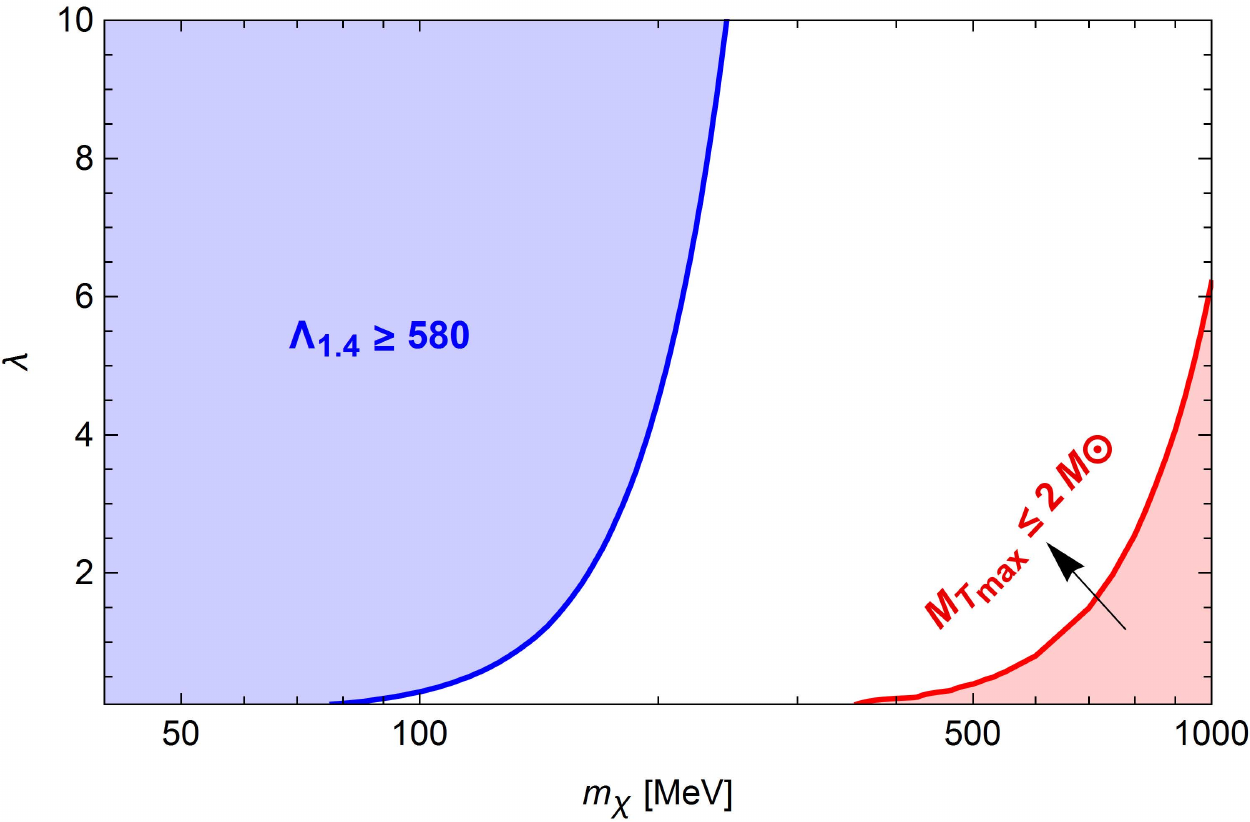}
    \caption{The  same as Fig. \ref{fig19}, but for DD2 EoS. Obviously, the $\lambda-m_{\chi}$ parameter space  of bosonic DM model  is weakly constrained by this BM EoS and $R_{1.4}$ value  is well consistent all over the region.}
    \label{fig20}
\end{figure}

In Figs. \ref{fig17} and \ref{fig18}, considering astrophysical limits of NSs obtained from NICER and LIGO/Virgo observations,  a scan has been done over  $F_{\chi}-m_{\chi}$ DM parameter space for IST and DD2 EoSs, respectively. The white regions indicate the allowed areas with respect to the $M_{T_{max}}\geq2M_\odot$,  $R_{1.4}\geq11$ km and $\Lambda_{1.4}\leq580$ constraints. As it is seen in Fig. \ref{fig17}, both mass and radius constraints are following the same trend, while the maximum mass of the DM admixed NS provides a more stringent constraint compared to the $R_{1.4}$. We find that in the low particle mass limit ($m_{\chi}\lesssim100$), the tidal deformability excludes the presence of sizable amount of DM within NSs, however,  towards more massive bosons the possible fraction reached to a constant value $4\%$, imposed by  both $M_{T_{max}}=2M_{\odot}$ and $R_{1.4}=11$ km lines. Satisfying our triple constraints lead to the highest possible fraction of DM to be $5\%$ at $m_{\chi}\approx120$ MeV. This fact demonstrates that  every observed NS could in principle include  low fractions of DM less than $5\%$.

From Fig. \ref{fig18}, where we applied DD2 EoS  as BM component, we find that the tidal deformability constraint excluded the whole sub-GeV bosons for $F_{\chi}\leq 2.5\%$ and the entire DM fractions for $m_{\chi}\lesssim 170$ MeV. This behaviour is due to the fact that the original tidal deformability value of DD2 EoS ($\Lambda_{1.4}\approx 680 $) is above the maximum observational limit ($\Lambda_{1.4}\approx 580 $). The DM core formation decreases  tidal deformability in higher boson masses and above a minimum fraction becomes consistent with $\Lambda_{1.4}$  limit, while for light bosons the halo formation causes the enhancement in tidal deformability which disfavours by GW observations. It is notable that $R_{1.4}$ does not impose any limitations on the given $F_{\chi}-m_{\chi}$ parameter space. There is a maximum allowed fraction $20\%$ around $m_{\chi}\approx 200$ MeV, thus for $F_{\chi}\in [2.5,20]\%$ one can find a possible range of $m_{\chi}$ which is in agreement with aforementioned bounds. As it is seen, the maximum mass limit excludes the right-up corner of the parameter space corresponding to DM core formation.

For the sake of completeness, in Figs. \ref{fig19} and \ref{fig20}, we present   a scan over self-coupling constant and boson mass for IST and DD2 EoSs considering different fractions $5\%$ and $10\%$ in upper and lower panels, respectively. We see from the upper panel of Fig. \ref{fig19} that there is a narrow allowed range of $\lambda-m_{\chi}$ parameter space for IST EoS thanks to the astrophysical constraints. However, the whole parameter space is excluded in the lower panel by increasing the DM fraction to $10\%$. These two figures reconfirm that the low fractions of DM less than $5\%$ are more favourable in the light of the latest measurements of NS properties. As it is evident from Fig. \ref{fig20}, we obtain a less restricted parameter space taking into account DD2 EoS as BM component. The tidal deformability constraint avoids light bosons ($m_{\chi}\lesssim 80$ MeV) coexisting with nuclear matter described by DD2 EoS in the whole considered range of coupling constant. By increasing DM fraction, the allowed region is reduced due to both $\Lambda_{1.4}$ and $M_{T_{max}}$. Similar to Fig. \ref{fig18}, $R_{1.4}$ gives no excluded region in $\lambda-m_{\chi}$ parameter space. Moreover, a density plot of the figures  shown in this section is presented in Figs. \ref{densityIST} and \ref{densityDD2}  of the appendix.

In summary, we conclude that  IST as a soft EoS, which satisfies the observational limits marginally, gives tighter constraints on bosonic DM parameter space compared to DD2 as a stiff EoS. The distribution of DM as a core or halo which influences the observational features is independent of BM fluids, however, owing to different values of NS properties arising from various BM EoSs, the obtained constraints for DM model will be changed. For both EoSs, we find that light particles are in favour of mass and radius constraints, however,  it can be seen that the tidal deformability significantly restricted available fractions in this regime. The maximum allowed fraction of DM is limited to $5\%$ for IST, while applying DD2 EoS leads to a range of $F_{\chi}$ between $2.5\%$ and $20\%$.

\section{{\bf CONCLUSION AND REMARKS}}
In this article, we have considered  bosonic particles with self-repulsive interaction to model DM admixed NSs with IST and DD2 EoSs as the BM component. The distribution of DM as a core or halo in mixed compact objects crucially depends on DM model parameters such as boson masses, self-coupling strength and also the amount of DM in NSs. While our two soft and stiff BM EoSs provide a  coverage of various observable features inferred from nuclear matter, it is seen that the general behaviour of DM core/halo formation is not sensitive to BM model.
The equilibrium configurations of DM admixed NSs are considered to probe DM model parameter space in the light of the latest NICER observations, where the total mass of the object is illustrated with respect to  the visible radius.  Owing to the fact that DM core formation reduces both the total mass and visible radius, we have shown that lighter bosons, larger coupling constants and lower DM fractions are in favor of the  NICER measurements for PSR J0030+0451 and PSR J0740+662.

Regarding the lower limit of radius $\sim11$ km for NSs with $1.4M_\odot$  obtained from joint analysis of NICER and LIGO/Virgo,
we focused on the variation of the visible and dark radius and their dependency upon $F_{\chi}$, $m_{\chi}$ and $\lambda$. It is turned out that by increasing DM fractions,   $R_{B}$ is mainly a decreasing function for DM core/halo formation, however, $R_{D}$ always rises. Moreover,  a shift  from DM core to DM halo could occur for all DM particles by increasing $F_{\chi}$ where the outermost radius of the mixed compact object changes from $R_{B}$ to $R_{D}$.  We found that for bosonic DM with larger $m_{\chi}$ and/or smaller $\lambda$, $R_{B}$ is reduced and going below  $11$ km  which can be used to further constrain the DM model. It was demonstrated that the variation rate of the visible radius of a DM admixed NS is within the sensitivity range of the upcoming X-ray telescopes such as STROBE-X \cite{STROBE-XScienceWorkingGroup:2019cyd}, ATHENA \cite{2012arXiv1207.2745B} and eXTP \cite{2019SCPMA..6229503W} (see Sec. \ref{sec4}).

We introduced the  pulse profile as a new observable in Sec. \ref{sec5},  to look for the evidences of the bosonic DM halo around NSs. The presence of DM around NSs via changing the geometry of space-time outside the BM radius and also  the compactness of the object will impact the trajectory of surface photons. The effect of DM model parameters and fraction on  the pulse profile has been investigated comprehensively which can be served as an independent probe for bosonic DM. It was shown that the deviation of the minimum fluxes for various $m_{\chi}$, $\lambda$ and $F_{\chi}$ in DM admixed NSs compared to pure NSs is a remarkable signature of  the DM halo. Our results  could be also included into  numerical methods for PPM and ray tracing from NS surface in X-ray telescopes giving rise DM admixed NS as a new promising possibility to interpret observations. Including current uncertainties in the measured EM pulse profiles  and the variation induced by the presence of DM, one could derive constraints for the DM model, this will be the subject of a future paper.

Finally, we perform a precise scan over bosonic mass, self-coupling constant and DM fraction  at given ranges. In this regard, combined multi-messenger astrophysical constraints of NSs from GW and EM measurements, $M_{T_{max}}\geq2M_\odot$, $R_{1.4}\geq11$ km and $\Lambda_{1.4}\leq 580$ are taken into account. It was seen that the maximum mass and radius limits restrict mainly massive bosons for which $M_{T_{max}}\geq2M_\odot$ gives wider exclusions. However, tidal deformability constrains mostly light bosons due to DM halo formation. We found  that there is a maximum allowed DM faction for IST ($\sim5\%$) and DD2 ($\sim20\%$) and each NS could contain a relatively low amount of DM without violating the astrophysical limits. Moreover, it was shown that the whole range of sub-GeV SIDM is excluded for $F_{\chi}\lesssim 2.5\%$ in DM admixed NS applying DD2 BM component. A scan over coupling constant and boson mass has been done for fixed $F_{\chi}$, it is seen that increasing the DM fraction tightens the allowed region of the parameter space. In the case of IST EoS, the presence of  $5\%$ of DM in NS provide a narrow allowed region in $\lambda-m_{x}$ space, which will be entirely excluded for $F_{\chi}=10\%$.

It should be mentioned that since IST EoS  gives marginal allowed values for $M_{\text{max}}$ and $R_{1.4}$, it is more sensitive to the effect of DM and consequently leads to more severe constraints on DM model parameter space.  The effect of  DM particles  may be viewed as an effective softening/stiffening of the EoS corresponding to the whole structure of  DM admixed NSs. 
In fact the presence of the DM component in the mixed object  can relax the constraints on BM EoSs for interpreting the latest mass-radius measurements to be consistent with them. This effect in principle represents similar behaviours to  nuclear matter EoSs at high density such as those happen in the hybrid, twin and strange quark stars including a deconfinement phase transition at the core of compact stars \cite{Christian:2021uhd,Sen:2022lig,Jimenez:2021nmr,Lenzi:2022ypb,Lopes:2023uxi,Ferreira:2022fjo,Yang:2023haz,Sen:2022pfr} or hyperon puzzle in NSs \cite{DelPopolo:2020pzh,Das:2020vng,Shahrbaf:2022upc,Shahrbaf:2023uxy}. 

The study presented in this article can be extended to include numerous BM/DM EoSs and in order to find the best case scenario, Bayesian analysis can also be done regarding the latest multi-messenger data. We tried to capture the general effects of SIDM on NS properties assuming a wide range of DM parameter space taking into account two substantially different BM EoSs. Note that uncertainties about the BM EoSs in high density regimes and the fact that the presence of DM could in principle mimic the behaviour of BM component, prevent us from obtaining a stringent conclusion in this matter.  However, considering exotic objects and applying several independent constraints simultaneously, may break the degeneracy underlying the internal structure of compact objects.

 It is worth noting that  DM admixed NSs can provide alternative explanations for exotic measurements which have been reported so far such as the secondary component in  the GW190814 event \cite{LIGOScientific:2020zkf}  with  $\sim2.6M_\odot$ \cite{Das:2021yny,Lee:2021yyn,DiGiovanni:2021ejn}. The NICER collaboration announced the radius of PSR J0740+6620 
with a gravitational mass of $2.08\pm 0.07 M_{\odot} $   to be $12.35\pm0.75 \text{km}$ \cite{Miller:2021qha}  
which is approximately similar to the corresponding radius for PSR J0030+0451 as $12.45\pm65\text{km}$ \cite{Miller:2021qha} while it is around 1.5 times less massive ($M\sim 1.4 \, M_{\odot}$). This observation  is debatable in terms of NSs theoretical approaches and also the corresponding mass-radius profiles \cite{Legred:2021hdx,Li:2021sxb,Christian:2021uhd,Drischler:2021bup,Lin:2023cbo} which could be  explained by the existence of DM admixed NS. More recently, HESS collaboration observed the lightest and smallest compact object so far, the supernova remnant HESS
J1731-347 with $M=0.77^{+0.20}_{-0.17} M_{\odot} $ and $R=10.4^{+0.86}_{-0.78}$ km \cite{2022NatAs}, which could be described by inclusion of DM in NSs \cite{Sagun:2023rzp,Routaray:2023txs}.

The upcoming  LIGO/Virgo/KAGRA results \cite{LIGOScientific:2021qlt}, as well as  LISA \cite{Baker:2019nia}, Einstein telescopes \cite{Maggiore:2019uih} and Cosmic Explorer which provide enormous binary merger detections \cite{Evans:2021gyd} will enable us to further investigate DM admixed NSs \cite{Evans:2023euw,Baryakhtar:2022hbu}. Note that the presence of DM core/halo should be precisely taken into account in numerical-relativity simulations to compute GW spectrum and waveforms during merger and postmerger phases of a binary system containing at least a DM admixed NS \cite{Ellis:2017jgp,Giudice:2016zpa,Bezares:2019jcb,Emma:2022xjs,Ruter:2023uzc,Bauswein:2020kor}.
In addition to the pulse profile which is mentioned in this paper, other observational properties can be applied to explore the  DM  features in or around stars. For instance, gravitational microlensing due to  the dark halo surrounding NS  could cause  measurable changes in the brightness of the lensed source which would give hints to discriminate them from other dense objects, and the presence of large halo may also affect Shapiro time delay. Our research can potentially have a significant impact  on the discovery of the DM admixed NSs with self-repulsive bosonic DM component. 
The upcoming astrophysical instruments thanks to  precise  measurements of the  compact object properties may shed light on the  nature of DM and the possibility of the existence of DM within NS.

\acknowledgments
S. S and D.R.K are grateful to Jürgen Schaffner-Bielich for reading the  manuscript and insightful comments. S.S would like to thank Remo Ruffini for supporting his visit at ICRANet Pescara as an adjunct professor where the last part of this work was done. 
S.S and D.R.K are thankful to Mahboubeh Shahrbaf, Violetta Sagun and Stefan Typel for helpful discussions and providing the data of DD2 EoS.
D.R.K is really grateful for fruitful discussions during Dark Matter in Compact Objects, Stars, and in Low Energy Experiments workshop in August 2022 at University of Washington.  S.S and D.R.K acknowledge the Sheikh Bahaei National High Performance Computing Center of the Isfahan University of Technology. D.R.K expresses his gratitude to James Lattimer, David Blaschke and Armen Sedrakian for practical and invaluable comments during The Modern Physics of Compact Stars and Relativistic Gravity 2023 conference. D.R.K appreciates Sajad Khalili for his worthwhile assistance in the numerical calculations  code. Finally, we express our  sincere appreciation to the referee for giving valuable and insightful comments which substantially helped to improve the quality of the paper. 

\bibliography{main}

\clearpage
\newpage
\maketitle
\onecolumngrid
\begin{center}
\textbf{\large Exploring DM Core/Halo Formation } \\ 
\vspace{0.05in}
{ \it \large Supplementary Material}\\ 
\vspace{0.05in}
{}

\end{center}

In this  supplementary material, we provide further details about  exploring the  DM model parameter space in order to determine  regions in which we have the formation of the DM core/halo.  We aim to find an approximately analytic relation for the core-halo border lines in order to specify the DM distribution within DM admixed NSs.  Since the chemical potential  in the proper frame is a constant quantity throughout the equilibrium sequence of the star, one can define the relativistic  enthalpy as:
\begin{eqnarray}\label{e18}
h=\int \frac{dP}{P+\rho}=\int \frac{d\mu}{\mu},
\end{eqnarray}
in our case, we can define two different enthalpies $h_{D}$ and $h_{B}$ for  DM and BM fluids,  respectively.  It was shown that for DM core (halo) formation, the central enthalpy of DM component  $h^{c}_{D}$ is smaller (larger) than the corresponding one of BM component  $h^{c}_{B}$ \cite{Miao:2022rqj}. Note  that the enthalpy is a decreasing function of radius  from the center to the surface of the star, in fact,  the rate of reduction is the same for both of the fluids in each radius intervals. This is the reason why we can consider enthalpy to determine the DM core-halo transition, the boundary line associated with the transition occurs at the radius $R\equiv R_{B}=R_{D}$ where $h\equiv h^{c}_{D}=h^{c}_{B}$. In order to compute the enthalpy corresponding to self-interacting bosonic DM, we rewrite the EoS (\ref{e1}) as below
\begin{equation}\label{e19}
P^{\ast}=\frac{1}{9\lambda}\left( \sqrt{1+3\lambda\rho^{\ast}}-1\right)^{2},
\end{equation}

where the pressure and the energy density are scaled as $P^{\ast}=P/m_{\chi}^{4}$ and $\rho^{\ast}=\rho/m_{\chi}^{4}$ to obtain a dimensionless EoS,  and the energy density takes the following form: 
\begin{eqnarray}\label{e20}
\rho^{\ast}&=&3P^{\ast}+2\sqrt{\frac{P^{\ast}}{\lambda}},
\end{eqnarray}
(see appendix \cite{Karkevandi:2021ygv} for more details), substituting Eqs. (\ref{e19}) and (\ref{e20}) into Eq. (\ref{e18}) enthalpy is obtained as 
\begin{eqnarray}
h=\int \frac{dP\ast}{4P^{\ast}+2\sqrt{\frac{P^{\ast}}{\lambda}}}=\frac{1}{2}\ln{(1+2\sqrt{P^{\ast}\lambda})},
\end{eqnarray}
then the energy density in terms of enthalpy is given by 
\begin{eqnarray}
\rho=\frac{m_{\chi}^{4}}{3\lambda}\Big[\frac{9}{2}\left((e^{2h}-1)+1\right)^{2}-1\Big]
\end{eqnarray}

\begin{figure*}\label{e21}
\includegraphics[width=2.5in]{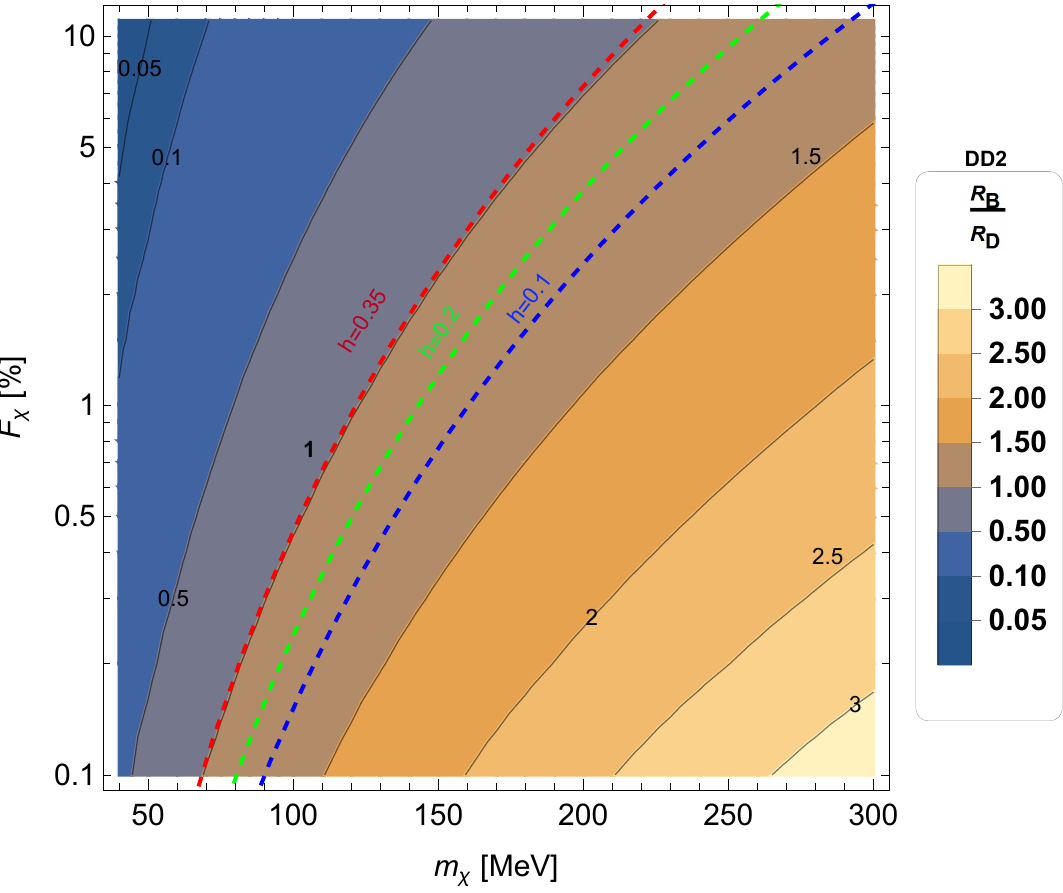}
\includegraphics[width=2.5in]{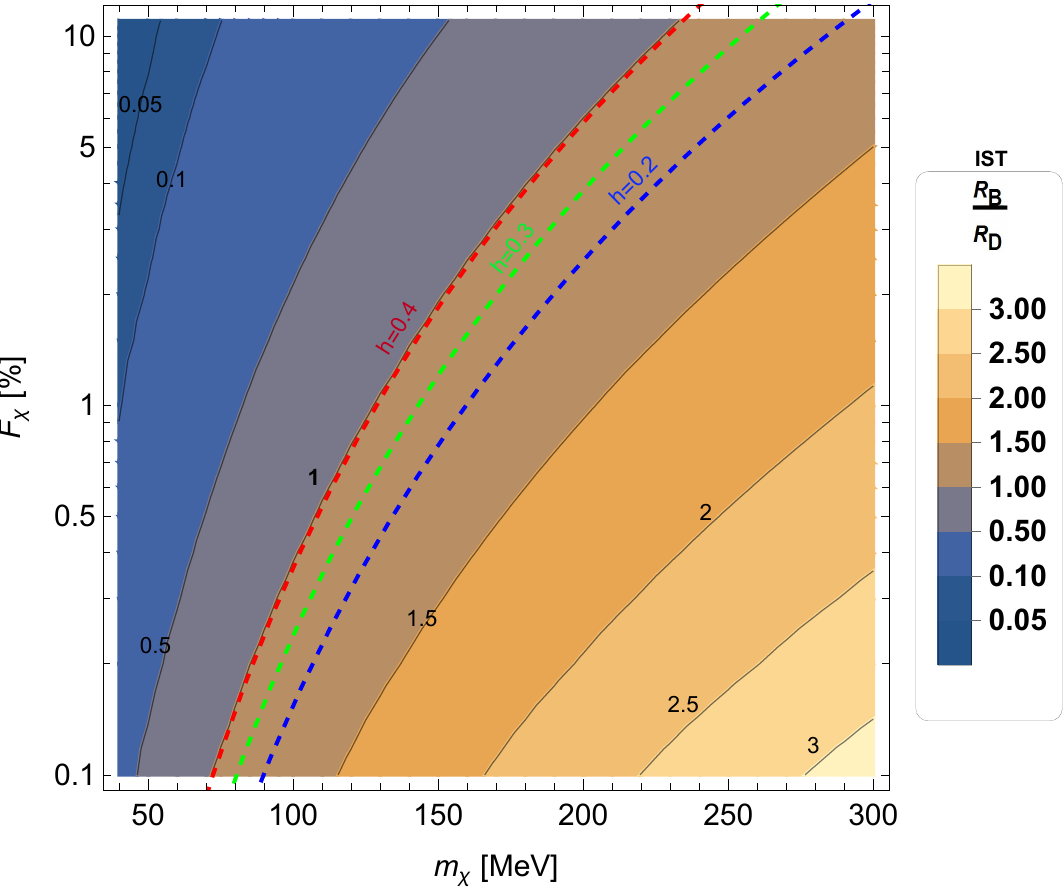}
  \caption{Comparision between analytic formula representing halo-core boundary lines and the result obtained from numerical solutions of two fluid TOV equations. In the left panel,  R is selected to be 13.15 \text{km} which is the BM radius of DD2 and in the right panel R is selected to be 11.37 \text{km} related to IST, different values of enthalpy is labeled with different colors to define core-halo boundary lines.}
\end{figure*}

The enclosed DM mass $M_{D}$ can be approximated as $M_{D}\approx 4\pi  R^{3}\rho/3$, therefore, for  given values of h, R and $M_{T}$ one can obtain a relation between  DM fraction $F_{\chi}=M_{D}/M_{T}$ and $\lambda$ and $m_{\chi}$ to define  marginal values of DM model parameters for core-halo transition.  While it is not straightforward  to determine enthalpy precisely for a DM admixed NS, we know that for a NS with  $1.4M_{\odot}$, selection a value between 0.1 - 0.4 for the central enthalpy would be a reasonable assumption \cite{Lindblom:2018rfr,Miao:2022rqj}. Assuming $h=0.2$ and an appropriate set of scaled parameters, the marginal fraction of DM for core-halo transition is given by:

\begin{eqnarray}\label{e23}
F_{\chi}\simeq 1.812 \times 10^{-9} \left(\frac{1}{\lambda}\right)\left( \frac{m_{\chi}}{300\text{MeV}}\right)^{4} \left(\frac{R}{12\text{km}} \right)^{3} \left(\frac{1.4M_{\odot}}{M_{T}}\right) .
\end{eqnarray}
It is instructive to compare the prediction of this formula with the numerical results presented in Fig. \ref{corehaloscan}, we have plotted core-halo boundary line in Fig. \ref{e21} for a range of enthalpy as labeled with different colors. Throughout the extensive analysis presented in this paper taking into account DD2 and IST as two BM EoSs, we have seen that the distribution of DM as a halo or core  is crucially dependent on the DM model parameters and its fraction and  is weakly influenced by BM EoSs. In other words, while the dark radius $R_D$ has a wide range variation depending on DM model parameters but $R_{B}$ has limited variations, this confirms again that the core-halo behavior is dictated mainly by DM component. Although estimation leads to Eq. (\ref{e23}) is based on some simple assumption, it reveals an important feature in DM admixed NSs.

\onecolumngrid
\begin{center}

\textbf{\large Precise Scan Over DM Model Parameters to Obtain Observable Features} \\ 
\vspace{0.05in}
{ \it \large Supplementary Material}\\ 
\vspace{0.05in}
{}

\end{center}

Here, we perform a precise scan over bosonic mass $m_{\chi}$, self-coupling constant  $\lambda$ and DM fraction $F_{\chi}$ to obtain multi-messenger observables such as the maximum mass $M_{T_{max}}$, the tidal deformability $\Lambda_{1.4}$ and the visible radius $R_{1.4}$ associated with $1.4 M_{\odot}$ DM admixed NS. It is seen that for light bosons increasing $\lambda$ and $F_{\chi}$ lead to an enhancement of the observable features $M_{T_{max}}$,  $\Lambda_{1.4}$. However, for massive bosons higher fractions cause more reduction in the properties of DM admixed NSs, while increasing the coupling constant results in larger values of these observables.
\begin{figure}[h]
    \centering
    \includegraphics[width=2.3in]{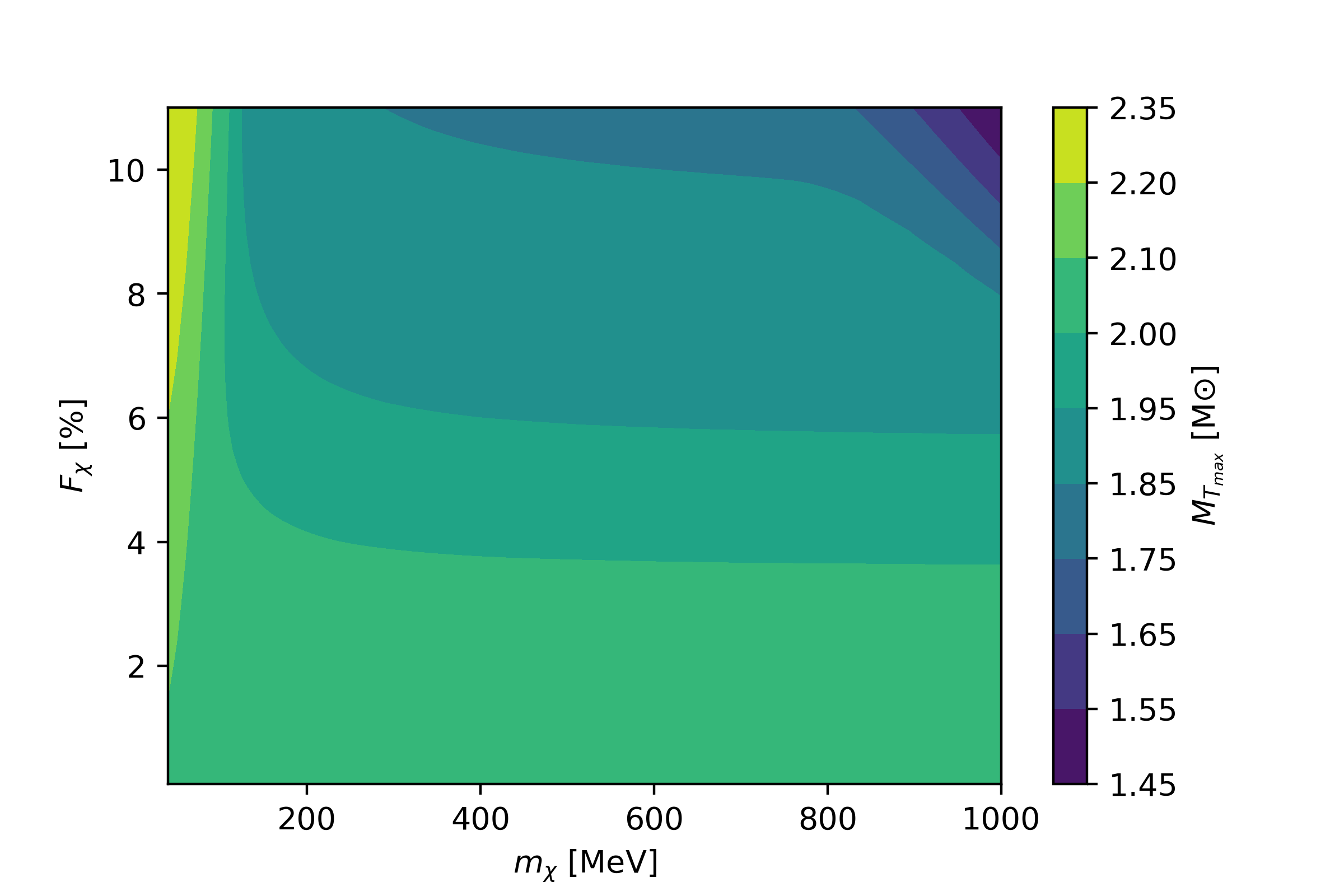}
\includegraphics[width=2.3in]{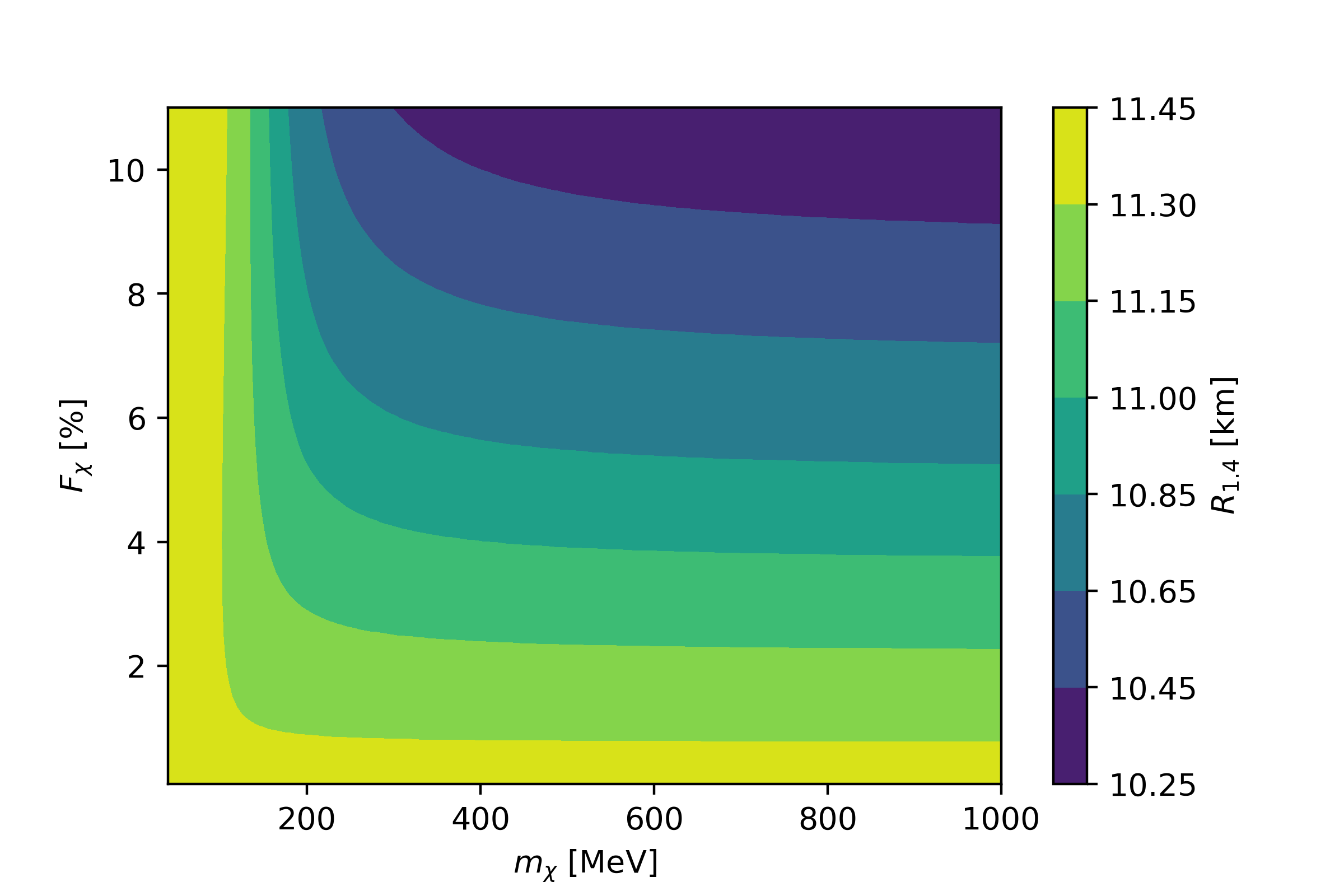}
\includegraphics[width=2.3in]{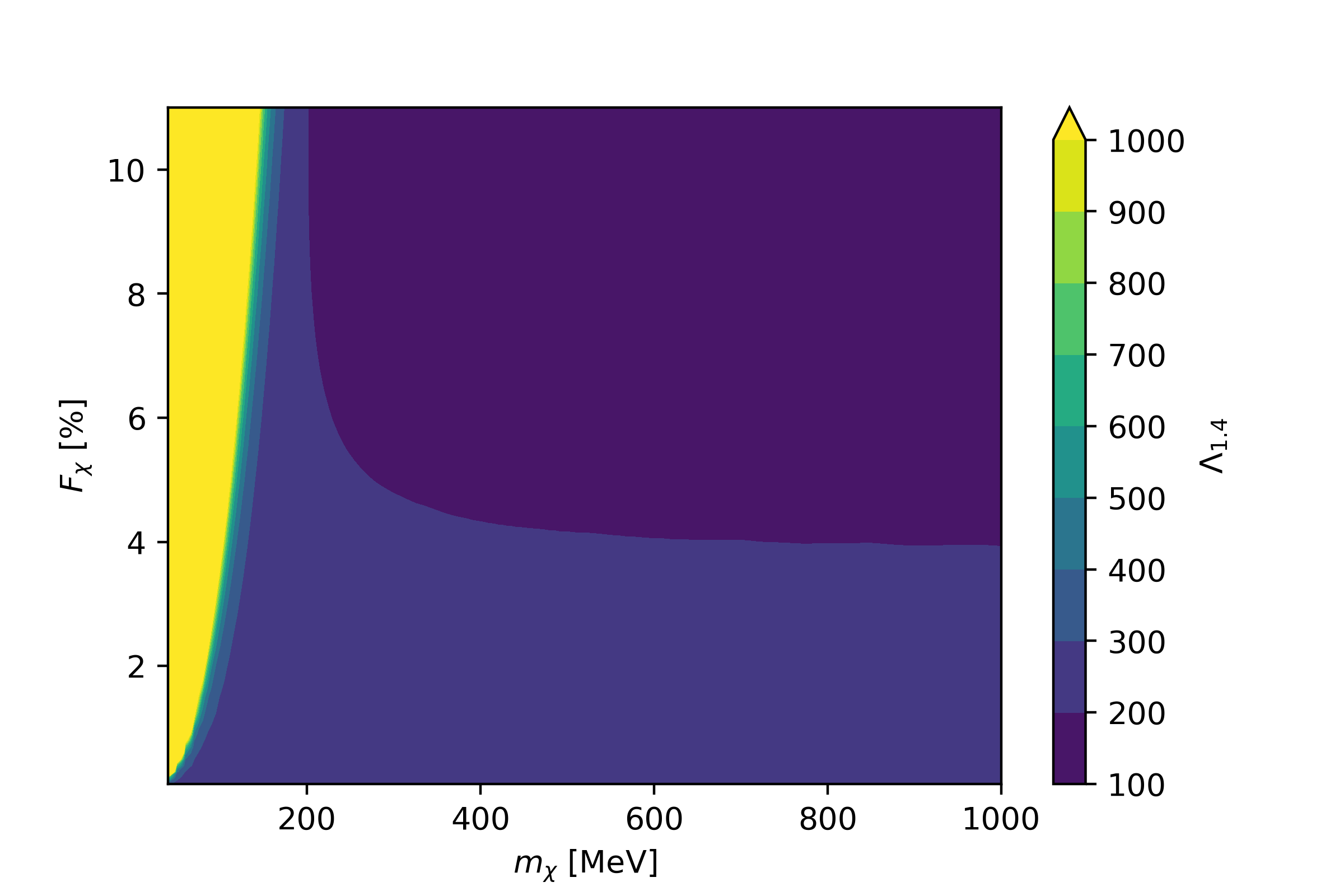}
\includegraphics[width=2.3in]{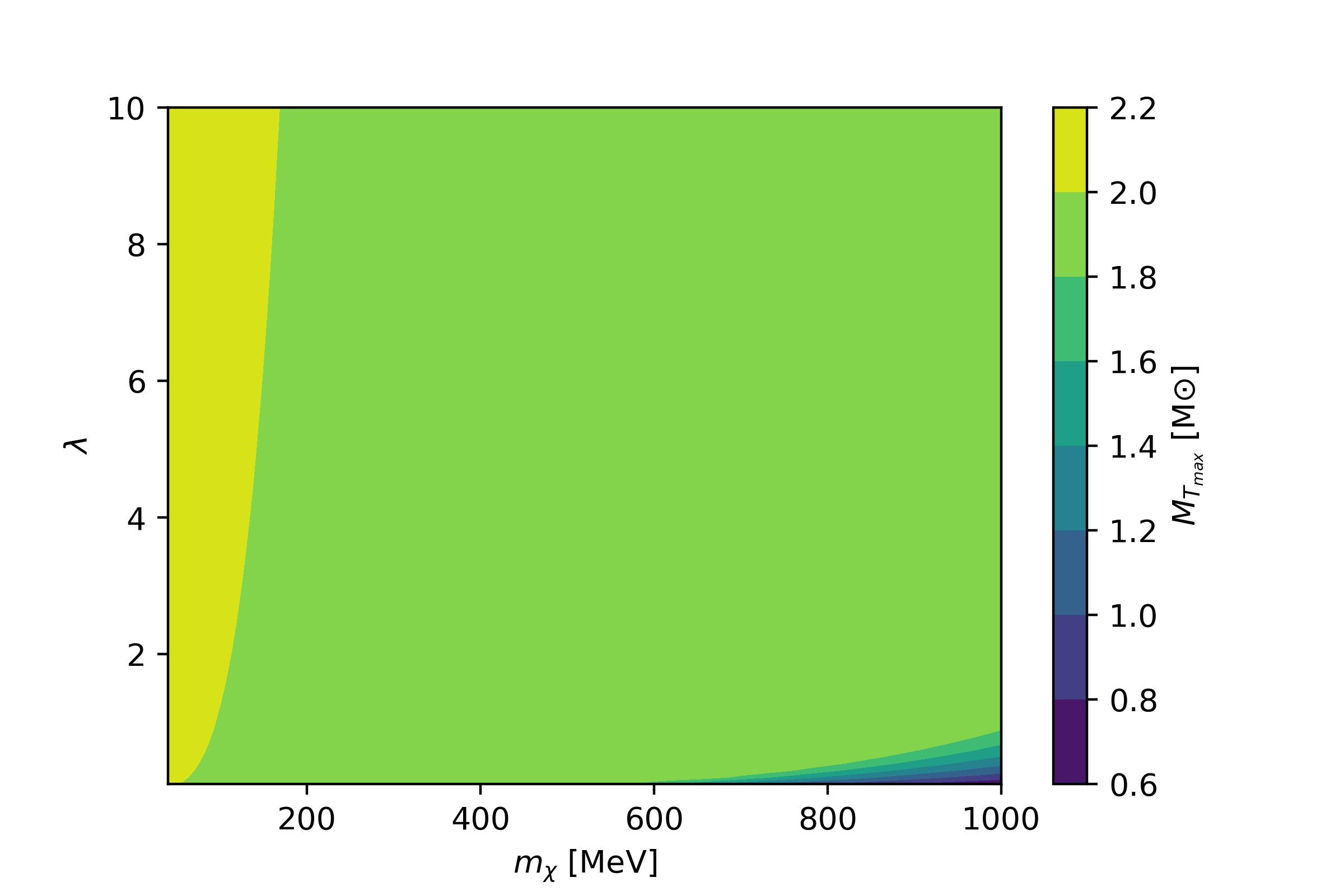}
\includegraphics[width=2.3in]{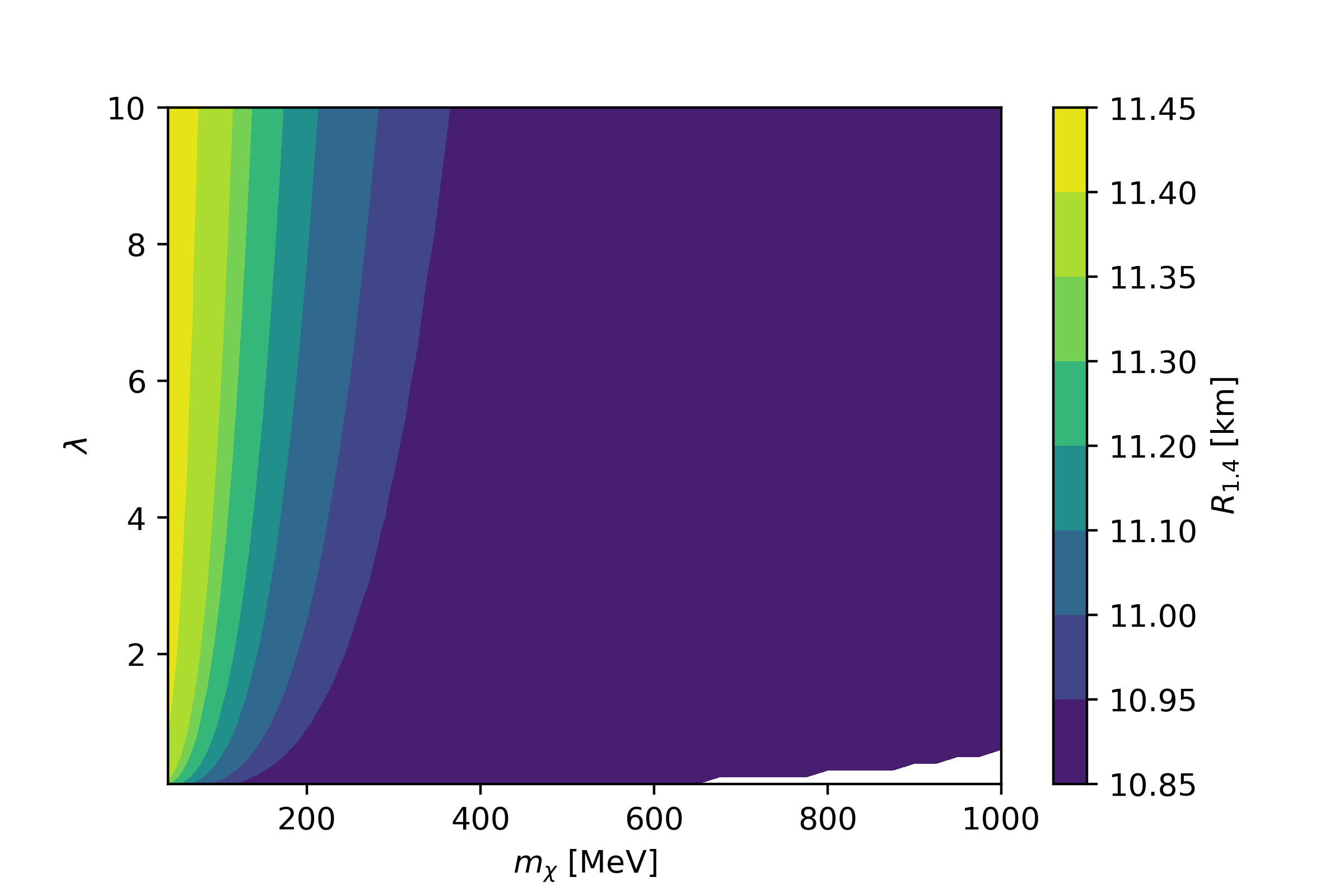}
\includegraphics[width=2.3in]{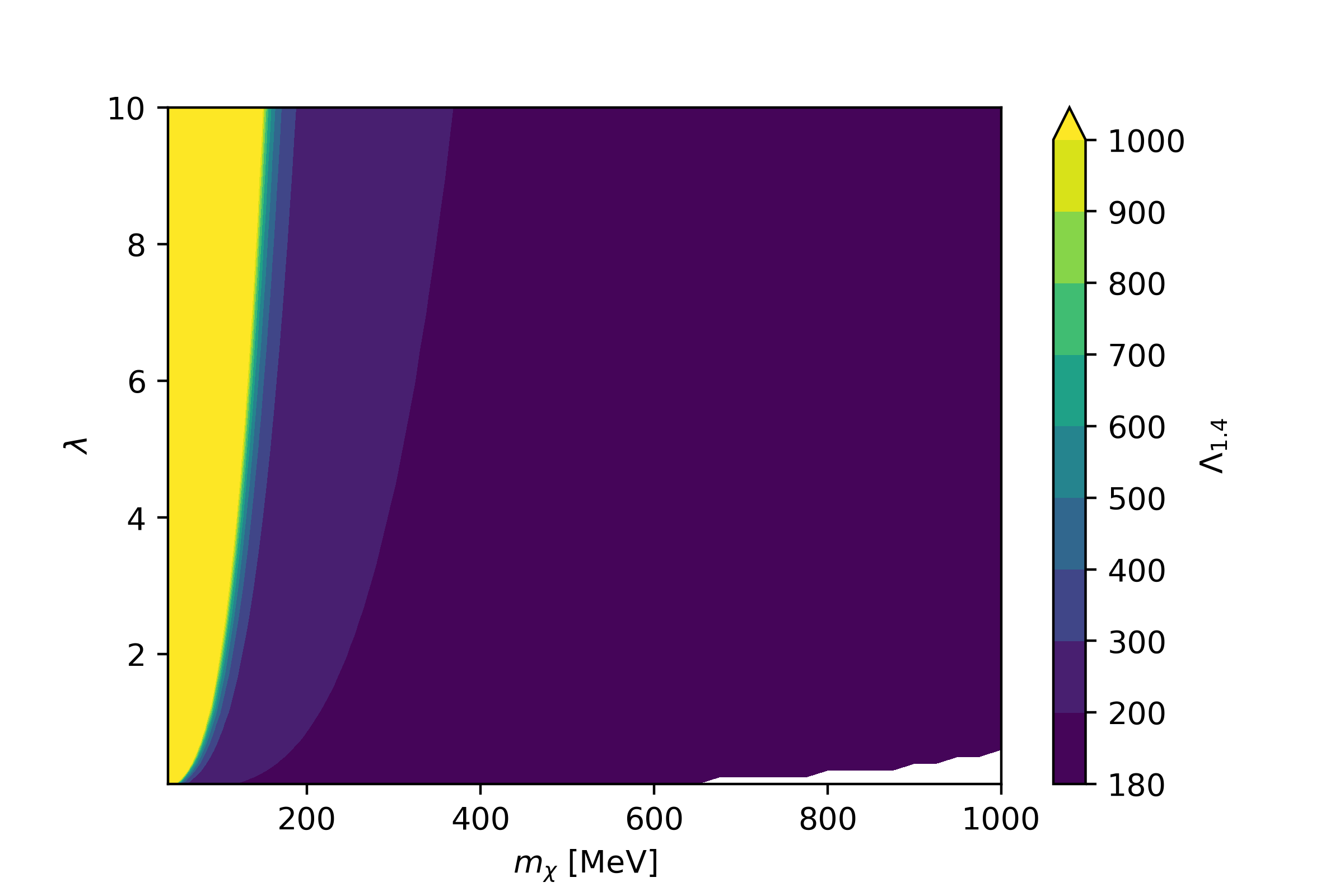}
    \caption{The parameter space $F_{\chi}-m_{\chi}$ at $\lambda=\pi$ is given  in the upper row  and in the lower row $\lambda-m_{\chi}$ parameter space for $F_{\chi}=5\%$ are explored for IST BM EoS. The density color represents the total maximum mass (left), visible radius (middle) and the tidal deformability (right) for a $1.4M_{\odot}$ DM admixed NS. Small white regions in right-down corners of $\lambda-m_{\chi}$ scan plots show those cases for which there is no $1.4M_{\odot}$ mixed object.}\label{densityIST}
\end{figure}

\begin{figure}[h] 
    \centering
    \includegraphics[width=2.3in]{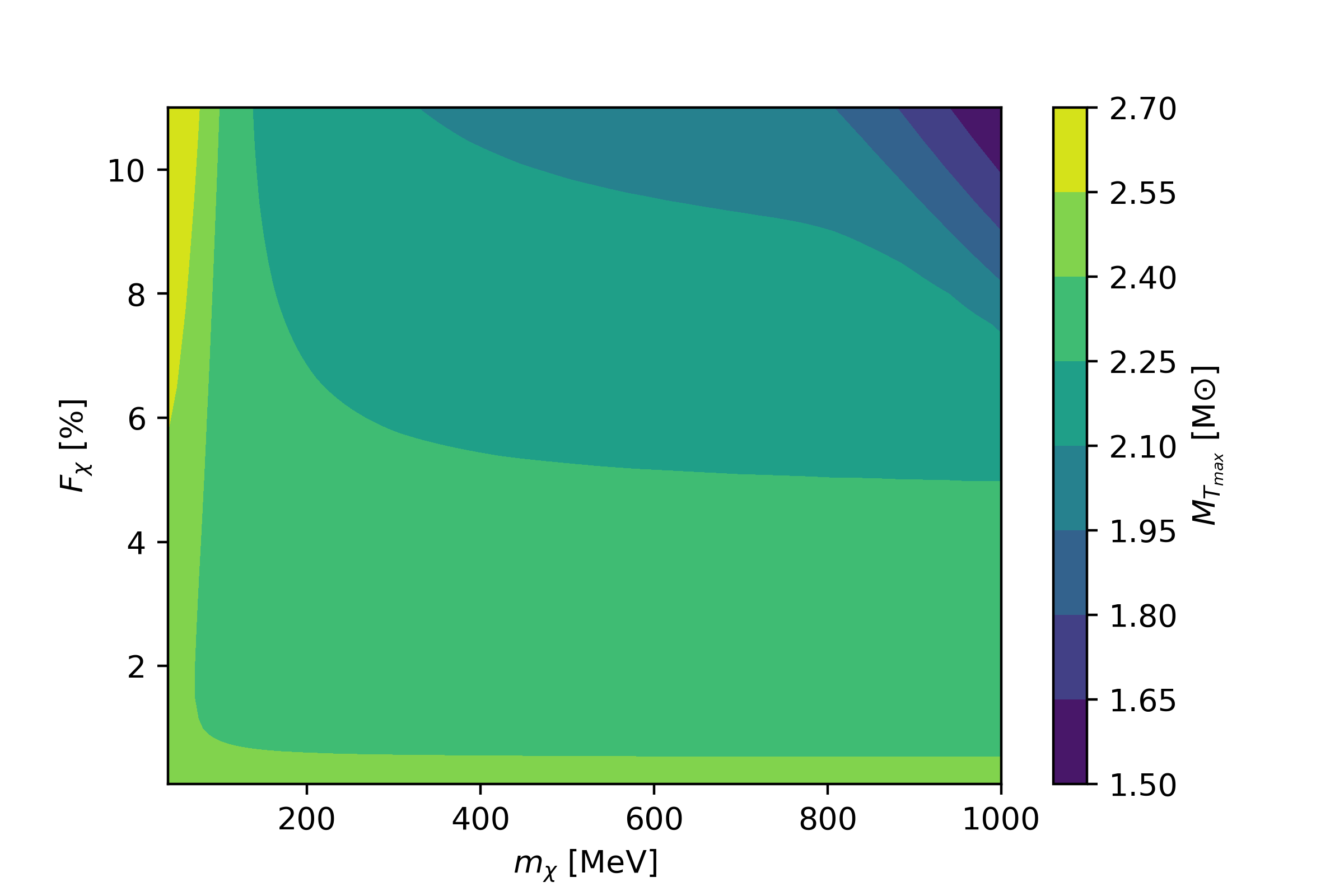}
\includegraphics[width=2.3in]{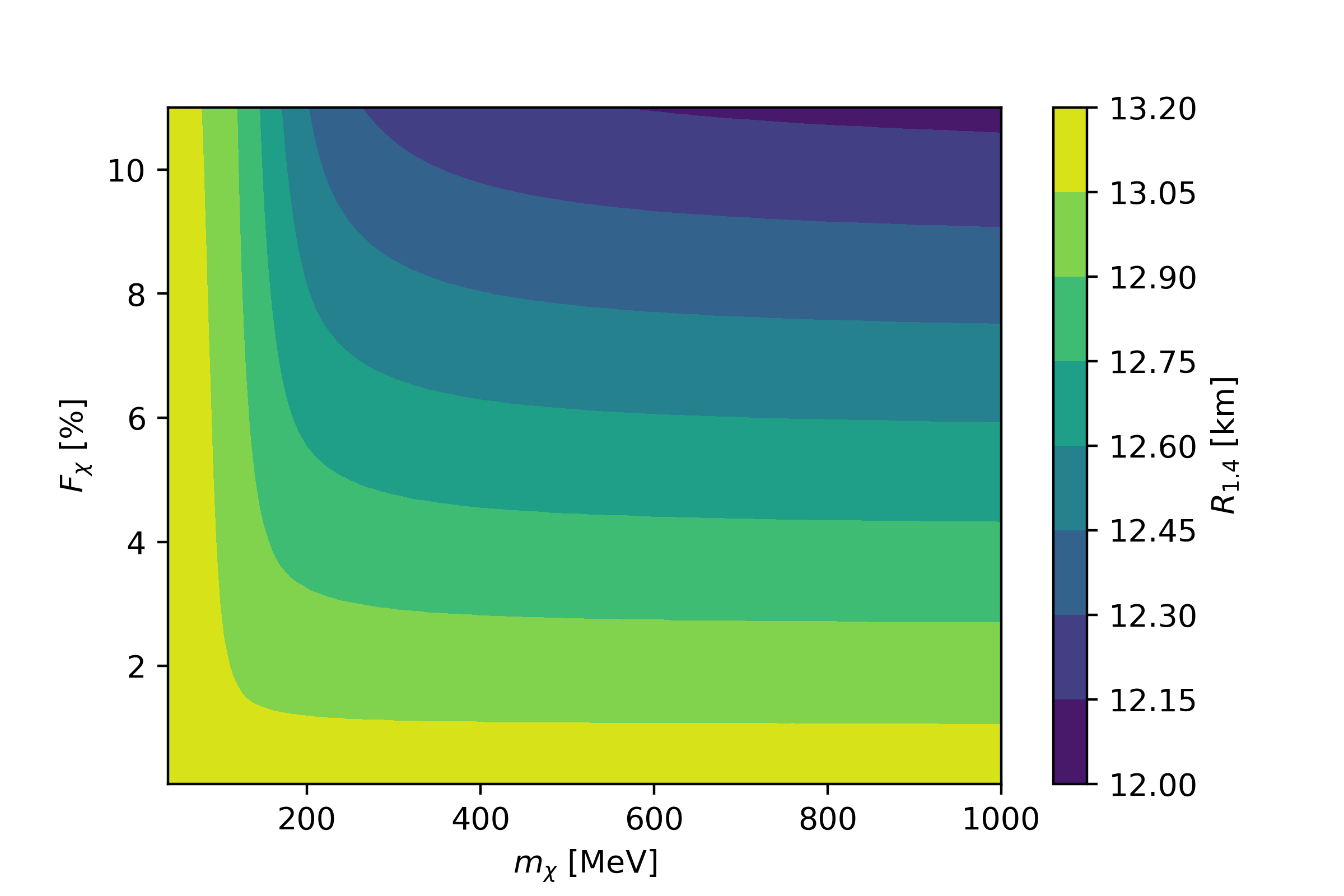}
\includegraphics[width=2.3in]{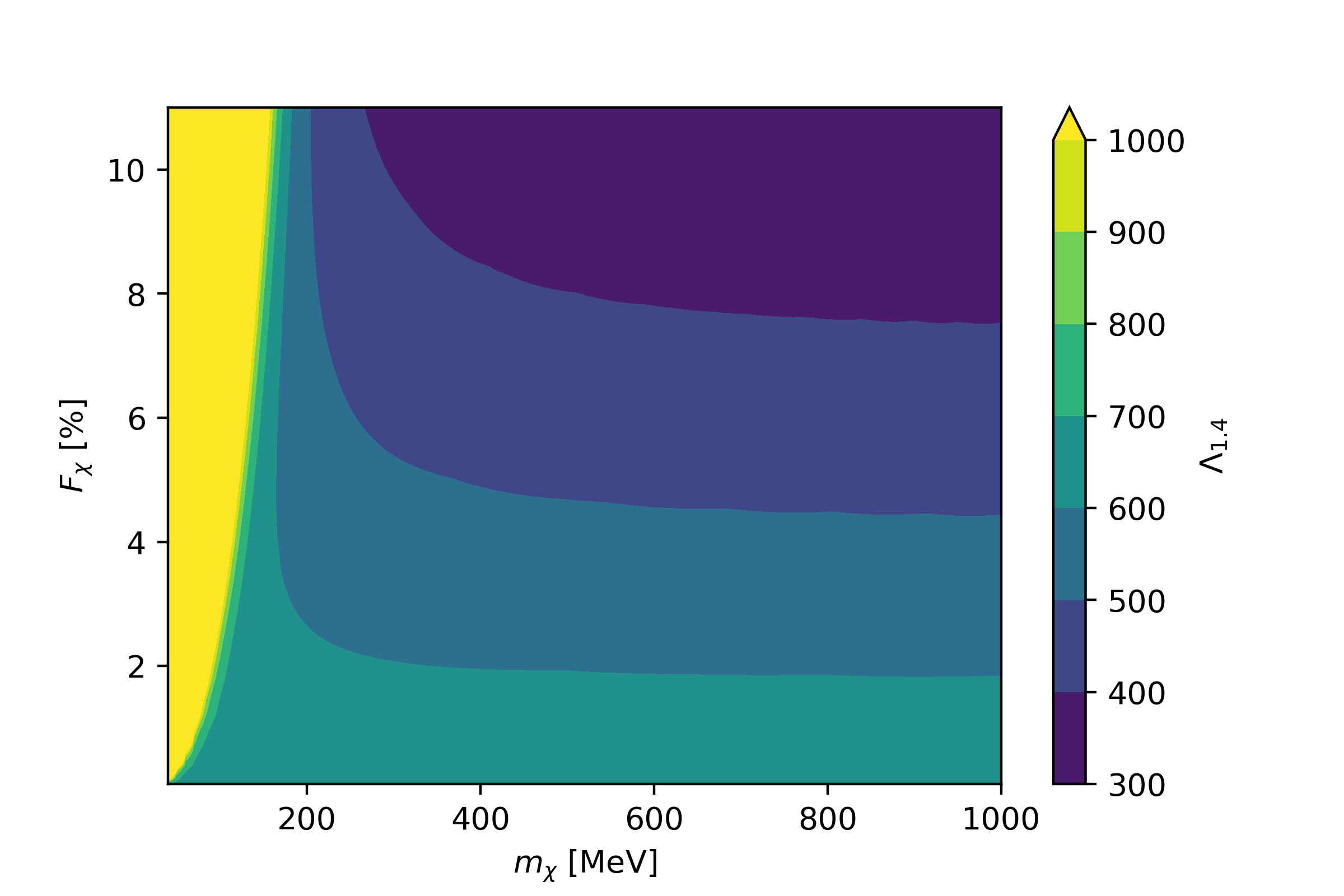}
\includegraphics[width=2.3in]{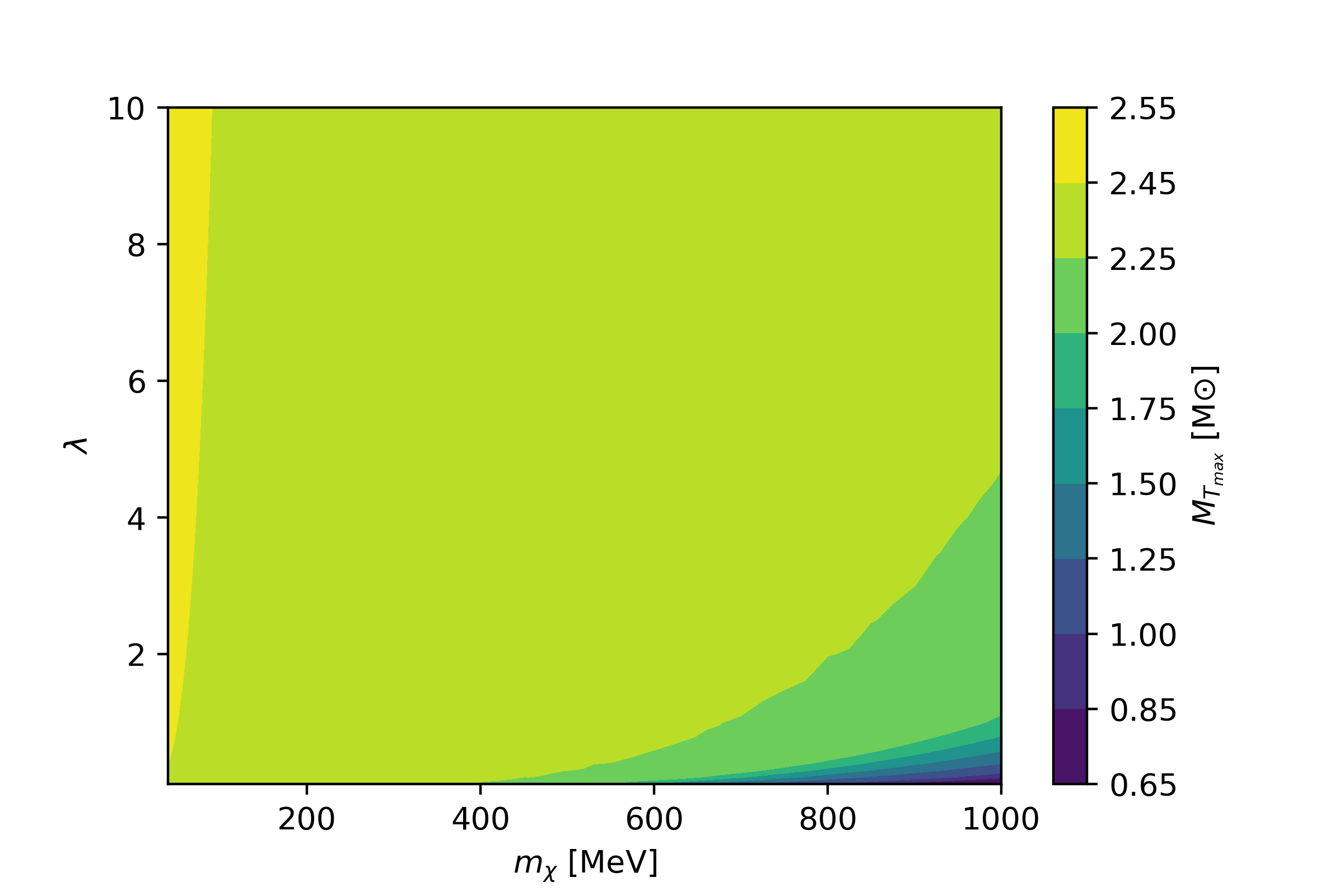}
\includegraphics[width=2.3in]{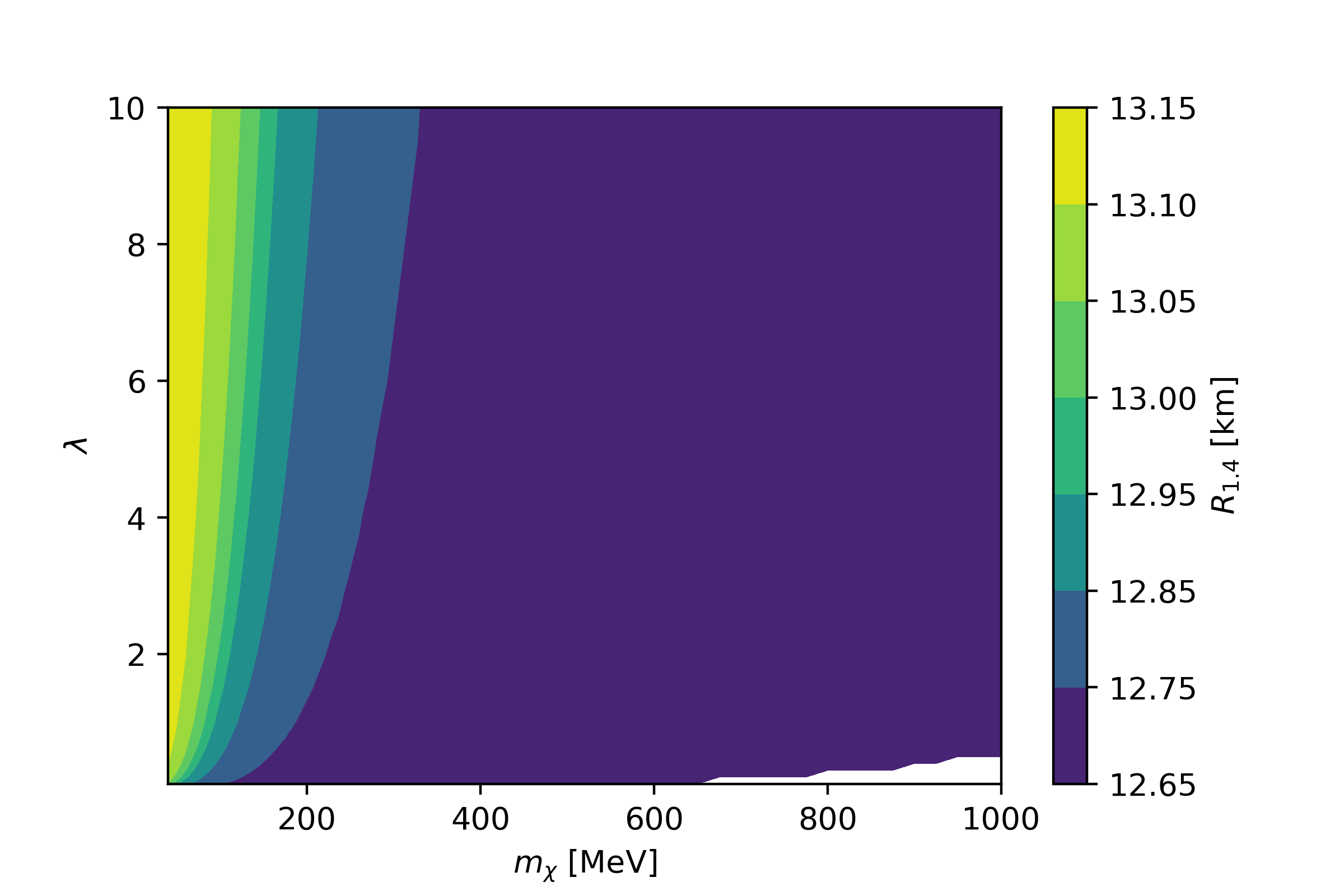}
\includegraphics[width=2.3in]{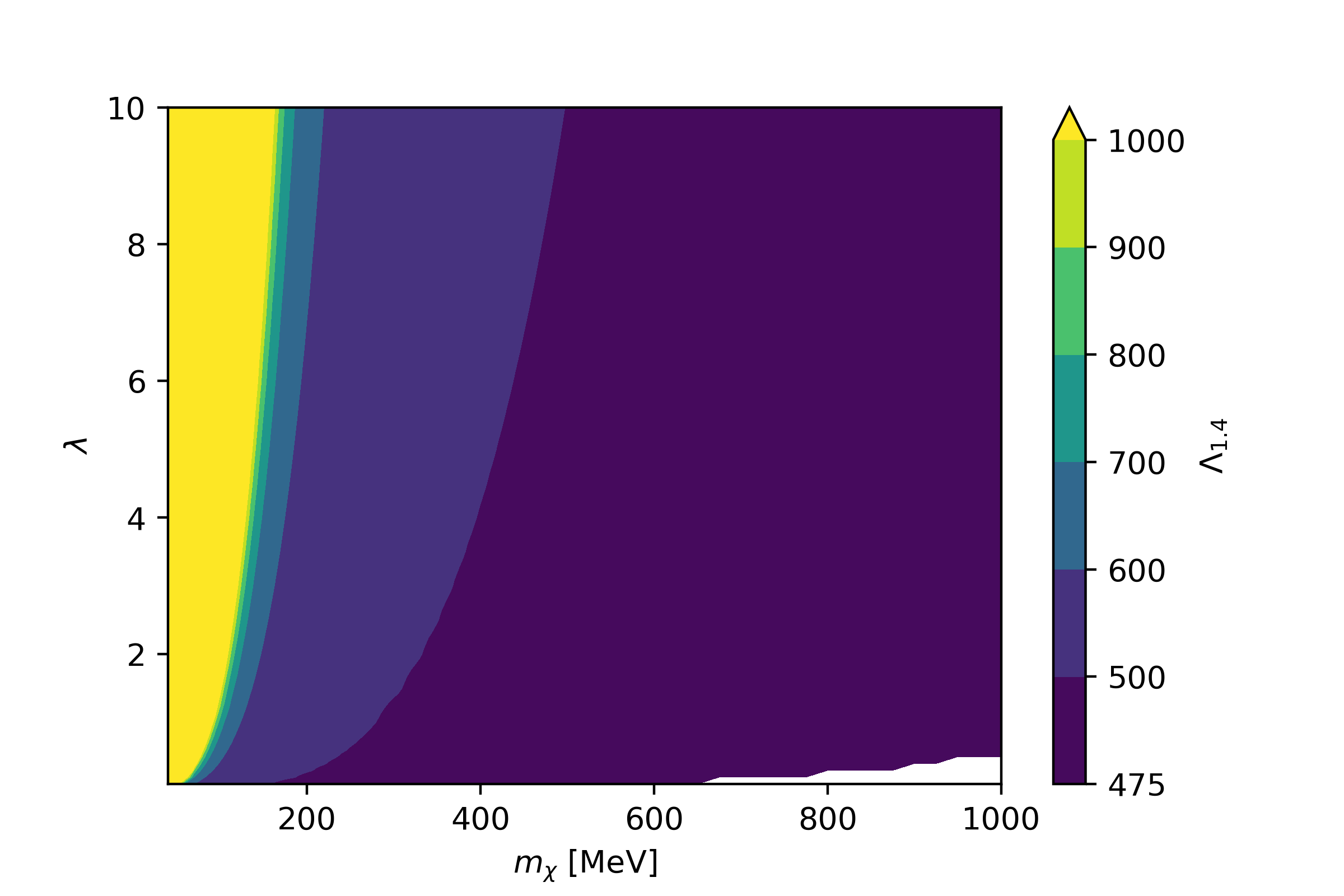}

    \caption{Similar to Fig. \ref{densityIST}, but for DD2 EoS.}\label{densityDD2}
\end{figure}

\end{document}